\documentclass[conf]{new-aiaa}
\usepackage[utf8]{inputenc}

\usepackage{bm}
\usepackage{url}
\usepackage{array}
\usepackage{float}
\usepackage{caption}
\usepackage{amsmath}
\usepackage{relsize}
\usepackage{hyperref}
\usepackage{booktabs}
\usepackage{multirow}
\usepackage{tabularx}
\usepackage{graphicx}
\usepackage{setspace}
\usepackage{makecell}
\usepackage{multicol}
\usepackage{algorithm}
\usepackage{longtable}
\usepackage{subcaption}
\usepackage{algpseudocode}
\usepackage[table, dvipsnames]{xcolor}
\usepackage[letterpaper, margin=1in]{geometry} % default margins on letter paper

\graphicspath{{Figures/}}

\definecolor{light-gray}{gray}{0.92}

\newcolumntype{M}[1]{>{\centering\arraybackslash}m{#1}}

\newcommand{\grayhline}{\arrayrulecolor[gray]{0.8}\hline\arrayrulecolor{black}}

\newcommand{\norm}[1]{\left\lVert #1 \right\rVert}

\usepackage{relsize}  % for \smaller
\newcommand{\code}[1]{%
  \begingroup
  \setlength{\fboxsep}{1pt}% small padding
  \raisebox{0.5pt}[0pt][0pt]{\fcolorbox{light-gray}{light-gray}{\texttt{\smaller #1}}}%
  \endgroup
}

\newcommand{\cg}{\cellcolor{YellowGreen}}
\newcommand{\cgr}{\cellcolor{Gray}}
\newcommand{\cgl}{\cellcolor{red!20!green!100!blue!20!}}

\usepackage{mathtools}

\DeclareCaptionLabelFormat{figlabel}{Figure~\thefigure#1}
\DeclareCaptionSubType*{figure}

\usepackage{fancyhdr}
\title{Flight-Ready Precise and Robust Carrier-Phase GNSS Navigation Software for Distributed Space Systems}

\author{Samuel Y. W. Low, Toby Bell and Simone D'Amico}

\begin{document}

\pagenumbering{arabic}

\setlength{\jot}{5pt} % Local adjustment
\setstretch{0.9}

\maketitle

\begin{center}
    \textit{Stanford University, Department of Aeronautics and Astronautics, Space Rendezvous Laboratory}
\end{center}

% ========================================
% 0. ABSTRACT
% ========================================

\vspace{3mm}
\begin{center}
\begin{minipage}{0.9\textwidth}
\textbf{This paper presents the full requirements analysis, design, development, and testing of high-precision navigation flight software for Distributed Space Systems (DSS) using carrier phase differential GNSS. Five main contributions are made. First, a survey of flown and upcoming DSS missions with stringent precision requirements is conducted, from which a thorough requirements analysis is distilled to guide development and testing. Second, a real-time navigation functional architecture is designed, and adopts a sparse and regularized Consider Kalman Filter with options for numerical stability in-flight. The filter rigorously accounts for uncertainties in process noise, measurement noise, and biases. It tracks float ambiguities with integer resolution where possible. The covariance correlation structure is preserved under all navigation modes, including contingencies and outages. Third, a lightweight, memoryless Fault Detection, Isolation, and Recovery module is developed to guard against anomalous measurements, providing statistical screening and ensuring robust navigation. Fourth, the software architecture is proposed for ease of integration, with strategies presented for modularity and computational efficiency tailored to constrained flight systems. Fifth, a comprehensive test campaign is conducted, mapped to a requirements verification matrix, spanning unit, interface, software-in-the-loop, and real-time hardware-in-the-loop tests, emphasizing gradual test fidelity for efficient fault isolation. Finally, flight-like results are demonstrated using the VISORS mission, due to the generalizability of the VISORS navigation operations, and the stringency which demands sub-centimeter relative position and sub-millimeter-per-second velocity accuracy. This architecture aims to serve as a reference for next-generation DSS missions adopting CDGNSS.}
\end{minipage}
\end{center}

\section*{Table of Abbreviations}

\renewcommand{\arraystretch}{1.2}
\begin{table}[H]
\centering
\scriptsize
\begin{tabular}{@{}p{0.07\textwidth} p{0.26\textwidth} p{0.04\textwidth} p{0.21\textwidth} p{0.03\textwidth} p{0.26\textwidth}@{}}
\grayhline
\textbf{CDGNSS} & Carrier Phase Differential GNSS &
\textbf{CEKF} & Consider Extended Kalman Filter &
\textbf{COM} & Center of Mass \\
\grayhline
\textbf{DDCP} & Double Difference Carrier Phase &
\textbf{DSS} & Distributed Space Systems &
\textbf{ECI} & Earth-Centered Inertial Frame \\
\grayhline
\textbf{EKF} & Extended Kalman Filter &
\textbf{EOP} & Earth Orientation Parameters &   
\textbf{FDIR} & Fault Detection, Isolation and Recovery \\
\grayhline
\textbf{GRAPHIC} & Group and Phase Ionospheric Calibration &
\textbf{IAR} & Integer Ambiguity Resolution &
\textbf{PCO} & Phase Center Offset \\
\grayhline
\textbf{PCV} & Phase Center Variations &
\textbf{RTN} & Radial-Tangential-Normal Frame &
\textbf{SDCP} & Single Difference Carrier Phase \\
\grayhline
\textbf{STM} & State Transition Matrix &
\textbf{UKF} & Unscented Kalman Filter &
\textbf{ZDCP} & Zero Difference Carrier Phase \\
\grayhline
\end{tabular}
\end{table} 
\vspace{-5mm}

% ========================================
% 1. INTRODUCTION
% ========================================

\section{Introduction}
\label{section1}

Precise relative navigation is a key enabler for emerging Distributed Space Systems (DSS), unlocking science not possible with monolithic spacecraft \cite{damico2010thesis}.
Distributed apertures, for instance, enable multi-static sensing, virtual telescopy, radio localization, gravimetry, and interferometry.
Carrier Phase Differential GNSS (CDGNSS), often with Integer Ambiguity Resolution (IAR) \cite{teunissen1994lambda}, have seen increasing adoption in mission concept studies.
Earlier missions like GRACE \cite{kroes2005grace} and TanDEM-X \cite{montenbruck2011tsx} used post-processed batch CDGNSS with IAR on the ground.
Recent advances support real-time onboard execution to meet stricter requirements \cite{damico2010dgps}.
PRISMA (2010) \cite{damico2013prisma} demonstrated real-time onboard CDGNSS with Single Difference Carrier Phase (SDCP) float ambiguity estimation, achieving $\leq$5cm and 1mm/s baseline errors.
Can-X 4/5 (2014) \cite{gnc2010canx45, kahr2018canx45} achieved $\leq$10cm in a CubeSat using an automotive-grade NovAtel OEMV-1G receiver.
These missions serve as pathfinders for low-cost, high-impact science using CDGNSS.
In future, the VISORS virtual telescope will employ L1-only CDGNSS with IAR for mm-level alignment across 40m baselines \cite{visors2021koenig, visors2023aas}, using a NovAtel OEM7 and laser rangefinder, for solar imaging.
The mDOT star shade mission applies CDGNSS with differential ionospheric correction across 500km, for host starlight suppression and exoplanet detection \cite{giralo2021mdot, koenig2015mdot, damicos2022mdot}.
Future pathfinders for spaceborne laser interferometry, like STARI \cite{monnier2024stari} and SILVIA \cite{ito2025silvia}, propose to fuse CDGNSS with optical metrologies for micro-arcsecond imaging of exoplanets and black hole accretion disks.
These missions, summarized in \autoref{tab:survey}, serve as the foundation of our requirements analysis.
The maturation of CDGNSS-based navigation parallels the growing availability and flight heritage of commercial-grade GNSS receivers, as seen in Can-X 4/5 \cite{kahr2018canx45} and VISORS \cite{visors2023aas}.
CDGNSS has been studied for DSS for over two decades \cite{corazzini1997dgps, inalhan2000dgps}, but adoption was initially hindered by costs and compute constraints.
Today, new mission concepts leveraging CDGNSS continue to be proposed \cite{scala2021gnss, shim2024precise, monnier2024stari, ito2025silvia}, as CDGNSS navigation architectures advance.
One example is the \textit{Distributed Multi-GNSS Timing and Localization} (DiGiTaL) flight software, which builds on PRISMA \cite{damico2013prisma} and processes multi-GNSS measurements using a hybrid EKF-UKF architecture.
DiGiTaL includes computational optimizations to the UKF that enable measurement updates every 30s with real-time onboard IAR—never before demonstrated in-flight \cite{giralo2019digital, giralo2021digital}.
Sensor fusion with DiGiTaL has been explored using vision-based angles-only measurements at far-range and image-based pose estimation at close-range \cite{kruger2024rpokit}.
Tight coupling of angles-only and range measurements within the integer search process has demonstrated improved IAR accuracy in noisy environments, potentially enabling IAR in HEO/GEO sidelobe-only scenarios \cite{low2024digital}.
For the $N$-spacecraft case, scalability of the software has been explored briefly \cite{giralo2021digital}, and search-based optimization of attitude pointing profiles have been proposed to support such navigation ops in larger swarms \cite{low2022pointing}.

\renewcommand{\arraystretch}{1.0}
\begin{table}[H]
\centering
\footnotesize
\caption{Survey of contemporary distributed space missions exploiting GNSS for precision applications}
\label{tab:survey}
\begin{tabular}{@{}M{0.05\textwidth} M{0.085\textwidth} M{0.07\textwidth} M{0.09\textwidth} M{0.09\textwidth} M{0.06\textwidth} M{0.27\textwidth} M{0.09\textwidth}@{}}
\toprule
\textbf{Year} & \textbf{Mission} & \textbf{Mass [kg]} & \textbf{Purpose} & \textbf{Orbit} & \textbf{Baseline} & \textbf{Methodology} & \textbf{Accuracy} \\
\midrule
2002 & \makecell{GRACE \\ \cite{kroes2005grace}}
& 487 × 2 & Gravimetric Modeling & Polar LEO, 500km & 170 to 270km & Ground post-processed CDGNSS with IAR, with K-Band range aiding & $\sim$ 1mm \\
\grayhline
2010 & \makecell{TanDEM-X \\ \cite{montenbruck2011tsx}} 
& 1230 \& 1330 & InSAR \& Elevation Mapping & Polar LEO, 514km & 250 to 500m & Ground post-processed CDGNSS with IAR, with ground satellite laser ranging & $\sim$ cm \\
\grayhline
2010 & \makecell{PRISMA \\ \cite{damico2012safe, damico2013prisma}} 
& 150 \& 40 & Formation Flying Demo & LEO, 700km & 100m to 2km & Real-time onboard CDGNSS float ambiguity estimation & $\leq$ 10cm \\
\grayhline
2014 & \makecell{Can-X 4/5 \\ \cite{gnc2010canx45, kahr2018canx45}} 
& 6.5 × 2 & Formation Flying Demo & Polar LEO, 650km & 50m to 1km & Real-time onboard CDGNSS float ambiguity estimation & $\leq$ 10cm \\
\grayhline
2018 & \makecell{GRACE-FO \\ \cite{kornfield2019gracefo, xia2021gracefo}} 
& 600 × 2 & Gravimetric Modeling & Polar LEO, 490km & 220km & Ground post-processed CDGNSS with IAR, with onboard laser interferometry & $\leq$ 1mm \\
\grayhline
2022 & \makecell{CPOD \\ \cite{bowen2015cpod, roscoe2018cpod}} 
& 4.5 × 2 & RPOD Tech Demo & LEO, $\sim$ 500km & 361m to 997km & Real-time onboard fusion of GPS pseudorange with crosslink range and camera measurements & $\sim$ meters \\
\grayhline
2024 & \makecell{PROBA-3 \\ \cite{llorente2013proba, ardaens2013proba3gps, enderle2019proba}} 
& 340 \& 200 & Distributed Telescopy & HEO, 600 to 60,530km & 200m to 2km & Ground post-processed GNSS at perigee; laser ranging, vision and shadow position sensors at apogee & $\sim$ mm \\
\grayhline
$\sim$ 2026 & \makecell{VISORS \\ \cite{visors2021koenig, visors2023aas}} 
& 10 \& 11 & Distributed Telescopy & Polar LEO, 500 - 600km & 40 to 200m & Real-time onboard CDGNSS with IAR, with laser ranging during observations & $\sim$ mm \\
\grayhline
$\sim$ 2027 & \makecell{SNUGLITE III \\ \cite{hwang2025snuglite3, shim2024precise, shim2024rtk}} & 3.6 × 2 & GPS-RO and Tech Demo & SSO, 500--600km & $\leq$ 1km & CDGNSS with IAR, GPS-only attitude determination, drag-only control & $\sim$ mm \\
\grayhline
Proposed & \makecell{mDOT \\ \cite{koenig2015mdot, damicos2022mdot, giralo2021mdot}}
& 246 total & Exoplanet Detection & LEO, $\sim$ 500km & 500km & Real-time onboard CDGNSS with IAR & $\sim$ mm \\
\grayhline
Proposed & \makecell{STARI \\ \cite{monnier2024stari}} 
& 10 × 2 & Interferometry Demo & LEO, $\sim$ 500km & 10 to 100m & Real-time onboard CDGNSS with IAR, possibly aided by LED-based image alignment measurements & $\sim$ mm \\
\grayhline
Proposed & \makecell{SILVIA \\ \cite{ito2025silvia}} 
& 100 × 3 & Interferometry Demo & LEO, 500 - 600km & $\sim$ 100m & Float-only CDGNSS (precise mode), laser and beam position sensors (ultra-precise) & $\sim$ mm - \textmu m \\
\bottomrule
\end{tabular}
\end{table}

Despite advances in CDGNSS-based navigation, challenges persist.
Emerging missions such as VISORS \cite{visors2021koenig, visors2023aas}, mDOT \cite{koenig2015mdot, damicos2022mdot}, and STARI \cite{monnier2024stari} demand unprecedented accuracy and robustness yet have not achieved the required performance \cite{monnier2019roadmap}.
A key gap in literature is the absence of an end-to-end architecture that:
(i) generalizes to future DSS missions,
(ii) rigorously models state, measurement, and dynamic uncertainties under off-nominal conditions,
(iii) ensures robustness to navigation errors and software faults,
(iv) remains computationally efficient for frequent measurement updates in flight, and
(v) closes the requirements loop through comprehensive testing.
This work formalizes a DiGiTaL v2 flight software to address these gaps.
\autoref{section1} reviews relevant missions for a generalizable requirements analysis.
\autoref{section2} refines requirements into Interfacing, Functional, and Performance categories.
\autoref{section3} presents a filter design that rigorously treats state, dynamics, and measurement uncertainty, introducing a regularized Joseph-like Consider Kalman Filter.
\autoref{section4} integrates a lightweight, memoryless FDIR module leveraging statistical guarantees to handle anomalies such as undetected carrier phase cycle slips.
\autoref{section5} details a modular, compute-optimized software architecture with full execution logic.
\autoref{section6} outlines a structured, requirements-driven test campaign spanning unit to real-time GNSS hardware-in-the-loop tests, demonstrated using the VISORS mission concept.
The VISORS campaign is chosen as it generalizes to future DSS missions and establishes a benchmark for flight software evaluation.
State-of-the-art models for dynamics, measurements, subsystem latencies, and body-frame uncertainties are incorporated.
Overall, this work delivers a traceable, closed-loop framework from requirements to compliance testing, advancing the readiness of CDGNSS-based navigation for next-generation DSS missions.

\pagebreak

% ========================================
% 2. NAVIGATION REQUIREMENTS
% ========================================

\section{Navigation Requirements}
\label{section2}

% ========================================
% 2.1 REQUIREMENTS GENERALIZATION
% ========================================

\subsection{Requirements Generalizations and Assumptions}

\begin{figure}[H]
	\centering
    \includegraphics[width=0.7\textwidth]{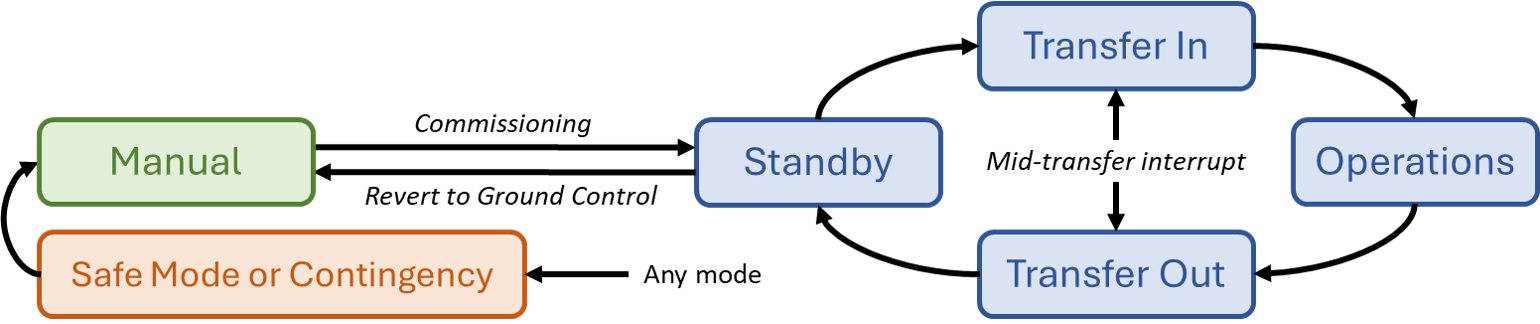}
    \caption{Generalized \textit{modes} that spacecraft in a distributed system may adopt.}
    \label{fig:modes}
\end{figure}

This section presents a generalized set of CDGNSS-based navigation requirements catered to a set of generalized operating modes applicable to missions surveyed. Clearly defined requirements are essential for guiding design decisions, ensuring validation, and avoiding costly issues later in development lifecycle \cite{kapurch2010nasa}. The first necessary assumption made is that the spacecraft in question are cooperative, communicable, and operating under main-lobe GNSS reception. For guidance, a general DSS mission may adopt the following modes in \autoref{fig:modes},

A \textit{mode} is a logical abstraction of a spacecraft's current state meant to fulfil objectives or situational needs at each phase of the mission. With regards to software, different modes can adopt different configuration parameters and tailor compute resources to each phase of the mission. With regards to ops, different modes are often also characterized by different operating baselines, relative orbital elements, and passive safety margins \cite{koenig2018safety}. The purpose of the abstraction in \autoref{fig:modes} is to provide a context in which we present requirements, and cater them in accordance to the generalized stringency of each mode. Broadly, Standby mode is purposed for subsystem health monitoring and software checkouts prior to Transfer, which is a relative orbit reconfiguration trajectory, into Operations, which is purposed for execution of the payload. Generally, navigation accuracy requirements and passive safety margins are most stringent during Operations, and less so during Standby. The transition between Manual to and from Standby also imposes unique requirements on the handling of mode transitions and performance, if the DSS is in acquisition or recovery (where the CDGNSS may not be available). Such mode-based operations mirror that seen in several recent or proposed missions such as VISORS \cite{visors2021koenig, visors2023aas}, mDOT \cite{koenig2015mdot, damicos2022mdot}, and PROBA-3 \cite{llorente2013proba}.

% ========================================
% 2.2 REQUIREMENTS: IDENTIFICATION
% ========================================

\subsection{Requirements Identification}

This survey of contemporary and emerging distributed space missions \autoref{tab:survey} forms the foundation for identifying navigation flight software requirements that are both mission-relevant and broadly generalizable. By analyzing commonalities across mission objectives, operating baselines, and CDGNSS-based navigation methodologies, a unified set of requirements was distilled to guide the design of a robust and adaptable navigation architecture.

\begin{figure}[H]
	\centering
    \includegraphics[width=1.0\textwidth]{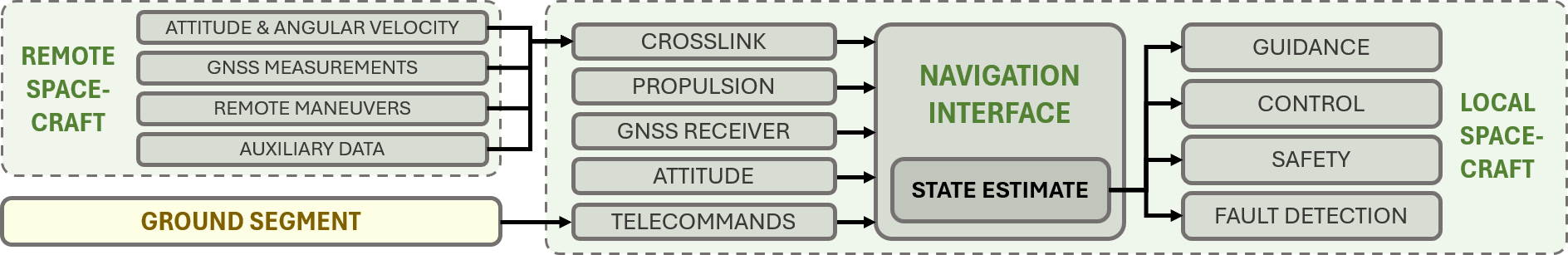}
    \caption{Assumed interface architecture applied in the context of identifying interface requirements.}
    \label{fig:interfaces}
\end{figure}

Requirements are segmented into Interfacing, Functional, and Performance in \autoref{tab:reqs}. The specification of interface and functional requirements implicitly assume an interface architecture depicted in \autoref{fig:interfaces}. Such a prescribed interface has been adopted for several missions that employ CDGNSS, such as PRISMA \cite{damico2010thesis}, Can-X 4/5 \cite{gnc2010canx45} \cite{kahr2018canx45}, DWARF \cite{giralo2020dwarf}. It is generalizable to future missions such as VISORS \cite{visors2023aas} and STARI \cite{monnier2024stari} as well.

\begin{figure}[H]
	\centering
    \includegraphics[width=0.9\textwidth]{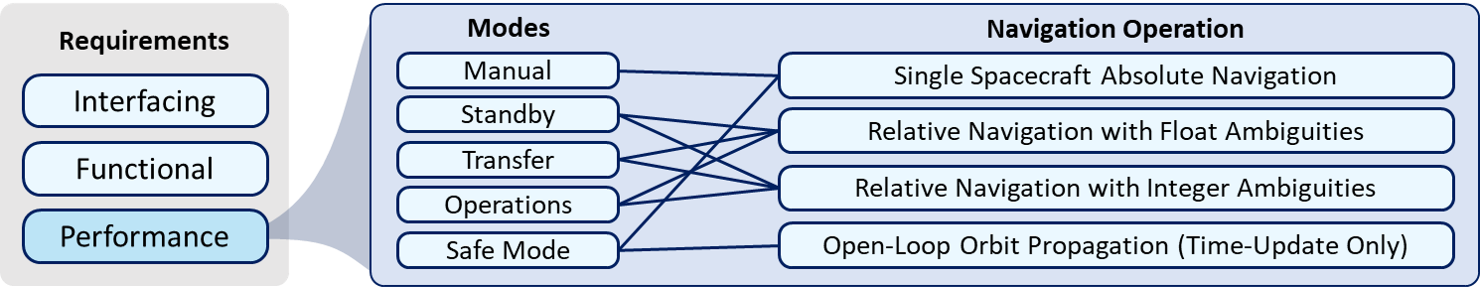}
    \caption{Mapping of navigation operating requirements to the spacecraft modes}
    \label{fig:modes-map}
\end{figure}

The specification of performance requirements are segmented according to the types of navigation operations in \autoref{fig:modes-map}. These are derived from navigation requirements from each mode of \autoref{fig:modes}. The line mappings are a suggestive (and not prescriptive) indication on whether requirements for each form of navigation operation are being driven by mode requirements.

\renewcommand{\arraystretch}{1.5} % Adjust row spacing
{\small
% ========================================
% 2.2.1 REQUIREMENTS: INTERFACES
% ========================================
\begin{longtable}{@{\extracolsep{\fill}}
>{\centering\arraybackslash}p{0.05\textwidth}
>{\arraybackslash}p{0.8\textwidth}
>{\centering\arraybackslash}p{0.1\textwidth}
@{}}
  \hline \hline
  & \textbf{Interfacing Requirements} & \textbf{Parameter} \\
  \hline \hline
  \endfirsthead
    R1.1\label{req:1.1} & 
    \textbf{Local GNSS:} Provide an interface for receiving local GNSS measurements at intervals $\tau_{meas}$, GNSS ephemeris at intervals $\leq$ 2 hours, and receiver health data, with validation via checksums. &
    $\tau_{meas}$ \\
    R1.2\label{req:1.2} & 
    \textbf{Local Attitude:} Provide an interface for receiving local attitude at intervals $\tau_{att}$. If no angular velocity provided for propagation, then $\geq 2$ attitude states are to be buffered for inter/extrapolation over short arcs. &
    $\tau_{att}$ \\
    R1.3\label{req:1.3} & 
    \textbf{Local Propulsion:} Provide an interface for receiving time-tagged local spacecraft maneuver plans to be buffered and applied during navigation filter time updates. &
    - \\
    R1.4\label{req:1.4} & 
    \textbf{Crosslink In/Out:} Provide an interface for receiving crosslink \textit{from} the remote spacecraft (and \textit{to}, if the remote also executes onboard navigation), at intervals $\tau_{link}$, with validation via checksums. Crosslink packets may include GNSS measurements, attitudes, maneuver plans, status bytes, and state estimates. &
    $\tau_{link}$ \\
    R1.5\label{req:1.5} & 
    \textbf{Ground Segment:} Provide an interface for receiving ground tele-commands \textit{e.g.} filter configurations, and auxiliary data such as GPST leap seconds and Earth Orientation Parameters (EOPs). &
    - \\
    R1.6\label{req:1.6} &
    \textbf{Outputs:} Provide an interface to return or call-back to the host software with state estimates at intervals $\tau_{update}$. If state estimates are anomalous, an incapability flag with status bytes should be returned instead. &
    $\tau_{update}$ \\
  \hline
\end{longtable}
\vspace{-8mm}
% ========================================
% 2.2.2 REQUIREMENTS: FUNCTION
% ========================================
\begin{longtable}{@{\extracolsep{\fill}}
>{\centering\arraybackslash}p{0.05\textwidth}
>{\arraybackslash}p{0.8\textwidth}
>{\centering\arraybackslash}p{0.1\textwidth}
@{}}
  & \textbf{Functional Requirements} \\
  \hline
    R2.1\label{req:2.1} &
    \textbf{State Covariance:} Maintain the state estimate with a \textit{positive-definite} covariance matrix that is reflective of the true state errors. &
    - \\
    R2.2\label{req:2.2} &
    \textbf{State Dynamics Fidelity:} Filter dynamics must provide sufficient fidelity in orbit perturbations without exceeding the margin of compute resources \textit{i.e.} the filter time update and measurement update can be still be completed within $\tau_{update}$, tested on a representative flight computer. &
    - \\
    R2.3\label{req:2.3} & 
    \textbf{Screening for Data Health:} The flight software shall...
    \begin{itemize}
        \item Reject measurements with poor $C/N_0$ ratios. A recommended minimum value is 45 dB-Hz \cite{gpstextbook2006}.
        \item Reject local/remote maneuvers and measurements with outdated time-tags, with respect to filter time.
        \item Reject observed ranges and range rates with impossible magnitudes that lie outside expected bounds.
        \item Reject stale GNSS ephemerides ($\geq$ 2 hours) or with bad Keplerian elements (\textit{e.g.} negative eccentricity).
        \item Reject bad attitude state representations \textit{e.g.} quaternions which violate unit norm constraints, and emit warning flags if attitudes received are not indicative of the pointing profile in the current mode.
    \end{itemize}
    &
    - \\
    R2.4\label{req:2.4} &
    \textbf{Managing Telecommands and Telemetry:} Navigation remains uninterrupted during telecommand uplink (\textit{e.g.} filter tuning parameters, leap-second update, or new EOPs), and telemetry downlink. &
    - \\
    R2.5\label{req:2.5} &
    \textbf{Managing Mode Switches:} Navigation remains uninterrupted in the event of a system-wide role or mode switch. A handshake should be conducted between instances of the navigation flight software onboard and with remote spacecraft, if necessary, to prevent role or mode inconsistencies across the formation. &
    - \\
    R2.6\label{req:2.6} &
    \textbf{Managing Measurement Latency:} Provide a decision logic for transitioning to local-only spacecraft absolute state estimation, if an extended outage from the remote crosslink is experienced beyond some time horizon $\tau_{outage}$. A decision logic for recovery back to precise relative state estimation should also be provided. &
    $\tau_{outage}$ \\
    R2.7\label{req:2.7} & 
    \textbf{Managing Attitude Latency:} Propagate body-to-inertial attitude coordinates to the time-tag of the measurements at filter updates. If angular velocities are absent, perform inter/extrapolation instead \cite{shoemake1985slerp}. &
    - \\
    R2.8\label{req:2.8} &
    \textbf{Fault Detection in Computations:} Design against numerical instabilities. Automated unit tests are recommended. Catch problems involving potential divisions by zero, negative eccentricities, square roots of negative real numbers, integer over/underflow, precision loss and ill-conditioned matrices. Conversions of units, state representations, coordinate frames, time units and time scales must be verified. &
    - \\
    R2.9\label{req:2.9} &
    \textbf{Cycle Slip Detection:} Provide a mechanism for screening cycle slips (if the receiver does not already provide cycle slip flags), and provide a remedy if slip exists. Recommended techniques in literature include (i) consistency checks between the carrier phase and Doppler measurements, (ii) detection of discontinuities in double-differenced carrier phase, or (iii)  impulses in triple-differenced carrier phase \cite{gpstextbook2006}. &
    - \\
  \hline
\end{longtable}
\vspace{-8mm}
% ========================================
% 2.2.3 REQUIREMENTS: PERFORMANCE
% ========================================
\begin{longtable}{@{\extracolsep{\fill}}
>{\centering\arraybackslash}p{0.05\textwidth}
>{\arraybackslash}p{0.8\textwidth}
>{\centering\arraybackslash}p{0.1\textwidth}
@{}}
  & \textbf{Performance Requirements} \\
  \hline
    R3.1\label{req:3.1} &
    \textbf{Absolute Navigation:} Maintain a single-point positioning $1\sigma$ error of $\leq e_{abs}$ m and $\Dot{e}_{abs}$ m/s. &
    $e_{abs}$, $\Dot{e}_{abs}$ \\
    R3.2\label{req:3.2} &
    \textbf{Relative Navigation (Pre-IAR):} Maintain a relative positioning $1\sigma$ error of $\leq \delta e_{float}$ m and $\delta \Dot{e}_{float}$ m/s at convergence of estimated single-differenced float ambiguities. &
    $\delta e_{float}$, \linebreak $\delta \Dot{e}_{float}$ \\
    R3.3\label{req:3.3} &
    \textbf{Relative Navigation (Post-IAR):} Maintain a relative positioning $1\sigma$ error of $\leq \delta e_{iar}$ m and $\delta \Dot{e}_{iar}$ m/s of single-differenced integer ambiguities using a $\geq 99\%$ probability of success metric. &
    $\delta e_{iar}$, \linebreak $\delta \Dot{e}_{iar}$ \\
    R3.4\label{req:3.4} &
    \textbf{Relative Navigation (Operations):} In addition to \hyperref[req:3.1]{R3.1} - \hyperref[req:3.3]{R3.3}, instrument or payload alignment requirements may be imposed along the longitudinal boresight axis of the instrument, or across the lateral plane perpendicular to the longitudinal, or both, given as the $1\sigma$ error $\delta e_{lon}$ and $\delta e_{lat}$. &
    $\delta e_{lon}$, \linebreak $\delta e_{lat}$ \\
    R3.5\label{req:3.5} &
    \textbf{Open-Loop Orbit Propagation:} If state updates are necessary during measurement outages, then the open-loop propagation of the relative trajectory must be within an error tolerance of $\delta e_{open}$ in the time horizon $\tau_{open}$. &
    $\delta e_{open}$, \linebreak $\tau_{open}$ \\
    R3.6\label{req:3.6} &
    \textbf{Robustness to Offset-from-Origin Biases:} If there exists a poorly observable bias $\Vec{p}$, between the chosen origin of the spacecraft body frame and the GNSS antenna phase center, then the performance requirements must be tolerant to the knowledge error of this bias given by $\Delta \Vec{p}_{req}$. &
    $\Delta \Vec{p}_{req}$ \\
    R3.7\label{req:3.7} &
    \textbf{Robustness to Over/Under-Actuation Maneuvers:} Requirements \hyperref[req:3.1]{R3.1} - \hyperref[req:3.5]{R3.5} must remain satisfied in the event of over or under-actuation between the known maneuvers and true executed maneuvers up to some margin $M_{man}$. &
    $M_{man}$ \\
    R3.8\label{req:3.8} &
    \textbf{Robustness to Measurement Noise Variations:} Requirements \hyperref[req:3.1]{R3.1} - \hyperref[req:3.5]{R3.5} must remain satisfied in the event of variations between modeled and actual measurement noise up to some margin $M_{meas}$. This variation is to be tested using a representative receiver under noise-configurable emulated GNSS signals. &
    $M_{meas}$ \\
  \hline \hline
\caption{Identified requirements of a generalized CDGNSS-based navigation flight software for DSS missions.}
\label{tab:reqs}
\end{longtable}
}
\vspace{-5mm}

Deciding on the Interfacing and Functional requirement parameters in \autoref{tab:reqs} may depend on the mission context and domain knowledge of other subsystems interfacing with navigation.
$\tau_{meas}$ is limited by available observation rate of the GNSS receiver;
$\tau_{att}$ is limited by the bus ADCS;
$\tau_{link}$ is limited by crosslink availability;
$\tau_{outage}$ requires domain knowledge on the expected crosslink latency;
and while $\tau_{update}$ can be set as an ops requirement, it is also limited by $\tau_{meas}$ and $\tau_{link}$.
Deciding on the Performance requirements parameters requires a careful assessment of the navigation error tolerance necessary to meet mission-specific requirements.
An example of requirements parametrization is drawn from the VISORS mission: critical considerations for science observations are the focal length control accuracy, image drift rate, and attitude pointing accuracies, all during telescopic alignment.
The science requirement is $\geq 20\%$ likelihood of success per observation attempt, and hence $\sim99\%$ likelihood of $\geq 1$ successful observation after 20 attempts \cite{visors2021koenig, visors2023aas}.
In total, these translate into overall Interfacing and Functional requirements parameters in \autoref{tab:req-visors-timing} and Performance requirements parameters in \autoref{tab:req-visors-performance}, specific to the VISORS mission:

\begin{table}[H]
\centering
\caption{Interfacing and Functional requirements drawn from the VISORS Navigation Flight Software}
\label{tab:req-visors-timing}
\begin{tabular}{ccccc}
  \hline
  $\tau_{meas}$ & $\tau_{att}$ & $\tau_{link}$ & $\tau_{update}$ & $\tau_{outage}$ \\
  \hline
  10s & 10s & 10s & 10s & 4s \\
  \hline
  \end{tabular}
\end{table}

\begin{table}[H]
\centering
\footnotesize
\caption{Performance Requirements drawn from the VISORS Navigation Flight Software}
\label{tab:req-visors-performance}
\begin{tabular}{ccccccccccccc}
  \hline
  $\Vec{e}_{abs}$ &
  $\Dot{\Vec{e}}_{abs}$ &
  $\delta \Vec{e}_{float}$ &
  $\delta \Dot{\Vec{e}}_{float}$ &
  $\delta \Vec{e}_{int}$ &
  $\delta \Dot{\Vec{e}}_{int}$ &
  $\delta e_{lon}$ &
  $\delta e_{lat}$ &
  $\delta e_{open}$ &
  $\tau_{open}$ &
  $\Delta \Vec{p}_{req}$ &
  $M_{man}$ &
  $M_{meas}$ \\
  \hline
  5m &
  5mm/s &
  10cm &
  1mm/s &
  1cm &
  100$\mu$m/s &
  1.5cm &
  1.75cm &
  1m &
  1 orbit &
  5cm & 
  $\pm 100 \%$ &
  $\pm 300 \%$ \\
  \hline
  \end{tabular}
\end{table}

\pagebreak

% ========================================
% 2.3 REQUIREMENTS: DISCUSSION
% ========================================

\subsection{Implications of Precision Requirements on the Receiver Phase Center and Spacecraft Attitude}

Stringent precision requirements mandate consideration of the receiver antenna Phase Center Offset (PCO) from the body Center Of Mass (COM), and the directionally dependent Phase Center Variations (PCV).
The PCO is the average point where GNSS signals are received from all signal directions, as opposed to the mechanical center \cite{gpstextbook2006}.
The estimated positions of the spacecraft typically refer to the COM for ease of reference in orbit and attitude control.

\begin{figure}[H]
	\centering
    \includegraphics[width=0.95\textwidth]{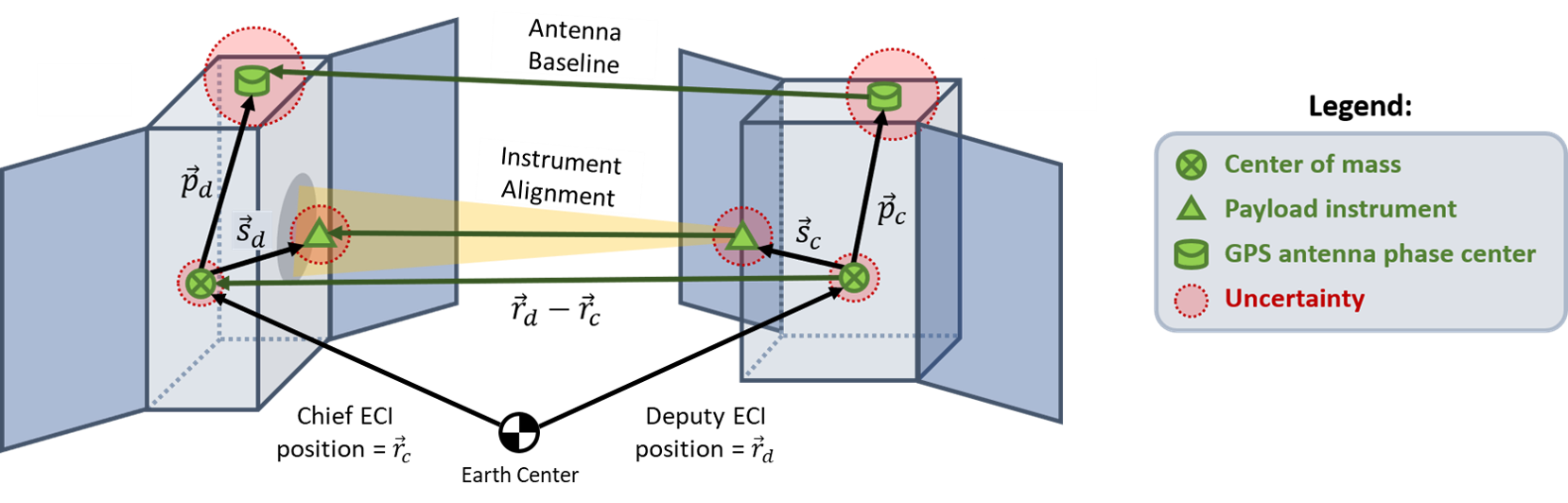}
    \caption{Illustration of body-frame component position uncertainties in the VISORS spacecraft (Source: \cite{visors2023aas}).}
    \label{fig:req-visors-pco}
\end{figure}

However, GNSS measurements are observed at the PCO.
Hence, the only observable point in the body frame through GNSS measurements is the PCO, whereas the COM has poor observability, as per \autoref{fig:req-visors-pco}.
The components of the COM-to-PCO vector $\Vec{p}$ are hence poorly observable, but must be accounted for in modeled CDGNSS measurements \cite{psiaki2007cdgps}.
A known \textit{a priori} vector $\Vec{p}_{static}$ from the body origin to a static antenna center is typically recorded from a mechanical model as an initial guess.
Then $\Vec{p}$ is modeled in body-frame coordinates as

\begin{equation}
    \Vec{p} = \Vec{p}_{static} + \Delta \Vec{p}
\end{equation}

where $\Delta \Vec{p}$ is assumed zero-mean Gaussian with cm-level uncertainty.
Ground calibration of the PCO and PCV in an anechoic chamber may reduce uncertainty.
However, calibration campaigns may be infeasible if the system integrator lacks the facilities and resources.
Often, DSS pathfinder missions employ cost-efficient platforms \textit{i.e.} CubeSats \cite{damico2010thesis, damico2013prisma, gnc2010canx45, kahr2018canx45, visors2021koenig, visors2023aas, monnier2024stari} using commercial grade components with limited calibration.
Even with calibration, time-varying uncertainties in COM-to-PCO evolve in flight.
For instance, non-rigid body dynamics (solar panels, fuel slosh etc) contribute to an evolving COM, while signal-in-space conditions introduce time-varying effects on the PCO.
Recent work suggests that highly agile attitude maneuvers improve the observability of $\Delta \Vec{p}$ if estimated directly via state augmentation \cite{gutsche2024pco}.
However, this assumes a static COM-to-PCO.
Furthermore, frequent attitude calibration maneuvers may be infeasible if attitude profiles are restricted.
Therefore, it is necessary to account for the influence of COM-to-PCO bias errors on navigation precision within the filter.
Error budgeting of body-frame bias errors in DSS are not commonly considered in literature.
An exception is the proposed Cal X-1 mission \cite{lowe2022errbudget}. Such accounting can be pursued in two ways.
First, additive or multiplicative measurement underweighting is one recommended practice in literature \cite{zanetti2010underweight, nasa2018filter}.
Second, treating bias errors as Consider parameters \cite{schmidt1966cekf} using a Consider Kalman Filter \cite{jazwinski1970filter, woodbury2010cekf} could accurately incorporate their uncertainties into the full state covariance.
This reduces filter over-confidence, prevent larger-than-expected prefit residuals, and hence protect against filter divergence \cite{schlee1967divergence}.
This is a critical motivation behind the choice of adopting the Consider Kalman Filter in \autoref{section3}.

Accounting for $\Vec{p}$ in the measurement model requires minimizing latency between the body-to-inertial attitude time and the filter time, as per \hyperref[req:2.7]{R2.7}.
This minimizes contributions to residuals due to the uncompensated lever arm motion of $\Vec{p}$.
A simple Euler step as per \autoref{eq:att-step} can mitigate much of the error.
Let the last known attitude packet at time $t_{att}$ be the set $[\omega(t_{att}), \mathbf{q}(t_{att})]$, where $\mathbf{q}(t_{att})$ is the body-to-inertial quaternion and $\omega(t_{att})$ is the body-to-inertial angular velocity expressed in the body frame.
Then, let $\Delta t_{att} \in \mathbb{R}$ be the difference between the current filter time and $t_{att}$,

\begin{equation}
\label{eq:att-step}
    \mathbf{q}( t_{att} + \Delta t_{att} ) = \delta \mathbf{q} \otimes \mathbf{q}(t_{att}) \ ,
    \qquad
    \delta \mathbf{q} = 
    \begin{bmatrix}
        \cos{\left( \frac{1}{2} \norm{\omega(t_{att})} \Delta t_{att} \right)} \\
        \hat{\omega} \sin{\left( \frac{1}{2} \norm{\omega(t_{att})} \Delta t_{att} \right)} \\
    \end{bmatrix}
\end{equation}

where $\hat{\omega} \in \mathbb{R}^3$ is the normalized direction of $\omega(t_{att})$ and $\otimes$ is the quaternion multiplier. If angular velocity data is absent, one may instead do quaternion interpolation \textit{i.e.} SLERP \cite{shoemake1985slerp}, as was implemented for the PRISMA mission \cite{damico2010thesis}.

\pagebreak

% ========================================
% 3. NAVIGATION ARCHITECTURE
% ========================================

\section{Navigation Functional Architecture}
\label{section3}

% ========================================
% 3.1 NAVIGATION ARCHITECTURE: OVERVIEW
% ========================================

\subsection{Overview of Architecture}

\begin{figure}[ht]
	\centering
    \includegraphics[width=1.0\textwidth]{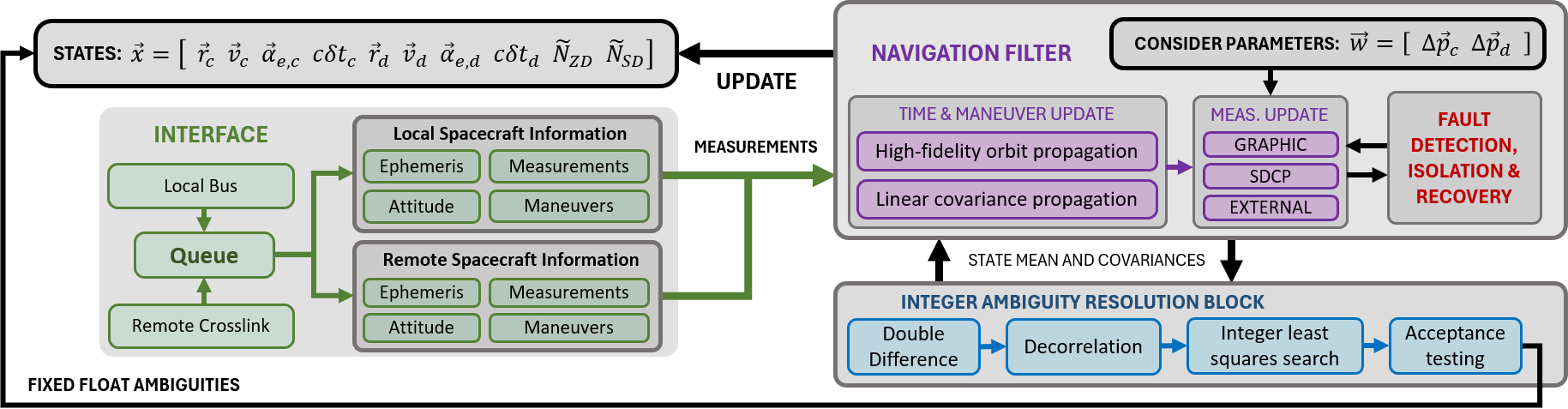}
    \caption{Navigation functional architecture based on CDGNSS with IAR, for a two-spacecraft state vector}
    \label{fig:architecture}
\end{figure}

The navigation architecture builds on the DiGiTaL flight software \cite{giralo2019digital, giralo2021digital}, with heritage from PRISMA \cite{damico2012safe, damico2013prisma}, and is hereafter referred to as DiGiTaL v2. Absolute position estimation uses Group and Phase Ionospheric Calibration (GRAPHIC) measurements \cite{yunck1993graphic}, which combine pseudorange and Zero-Difference Carrier Phase (ZDCP) to mitigate ionospheric errors. Relative baseline estimation uses Single-Difference Carrier Phase (SDCP) measurements shared over a crosslink. The architecture in \autoref{fig:architecture} comprises four blocks: (i) the data interface, (ii) navigation filter, (iii) IAR, and (iv) FDIR.

The state vector per spacecraft contains the inertial position $\Vec{r}_i$ and velocity $\Vec{v}_i$ of the spacecraft COM, empirical accelerations $\Vec{\alpha}_{e,i}$, and receiver clock offsets $c \delta t_i$. 48 channels total are set for estimating ZDCP and SDCP float ambiguities $\Tilde{N}_{ZD}$ and $ \Tilde{N}_{SD}$. Receiver clock offsets and empirical accelerations are modeled temporally as first-order Gauss-Markov processes.

\begin{equation}
    \Vec{x} = \left[ \
    \Vec{r}_c, \ \Vec{v}_c, \ 
    \Vec{\alpha}_{e,c}, \ c \delta t_c, \
    \Vec{r}_d, \ \Vec{v}_d, \ 
    \Vec{\alpha}_{e,d}, \ c \delta t_d, \
    \Tilde{N}_{ZD}, \Tilde{N}_{SD} \ \right]
    \ , \quad
    \Vec{w} = \left[ \ \Delta \Vec{p}_c, \ \Delta \Vec{p}_d \ \right]
    \ , \quad
    \mathbf{\Sigma} =
    \begin{bmatrix}
        \mathbf{\Sigma}^{(x)} & \mathbf{\Sigma}^{(xw)} \\
        \mathbf{\Sigma}^{(wx)} & \mathbf{\Sigma}^{(w)} \\
    \end{bmatrix}
\label{eq:state-vector}
\end{equation}

For a two-spacecraft formation (chief $c$ and deputy $d$), the state mean and consider parameters have dimensions $\Vec{x} \in \mathbb{R}$$^{68}$ and $\Vec{w} \in \mathbb{R}$$^{6}$. The full covariance block matrix, including consider parameters, comprises the state covariance $\mathbf{\Sigma}^{(x)}$; consider covariance $\mathbf{\Sigma}^{(w)}$; and cross-covariances between states and consider parameters $\mathbf{\Sigma}^{(xw)} = \mathbf{\Sigma}^{(wx)\top}$.
Ambiguities are indexed to the rear in \autoref{eq:state-vector} to allow for further computational optimization as shown in \autoref{section5}.

% ========================================
% 3.2 NAVIGATION ARCHITECTURE: FILTER SPECS
% ========================================

\subsection{Stable and Regularized Consider Extended Kalman Filter}

The Consider Extended Kalman Filter (CEKF) propagates the uncertainties of $\Delta \Vec{p}_c, \Delta \Vec{p}_d$ into the state covariance \cite{schmidt1966cekf, jazwinski1970filter}. However, the standard minimum covariance form \cite{woodbury2010cekf} can pose numerical stability issues, detrimental to real-time flight.
The key contribution in this section is the modification of the CEKF into a Positive-Definite (PD) preserving Joseph form \cite{low2025cekf}, with regularization of the innovation covariance for additional stability.
Modifications are in \autoref{eq:meas-update-1} and \autoref{eq:meas-update-6}.
The time update equations span from \autoref{eq:time-update-1} to \autoref{eq:time-update-7}.
Let $f$ be the non-linear orbit propagation force model, where $\Dot{\Vec{x}}=f(t, \Vec{x}, \mathbf{M})$ includes higher order perturbations; $\mathbf{M} \in \mathbb{R}^{r \times 3}$ is a maneuver buffer of size $r$; $\mathbf{\Phi}_{t-1}$ is the State Transition Matrix (STM) at time $t-1$; and $\mathbf{Q}_t$ is the process noise matrix at time $t$. Then, 

\begin{align}
    \Vec{x}_{t|t-1} & = \text{Propagate}
    \left( t, \Vec{x}_{t-1|t-1}, \mathbf{M}, f \right)
    \label{eq:time-update-1} \\
    \mathbf{\Sigma}_{t|t-1}^{(x)} & =
    \mathbf{\Phi}_{t-1}^\top \mathbf{\Sigma}_{t-1|t-1}^{(x)} \mathbf{\Phi}_{t-1} + \mathbf{Q}_t
    \label{eq:time-update-2} \\
    \mathbf{\Sigma}_{t|t-1}^{(xw)} & =
    \mathbf{\Sigma}_{t|t-1}^{(wx)\top} =
    \mathbf{\Phi}_{t-1}^\top \mathbf{\Sigma}_{t-1|t-1}^{(xw)}
    \label{eq:time-update-3}
\end{align}

Empirical accelerations are decayed based on a first-order Gauss-Markov \cite{damico2010thesis, nasa2018filter, stacey2021adaptive, giralo2021digital}. Attitudes are propagated to match the filter time, so as to minimize residuals from uncompensated rotations of the COM-to-PCO lever-arm,

\begin{align}
    \Vec{\alpha}_{e,i} & =
    \Vec{\alpha}_{e,i} \cdot e^{-\Delta t / \tau_{\alpha_e}}
    \ , \quad \forall i \ \textcolor{gray}{\text{\small spacecraft}} 
    \label{eq:time-update-6} \\
    \mathbf{q}(t) & =
    \delta \mathbf{q} \otimes \mathbf{q}(t_{att})
    \ , \quad
    \textcolor{gray}{\text{\small see \autoref{eq:att-step}}} 
    \label{eq:time-update-7}
\end{align}

where $\mathbf{q}(t_{att})$ is the last known attitude at ADCS time $t_{att}$. Attitude time-tags $t_{att}$ marked by the ADCS must also be converted to the same time scale as the filter.
An analytical form of $\mathbf{\Phi}_{t-1}$ is difficult to achieve in practice, and hence the STM can be approximated via the variational equations so as to capture perturbations of $f$ through the STM.
This is reviewed in Ch.7 of Montenbruck and Gill \cite{montenbruck2010satellite}.
For example, given a state transition through time $t_0 \longrightarrow t$ in position-velocity space $\mathbf{\Phi}_{rv}(t,t_0)$, and given the Jacobian $\mathbf{F}$ of the force model $f$ with respect to $\Vec{r}, \Vec{v}$, the variational equations provide the STM rate of change:

\begin{equation}
    \frac{d}{dt} \mathbf{\Phi}_{rv}(t,t_0)
    = \mathbf{F}(t) \cdot \mathbf{\Phi}_{rv}(t,t_0)
    \quad , \quad \text{where} \quad
    \mathbf{F}(t) =
    \begin{bmatrix}
        \mathbf{0}_{3 \times 3} &
        \mathbf{I}_{3 \times 3} \\
        \frac{\partial f(\vec{r}, \vec{v}, t)}{\partial \vec{r}(t)} & \frac{\partial f(\vec{r}, \vec{v}, t)}{\partial \vec{v}(t)}
    \end{bmatrix}_{6 \times 6}
    \label{eq:variational}
\end{equation}

The partial derivatives in $\mathbf{F}(t)$ are approximated using difference quotients rather than solved analytically.
By the variational equations of \autoref{eq:variational} we can obtain the example STM $\mathbf{\Phi}_{rv}(t,t_0)$ by propagating the initial condition $\mathbf{\Phi}_{rv}(t_0,t_0) = \mathbf{I}_{6 \times 6}$.
Similarly, numerical difference quotients are used to evaluate partial derivatives within the sensitivity $\mathbf{\Phi}_{rv,\alpha}(t)$ of the position-velocity states with respect to empirical accelerations,

\begin{equation}
    \frac{d}{dt} \, \mathbf{\Phi}_{rv,\alpha}(t) =
    \mathbf{F}(t) \cdot 
    \mathbf{\Phi}_{rv,\alpha}(t) + \mathbf{G}(t)
    \quad , \quad \text{where} \quad
    \mathbf{G}(t) =
    \begin{bmatrix}
        \mathbf{0}_{3 \times 3} \\
        \frac{\partial 
        \vec{f}(\vec{r}, \vec{v},
        \vec{\alpha}_e, t)}{\partial \vec{\alpha}_e}
    \end{bmatrix}_{6 \times 3}
\end{equation}

where the initial condition of the sensitivity $\mathbf{\Phi}_{rv,\alpha}(t_0) = \mathbf{0}_{6 \times 3}$. 
Since empirical accelerations follow a first order Gauss Markov process modelled by a decaying transition with a correlation time constant $\tau_{\alpha_e}$, their STM is simply $\mathbf{\Phi}_{\alpha}=\textbf{diag} \left[ \phi_R, \phi_T, \phi_N \right]_{3 \times 3}$ and $\phi_k=e^{-(t-t_0)/\tau_{\alpha_e}}$ for $k \in [R,T,N]$.
Empirical accelerations are assumed independent across the Radial-Tangential-Normal (RTN) axes of the spacecraft.
For the full filter state, superscripts $(i)$ and $(j)$ indicate each spacecraft.
The total STM involving positions, velocities and empirical accelerations for both spacecraft is the concatenation given in \autoref{eq:proc-noise-stm},

\begin{equation}
    \mathbf{\Phi}_{rv\alpha} = 
    \begin{bmatrix}
        \mathbf{\Phi}_{rv\alpha}^{(i)} & \mathbf{0}_{9 \times 9} \\
        \mathbf{0}_{9 \times 9} & \mathbf{\Phi}_{rv\alpha}^{(j)}
    \end{bmatrix}, \quad \text{where} \quad
    \mathbf{\Phi}_{rv\alpha}^{(i)} = 
    \begin{bmatrix}
        \mathbf{\Phi}_{rv}^{(i)} & \mathbf{\Phi}_{rv,\alpha}^{(i)} \\
        \mathbf{0}_{3 \times 6} & \mathbf{\Phi}_{\alpha}^{(i)}
    \end{bmatrix} \quad \text{and} \quad
    \mathbf{\Phi}_{rv\alpha}^{(j)} = 
    \begin{bmatrix}
        \mathbf{\Phi}_{rv}^{(j)} & \mathbf{\Phi}_{rv,\alpha}^{(j)} \\
        \mathbf{0}_{3 \times 6} & \mathbf{\Phi}_{\alpha}^{(j)}
    \end{bmatrix}
	\label{eq:proc-noise-stm}
\end{equation}

Regarding the COM-to-PCO consider parameters $\Delta \Vec{p}_c, \Delta \Vec{p}_d$, these are left expressed in the body-frame. Hence their STM is identity \textit{i.e.} static in the body frame.
Else, if expressed in a non-body frame, their STM must reflect the time-varying rotation from the body to the frame of relevance.
Next, the measurement update steps are presented. Given an observed measurement $\Vec{z}_t$ of size $m$, the innovation, or prefit residual, $\Vec{\upsilon}_t \in \mathbb{R}^m$ is

\begin{equation}
    \Vec{\upsilon}_t = \Vec{z}_t - h \left( 
    \Vec{x}_{t|t-1}, \ \Vec{w}_{t|t-1}, \ \mathbf{q}(t) \right)
\end{equation}

where $\Vec{x}_{t|t-1} \in \mathbb{R}^n$, $\Vec{w}_{t|t-1} \in \mathbb{R}^\ell$, and $h: \mathbb{R}^{n+\ell} \times \mathbb{H} \to \mathbb{R}^m$ is the non-linear measurement model; then $\mathbf{H}_t^{(x)} \in \mathbb{R}^{m \times n}$ and $\mathbf{H}_t^{(w)} \in \mathbb{R}^{m \times \ell}$ are the first-order measurement sensitivities with respect to states $\Vec{x}$ and consider parameters $\Vec{w}$; $\mathbf{R}_t \in \mathbb{R}^{m \times m}$ is the modeled measurement covariance at $t$, positive definite at construction. A review of GRAPHIC and SDCP models can be found in the references \cite{yunck1993graphic, gpstextbook2006, damico2010thesis, giralo2021digital}. The innovation covariance, with consider parameters $\Vec{w}$ included, is expressed as

\begin{equation}
    \mathbf{V}_t =
    \underbrace{
    \mathbf{H}_t^{(x)} \mathbf{\Sigma}_{t|t-1}^{(x)} \mathbf{H}_t^{(x)\top}
    }_{\textcolor{gray}{\text{positive definite form}}}
    + \underbrace{
    \mathbf{H}_t^{(x)} \mathbf{\Sigma}_{t|t-1}^{(xw)} \mathbf{H}_t^{(w)\top}
    + \mathbf{H}_t^{(w)} \mathbf{\Sigma}_{t|t-1}^{(wx)} \mathbf{H}_t^{(x)\top}
    }_{\textcolor{gray}{\text{not guaranteed positive definite numerically}}}
    + \underbrace{
    \mathbf{H}_t^{(w)} \mathbf{\Sigma}_{t|t-1}^{(w)} \mathbf{H}_t^{(w)\top}
    }_{\textcolor{gray}{\text{positive definite form}}}
    + \mathbf{R}_t
    \label{eq:cov-innovation}
\end{equation}

where it must be noted that in \autoref{eq:cov-innovation}, the sum of two matrices that are transposes of each other will result in symmetric but not necessarily PD. Ideally, $\mathbf{V}_t$ should always be PD if factorized in block form. However, in practice, numerical errors from finite precision and rounding accumulations can compound. Hence, the innovation covariance $\mathbf{V}_t$ for the conventional CEKF is might not guarantee PD. Since the optimal Kalman gain $\mathbf{K}_t$ scales inversely with the uncertainty of the innovation, a catastrophic singularity may occur mid-flight if the inverse of $\mathbf{V}_t$ does not exist. In light of this, regularizing the innovation with an extra term ${\mathbf{\overline{V}}}$ is proposed in \autoref{eq:meas-update-1}, in order to numerically guarantee the existence of the inverse in \autoref{eq:meas-update-1} during actual computations. 

\vspace{-5mm}
\begin{align}
    \mathbf{K}_t & = 
    \left(
    \mathbf{\Sigma}_{t|t-1}^{(x)} \mathbf{H}_t^{(x)\top}
    + \mathbf{\Sigma}_{t|t-1}^{(xw)} \mathbf{H}_t^{(w)\top}
    \right)
    \left( \mathbf{V}_t
    + {\mathbf{\overline{V}}}
    \right)^{-1}
    \label{eq:meas-update-1}
\end{align}

The term ${\mathbf{\overline{V}}}$ is added to possibly non-PD terms such that $\mathbf{V}_t + \mathbf{\overline{V}}$ is PD and hence invertible. The construction of ${\mathbf{\overline{V}}}$ that makes the smallest possible adjustment to the terms within the inverse in \autoref{eq:meas-update-1}, as measured by the trace, can be found as follows. First, the sum of matrices that is non-guaranteed PD is real and symmetric and hence diagonalizable,

\begin{equation}
    \mathbf{\Psi \Lambda \Psi^\top} = 
    \text{eig} \left(
    \mathbf{H}_t^{(x)} \mathbf{\Sigma}_{t|t-1}^{(xw)} \mathbf{H}_t^{(w)\top}
    + \mathbf{H}_t^{(w)} \mathbf{\Sigma}_{t|t-1}^{(wx)} \mathbf{H}_t^{(x)\top}
    \right)
    \label{eq:meas-update-2}
\end{equation}

Then, for diagonal indices $i \in [1,m]$ along $\mathbf{\Lambda}$, where $m$ is the number of measurements, if a negatively valued $\lambda_i$ is found, the corresponding value of $\gamma_i$ is set to $-\lambda_i$; otherwise, it is set to zero since the eigenvalues are already positive,

\begin{equation}
    {\mathbf{\overline{V}}} =
    \mathbf{\Psi} \cdot
    \text{diag} \left(
    [ \gamma_1 \ , \gamma_2 \ , \cdots \ , \gamma_m ]
    \right) \cdot \mathbf{\Psi}^\top
    \quad \textcolor{gray}{\text{\small where}} \
    \gamma_i = \text{max}(0, -\lambda_i)
    \textcolor{gray}{\text{ and }}\lambda_i
    \textcolor{gray}{\text{ are diagonals of }} \mathbf{\Lambda}
    \label{eq:meas-update-3}
\end{equation}

thereby ensuring that $\mathbf{\Psi \Lambda \Psi^\top} + {\mathbf{\overline{V}}}$ will have only non-negative eigenvalues. In practice, eigen-decomposition at each measurement update is costly for flight. Independent measurements per transmitter is often assumed in GNSS \cite{gpstextbook2006}. This allows measurements to be processed sequentially as scalars ($m=1$). Hence, ${\mathbf{\overline{V}}} \in \mathbb{R}$, and \autoref{eq:meas-update-3} simplifies to:

\begin{equation}
    \mathbf{\overline{V}} =
    \text{max} \left(
    0 \ , \
    -\left(
    \mathbf{H}_t^{(x)} \mathbf{\Sigma}_{t|t-1}^{(xw)} \mathbf{H}_t^{(w)\top} + \mathbf{H}_t^{(w)} \mathbf{\Sigma}_{t|t-1}^{(wx)} \mathbf{H}_t^{(x)\top}
    \right) \right) \qquad \textcolor{gray}{\text{(scalar)}}
\end{equation}

The measurement update of the \textit{a posteriori} state mean applies the sub-optimal Kalman gain,

\begin{equation}
    \Vec{x}_{t|t} =
    \Vec{x}_{t|t-1} + \mathbf{K}_t \Vec{\upsilon}_t
    \label{eq:meas-update-4}
\end{equation}

Then, the Joseph form measurement update guarantees the \textit{a posteriori} state covariance in \autoref{eq:meas-update-6} is PD,

\vspace{-5mm}
\begin{align}
    \mathbf{U}_t & =
    \begin{bmatrix}
        [\mathbf{K}_t \mathbf{H}_t^{(x)} - \mathbf{I}_{n \times n}] \ , &
        [\mathbf{K}_t \mathbf{H}_t^{(w)}]
    \end{bmatrix}
    \label{eq:meas-update-5} \\
    \mathbf{\Sigma}_{t|t}^{(x)} & =
    \mathbf{U}_t \mathbf{\Sigma}_{t|t-1} \mathbf{U}_t^\top
    + \mathbf{K}_t \mathbf{R}_t \mathbf{K}_t^\top
    \quad 
    \textcolor{gray}{\text{\small(preserves positive definiteness)}}
    \label{eq:meas-update-6} \\
    \mathbf{\Sigma}_{t|t}^{(xw)} & =
    (\mathbf{I}_{n \times n} - \mathbf{K}_t \mathbf{H}_t^{(x)})
    \mathbf{\Sigma}_{t|t-1}^{(xw)}
    - \mathbf{K}_t \mathbf{H}_t^{(w)} \mathbf{\Sigma}^{(w)}
    \label{eq:meas-update-7}
\end{align}

For brevity, the full derivation of the CEKF Joseph form, as well as an analysis of its stability and runtimes, is published as an internal note \cite{low2025cekf}.
No optimality conditions are imposed on $\mathbf{K}_t$.
An important trade-off is that in the Joseph form, most terms are dense, and hence sparse matrix operations cannot be fully exploited for speed, in exchange for numerical stability. 

% ========================================
% 3.3 NAVIGATION ARCHITECTURE: PROCESS NOISE
% ========================================

\subsection{Process Noise Modeling}

To achieve required orbit prediction accuracies, including during data gaps, the process noise model $\mathbf{Q}_t$ must capture orbit propagation uncertainties and unmodeled dynamics while preserving state correlation structures \cite{nasa2018filter, carpenter2005navigation, jonhow2007sma}. When the spectral noise density accurately reflects these correlations, the resulting process noise preserves them appropriately. An exponential spatial correlation model is adopted to capture inter-spacecraft dynamics coupling,

\begin{equation}
    \beta_{ij} = \exp \left( -d_{ij} / D_0 \right)
    \label{eq:corr-space}
\end{equation}

which encodes correlations between spacecraft $i$ and $j$ as a function of their separation $d_{ij}$, where $D_0$ is a tunable correlation length scale.
This model is a spatial analog to the first-order Gauss-Markov process in time \cite{gallager2013stochastic}. It intuits that the dynamics of nearby spacecraft are more strongly correlated than those farther apart. Let $\sigma_R, \sigma_T, \sigma_N$ be the scalar components of the dynamics spectral noise density, decoupled along each axis of the RTN frame. Consequently, the spectral noise density matrices for individual and joint spacecraft dynamics, $\mathbf{S}\alpha$ and $\mathbf{S}\alpha^{(ij)}$, are defined as

\begin{equation}
    \mathbf{S}_\alpha^{(ij)} \equiv
    \begin{bmatrix}
        \phantom{\beta_{ij}} \mathbf{S}_\alpha &
        \beta_{ij} \mathbf{S}_\alpha \\
        \beta_{ij} \mathbf{S}_\alpha &
        \phantom{\beta_{ij}} \mathbf{S}_\alpha \\
    \end{bmatrix}
    \ , \ \text{where} \quad
    \mathbf{S}_\alpha \equiv \text{diag} \left(
    \begin{bmatrix}
        \sigma_R &
        \sigma_T &
        \sigma_N \\
    \end{bmatrix}
    \right)
	\label{eq:proc-noise-density}
\end{equation}

and both $\mathbf{S}_\alpha$ and $D_0$ are user-defined. The joint process noise covariance for the Cartesian states of spacecraft $i$ and $j$, denoted $\mathbf{Q}_{rv\alpha}$, is analytically derived by mapping the spectral noise density onto the Cartesian states via the process noise mapping matrix $\mathbf{\Gamma}$, and propagating it through the Cartesian state transition matrix $\mathbf{\Phi}_{rv\alpha}(t,\tau)$ \cite{stacey2022noise}, as shown in \autoref{eq:proc-noise-integral}. If the STM is integrable and across a short time interval $\Delta t$, \autoref{eq:proc-noise-integral} can be approximated in \autoref{eq:proc-noise-raw},

\vspace{-5mm}
\begin{align}
    \mathbf{Q}_{rv\alpha} & = \int_{t_0}^{t_0 + \Delta t} \mathbf{\Phi}_{rv\alpha} (t_0, \tau) \cdot \mathbf{\Gamma}(\tau - t_0) \cdot \mathbf{S}_\alpha^{(ij)} \cdot \mathbf{\Gamma}(\tau - t_0)^\top \cdot \mathbf{\Phi}_{rv\alpha} (t_0, \tau)^\top \, d\tau
    \label{eq:proc-noise-integral} \\
    & \approx \mathbf{\Phi}_{rv\alpha} (t_0, t_0 + \Delta t) \cdot \mathbf{\Gamma}(\Delta t) \cdot \mathbf{S}_\alpha^{(ij)} \cdot \mathbf{\Gamma}(\Delta t)^\top \cdot \mathbf{\Phi}_{rv\alpha} (t_0, t_0 + \Delta t)^\top
	\label{eq:proc-noise-raw}
\end{align}

The joint-spacecraft process noise mapping matrix $\mathbf{\Gamma}$ for spacecraft $i$ and $j$ over a time interval $\Delta t$ is simply

\begin{equation}
    \mathbf{\Gamma}(\Delta t) = 
    \begin{bmatrix}
        \mathbf{C}^{(i)}
        \widetilde{\mathbf{\Gamma}}
        (\Delta t) &
        \mathbf{0}_{9 \times 3} \\
        \mathbf{0}_{9 \times 3} &
        \mathbf{C}^{(j)}
        \widetilde{\mathbf{\Gamma}}
        (\Delta t) \\
    \end{bmatrix}
    \ , \quad \text{where} \quad 
    \widetilde{\mathbf{\Gamma}}(\Delta t) =
    \begin{bmatrix}
    \frac{1}{2} \Delta t^2 \cdot \mathbf{I}_{3 \times 3} \\
    \Delta t \cdot \mathbf{I}_{3 \times 3} \\
    (1-\theta^2) \cdot \mathbf{I}_{3 \times 3}
    \end{bmatrix}
	\label{eq:proc-noise-map}
\end{equation}

where $\theta = e^{-\Delta t / \tau_{\alpha_e}}$ is the time-decay of the empirical accelerations, and $\mathbf{C}^{(i)}, \mathbf{C}^{(j)}$ ensure that coordinate frames are handled correctly. In the two-spacecraft state vector of \autoref{eq:state-vector}, positions and velocities are in Earth-Centered Inertial (ECI), while empirical accelerations are in RTN. Let the RTN-to-ECI rotation be $\left[ {}^{ECI}\mathbf{R}^{(i)}_{RTN} \right] \in \mathbb{R}^{3 \times 3}$, and let $\left[ \mathbf{\omega}^\times \right]$ be the angular velocity of RTN with respect to ECI, in skew-symmetric form, in the RTN basis. Then, the coordinate frame mapping for spacecraft $i$ is

\begin{equation}
    \mathbf{C}^{(i)} =
    \begin{bmatrix}
    \left[ {}^{ECI}\mathbf{R}^{(i)}_{RTN} \right] &
    \mathbf{0}_{3 \times 3} &
    \mathbf{0}_{3 \times 3} \\
    \left[ {}^{ECI}\mathbf{R}^{(i)}_{RTN} \right]
    \left[ \mathbf{\omega}^\times \right] &
    \left[ {}^{ECI}\mathbf{R}^{(i)}_{RTN} \right] &
    \mathbf{0}_{3 \times 3} \\
    \mathbf{0}_{3 \times 3} &
    \mathbf{0}_{3 \times 3} &
    \mathbf{I}_{3 \times 3}
    \end{bmatrix}
	\label{eq:proc-noise-coord-map}
\end{equation}

The analytical process noise can now be evaluated by substituting \ref{eq:proc-noise-coord-map} into \ref{eq:proc-noise-map}, and then \ref{eq:proc-noise-density}, \ref{eq:proc-noise-stm}, \ref{eq:proc-noise-map} into \ref{eq:proc-noise-raw}. The result is:

\vspace{-5mm}
\begin{align}
    & \mathbf{Q}_{rv\alpha} =
    \left[
    \begin{array}{cc}
        \mathbf{Q}^{(i)}_{rv\alpha} &
        \mathbf{Q}^{(ij)}_{rv\alpha} \\
        \mathbf{Q}^{(ji)}_{rv\alpha} &
        \mathbf{Q}^{(j)}_{rv\alpha} \\
    \end{array}
    \right] \\
    \text{where} \qquad & 
    \begin{aligned}
        \mathbf{Q}^{(i)}_{rv\alpha} &= \boldsymbol{\Phi}_{rv\alpha}^{(i)} \cdot \mathbf{C}^{(i)} \cdot \tilde{\boldsymbol{\Gamma}}(\Delta t) \cdot \mathbf{S}_{\alpha} \cdot \tilde{\boldsymbol{\Gamma}}(\Delta t)^{\top} \cdot \mathbf{C}^{(i)\top} \cdot \boldsymbol{\Phi}_{rv\alpha}^{(i)\top} \\
        \mathbf{Q}^{(j)}_{rv\alpha} &= \boldsymbol{\Phi}_{rv\alpha}^{(j)} \cdot \mathbf{C}^{(j)} \cdot \tilde{\boldsymbol{\Gamma}}(\Delta t) \cdot \mathbf{S}_{\alpha} \cdot \tilde{\boldsymbol{\Gamma}}(\Delta t)^{\top} \cdot \mathbf{C}^{(j)\top} \cdot \boldsymbol{\Phi}_{rv\alpha}^{(j)\top} \\
        \mathbf{Q}^{(ij)}_{rv\alpha} &= \mathbf{Q}_{rv\alpha}^{(ji)\top} = \boldsymbol{\Phi}_{rv\alpha}^{(i)} \cdot \mathbf{C}^{(i)} \cdot \tilde{\boldsymbol{\Gamma}}(\Delta t) \cdot (\beta_{ij} \mathbf{S}_{\alpha}) \cdot \tilde{\boldsymbol{\Gamma}}(\Delta t)^{\top} \cdot \mathbf{C}^{(j)\top} \cdot \boldsymbol{\Phi}_{rv\alpha}^{(j)\top} \\
    \end{aligned}
\end{align}

The full process noise matrix for the two-spacecraft state vector of \autoref{eq:state-vector}, including receiver clock biases tracked for $k$ GNSS constellations, and $c$ channels each for ZDCP and SDCP float ambiguities, has the following structure:

\begin{equation}
    \mathbf{Q_t} =
    \setlength{\arraycolsep}{2pt} % default is 5pt
    \begin{bmatrix}
        \mathbf{Q}^{(i)}_{rv\alpha} &
        \textcolor{lightgray}{\mathbf{0}_{9 \times k}} &
        \mathbf{Q}^{(ij)}_{rv\alpha} &
        \textcolor{lightgray}{\mathbf{0}_{9 \times k}} &
        \textcolor{lightgray}{\mathbf{0}_{9 \times c}} &
        \textcolor{lightgray}{\mathbf{0}_{9 \times c}} \\

        \textcolor{lightgray}{\mathbf{0}_{k \times 9}} &
        \mathbf{Q}^{(i)}_{c \delta t} &
        \textcolor{lightgray}{\mathbf{0}_{k \times 9}} &
        \textcolor{lightgray}{\mathbf{0}_{k \times k}} &
        \textcolor{lightgray}{\mathbf{0}_{k \times c}} &
        \textcolor{lightgray}{\mathbf{0}_{k \times c}} \\

        \mathbf{Q}^{(ji)}_{rv\alpha} &
        \textcolor{lightgray}{\mathbf{0}_{9 \times k}} &
        \mathbf{Q}^{(j)}_{rv\alpha} &
        \textcolor{lightgray}{\mathbf{0}_{9 \times k}} &
        \textcolor{lightgray}{\mathbf{0}_{9 \times c}} &
        \textcolor{lightgray}{\mathbf{0}_{9 \times c}} \\
        
        \textcolor{lightgray}{\mathbf{0}_{k \times 9}} &
        \textcolor{lightgray}{\mathbf{0}_{k \times k}} &
        \textcolor{lightgray}{\mathbf{0}_{k \times 9}} &
        \mathbf{Q}^{(j)}_{c \delta t} &
        \textcolor{lightgray}{\mathbf{0}_{k \times c}} &
        \textcolor{lightgray}{\mathbf{0}_{k \times c}} \\
        
        \textcolor{lightgray}{\mathbf{0}_{c \times 9}} &
        \textcolor{lightgray}{\mathbf{0}_{c \times k}} &
        \textcolor{lightgray}{\mathbf{0}_{c \times 9}} &
        \textcolor{lightgray}{\mathbf{0}_{c \times k}} &
        \mathbf{Q}_{N_{ZD}} &
        \textcolor{lightgray}{\mathbf{0}_{c \times c}} \\
        
        \textcolor{lightgray}{\mathbf{0}_{c \times 9}} &
        \textcolor{lightgray}{\mathbf{0}_{c \times k}} &
        \textcolor{lightgray}{\mathbf{0}_{c \times 9}} &
        \textcolor{lightgray}{\mathbf{0}_{c \times k}} &
        \textcolor{lightgray}{\mathbf{0}_{c \times c}} &
        \mathbf{Q}_{N_{SD}} \\
    \end{bmatrix}
    \setlength{\arraycolsep}{5pt} % reset to default
    \label{eq:proc-noise-full}
\end{equation}

This analytical process noise model provides a physically grounded approach aligned with the filter’s estimate of the required dynamic noise compensation, while preserving the correlation structure between the Cartesian states. It also mitigates manual tuning of process noise parameters. Substantial improvements in state prediction accuracy during data gaps, as compared to the manually-tuned diagonal process noise of \cite{giralo2019digital}, will be demonstrated in \autoref{section6}.

\pagebreak

% ========================================
% 3.4 NAVIGATION ARCHITECTURE: MEASUREMENT NOISE
% ========================================

\subsection{Measurement Noise Modeling}

\begin{figure}[H]
    \centering
    \includegraphics[width=0.85\linewidth]{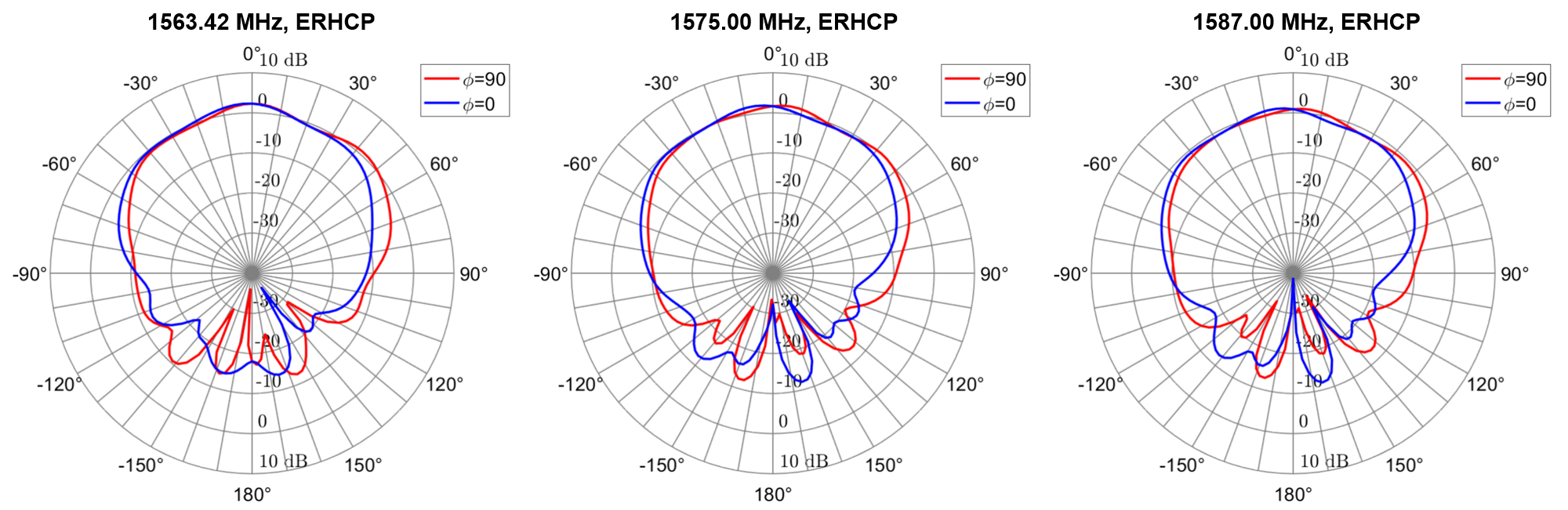}
    \caption{Calibrated antenna gain of the 1.9G1215P-12S-2-OSR model with orientation $\phi$ (Credits: ANTCOM)}
    \label{fig:ant-pattern}
\end{figure}

Measurement noise modeling $\mathbf{R}_t$ is critical in deciding how the filter weighs the uncertainty of observations against predicted states. Since correlated measurement biases are already canceled in the GRAPHIC and SDCP measurements, while remaining biases are actively estimated in the state, the filter is primarily concerned with modeling the thermal noise.
This section outlines the process of thermal noise modeling, using a Novatel OEM628 GNSS receiver, and an ANTCOM 1.9G1215P antenna.
Both units are near-identical representations of the flight units onboard the VISORS spacecraft, with flight heritage. The antenna gain pattern is made available to the flight software via a parametric fit of the actual calibrated data, as seen in \autoref{fig:ant-pattern}. Antenna gain values can be sampled from this parametric fit based on the elevation-azimuth. With the gain values, the Carrier-to-Noise Ratio $C_{N_0}$ can then be computed through a link budget for various elevation angles in \autoref{tab:link-budget},

\vspace{-5mm}

\begin{table}[H]
\renewcommand{\arraystretch}{1.0}
\centering
\small
\caption{Link budget analysis and thermal noise tabulation for the Novatel OEM628 and an ANTCOM 1.9G1215P antenna, at GPS L1 $f = 1575.42$ MHz. Elevation is measured from the antenna plane. Free space path loss assumes a fixed slant range of 20,000km. Carrier signal strength ($C$) [dBW] is the sum of gains and losses from the antenna, receiver, GPS transmitter, and the free space path.}
\label{tab:link-budget}
\begin{tabular}{|c|c|c|c|c|c|}
    \hline
    & & \bfseries Zenith & \bfseries Roll-Off & \bfseries Horizon & \bfseries Nadir \\
    \hline
    \small Elevation
    & ( $\cdot$ )$^\circ$ & 90 & 60 & 0 & -90 \\
    \hline
    \small Antenna Gain
    & dBW & 1.7 & -0.6 & -10.9 & -15.0 \\
    \hline
    \small Receiver LNA Gain
    & dBW & \multicolumn{4}{c|}{30.0} \\
    \small Receiver Circuit and Polarization Loss
    & dBW & \multicolumn{4}{c|}{-4.0} \\ 
    \hline
    \small GPS Antenna Gain
    & dBW & \multicolumn{4}{c|}{13.5} \\
    \small GPS Transmit Power
    & dBW & \multicolumn{4}{c|}{14.25} \\
    \small GPS Transmit Loss
    & dBW & \multicolumn{4}{c|}{-1.25} \\
    \hline
    \small Free Space Path Loss
    & dBW & \multicolumn{4}{c|}{-182.419} \\
    \small Atmospheric Losses
    & dBW & \multicolumn{4}{c|}{-0.10} \\
    \hline
    \small Noise Spectral Density ($N_0$)
    & dBW/Hz & \multicolumn{4}{c|}{-169.919} \\
    \hline
    \small Carrier Signal Strength ($C$)
    & dBW & -125.3185 & -127.6185 & -137.9185 & -142.0185 \\
    \small Carrier-to-Noise Ratio $(C_{N_0})$
    & dBW $\cdot$ Hz & 41.7 & 39.4 & 29.1 & 25.0 \\
    \hline
    \small \bf Pseudorange Thermal Noise $\sigma_\rho$
    & m & 0.149 & 0.194 & 0.634 & 1.016 \\
    \small \bf Carrier Phase Thermal Noise $\sigma_\phi$
    & mm & 0.682 & 0.887 & 2.909 & 4.664 \\
    \hline
\end{tabular}
\end{table}

The signal-to-noise ratio $S_{NR}$ relates to $C_{N_0}$ [dB-Hz] via a log-to-linear scale conversion $S_{NR} = 10^{C_{N_0}/10}$. An analytical thermal noise model formulated by Psiaki and Mohiuddin \cite{psiaki2007cdgps} for undifferenced carrier phase $\sigma_{\phi}$ in \autoref{eq:noise-cp} and raw pseudorange $\sigma_{\rho}$ in \autoref{eq:noise-pr-raw} is referenced below for convenience,

\begin{equation}
    \sigma_{\phi} = \frac{\lambda}{2 \pi}
    \sqrt{\frac{B_{PLL}}{2 S_{NR}}}
    \qquad \text{[meters]}
    \label{eq:noise-cp}
\end{equation}

\begin{equation}
    \sigma_{\rho} = c t_c \sqrt{
    \frac{B_{DLL}t_{eml}}{2 t_c S_{NR}}
    \left(1 + \frac{1}{t_{acc} S_{NR}} \right)}
    \qquad \text{[meters]}
    \label{eq:noise-pr-raw}
\end{equation}

where $B_{PLL}$ and $B_{DLL}$ are the receiver loop bandwidths for the receiver Phase-Lock Loop (PLL) and Delay-Lock Loop (DLL) respectively; $t_{eml}$ denotes the early-minus-late correlator spacing in seconds; $t_{acc}$ refers to the pre-detection integration time over which signal energy is accumulated before discriminator evaluation; $t_c = 1/f_c$ is the chipping period of the pseudorandom code, where $f_c = 1.023$ MHz for GPS L1 C/A; $c$ and $\lambda$ are the speed of light and wavelength. Typically, $B_{PLL}$ and $B_{DLL}$ may be disclosed by the manufacturer, but often, very specific design parameters such as $t_{eml}$ and $t_{acc}$ may not be. In the case of the Novatel OEM628, $B_{DLL} = 0.0076$ Hz and $B_{PLL} = 15$ Hz. It is assumed that $t_{eml} = $ 1-chip period, which is reasonable for modern receivers, and that the accumulation interval $t_{acc}$ is sufficiently long so that $t_{acc} S_{NR} \ggg 1$. This simplifies the pseudorange thermal noise model to \autoref{eq:noise-pr},

\begin{equation}
    \sigma_{\rho} \approx \frac{c}{f_c}
    \sqrt{\frac{B_{DLL}}{2S_{NR}}}
    \qquad \text{[meters]}
    \label{eq:noise-pr}
\end{equation}

The filter may then discern thermal noise \ref{eq:noise-cp} and \ref{eq:noise-pr} using a look-up table of the antenna gain values, which are in turn based on the azimuth-elevation of the arriving signal.

% ========================================
% 3.5 NAVIGATION: IAR BLOCK
% ========================================

\subsection{Integer Ambiguity Resolution}

\begin{figure}[htp]
	\centering
    \includegraphics[width=1.0\textwidth]{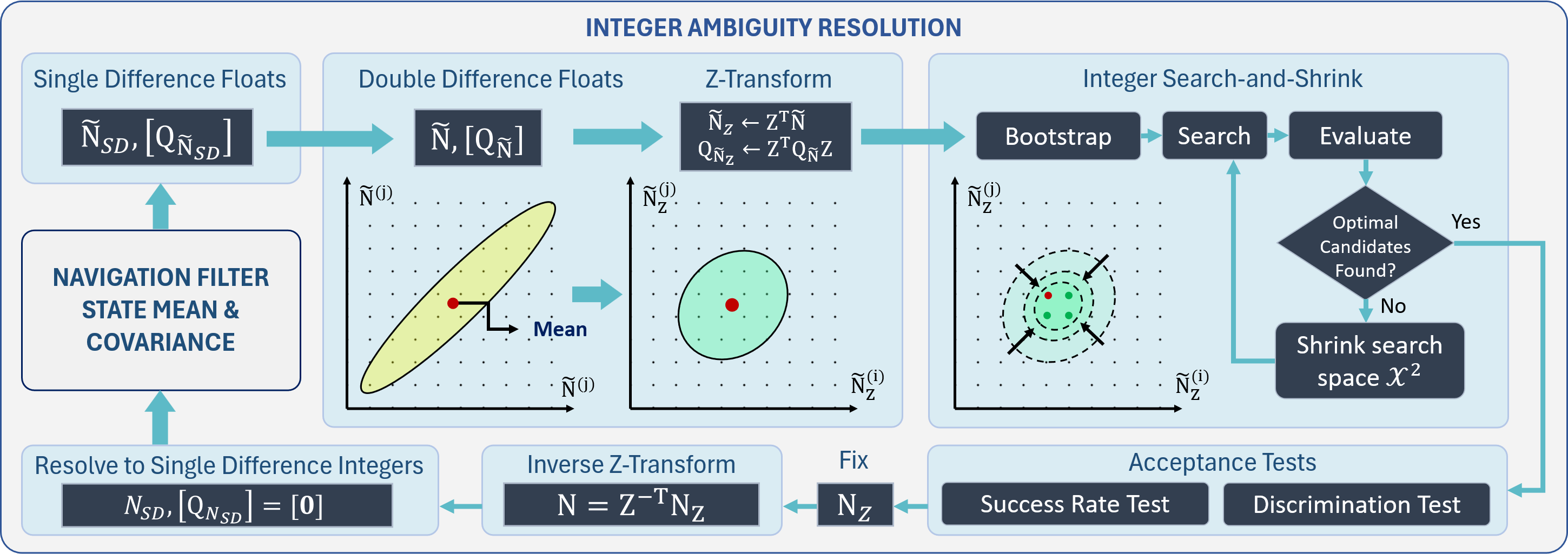}
    \caption{Block diagram of the Integer Ambiguity Resolution (IAR) block, expanded from \autoref{fig:architecture}.}
    \label{fig:iar-block}
\end{figure}

The topic of IAR is well-researched in literature. Hence, the IAR module, based on the Modified LAMBDA algorithm \cite{teunissen1994lambda, chang2005mlambda} is only briefly outlined in \autoref{fig:iar-block}, for the completion of the full presentation.

First, SDCP floats $\Tilde{N}_{SD}$ are double-differenced to DDCP floats $\Tilde{N}$. Components of the $\Tilde{N}$ vector are highly correlated due to the common SDCP reference measurement. Decorrelation using the integer-preserving Z-transform \cite{teunissen1994lambda} reduces the size of the search space prior to integer search. The Z-transform matrix $\mathbf{Z}$ is an integer approximation of the lower triangular matrix $\mathbf{L}$, which comes from the $\mathbf{Q}_{\Tilde{N}} = \mathbf{LDL}^\top$ decomposition. $\mathbf{Z}$ is derived by imposing the integer constraints given in \cite{teunissen1995invertible}, so that

\begin{equation}
  \mathbf{Q}_{\Tilde{N}_z} =
  \mathbf{Z^\top LDL^\top Z}
  \approx \mathbf{D}
  \label{eq:iar-ztransform}
\end{equation}

The Z-transformed DDCP float ambiguities $\Tilde{N}_z = \mathbf{Z^\top} \Tilde{N}$. The result is a transformed mean $\Tilde{N}_z$ and covariance $\mathbf{Q}_{\Tilde{N}_z}$ that is almost decorrelated, though not completely due to the integer constraints. The initial guess of integer candidates is done via recursive conditional rounding \cite{teunissen1998success}. The rounded integers allow bounding of the initial search width in the discrete integer space, which is successively shrunk during integer search to minimize the computational costs of evaluating sub-optimal candidates. For each integer candidate $N_z$, the integer least squares objective function to minimize over all candidates is

\begin{equation}
  \min_{N_z} || N_z - 
  \Tilde{N}_z ||^2_{Q^{-1}_{\Tilde{N}_z}}
  \label{eq:iar-cost}
\end{equation}

When the best candidate can no longer be improved, it is only resolved into an integer if it passes recommended acceptance tests, such as the Success Rate and Discrimination Tests \cite{teunissen1998success}, or meets an Ambiguity Dilution of Precision (ADOP) \cite{teunissen1997adop} threshold. This paper adopts the Success Rate Test with a $99\%$ threshold and the Discrimination Test.

\pagebreak

% ========================================
% 4. NAVIGATION FDIR
% ========================================

\section{Navigation Fault Detection, Isolation, and Recovery}
\label{section4}

\begin{figure}[H]
	\centering
    \includegraphics[width=1.0\textwidth]{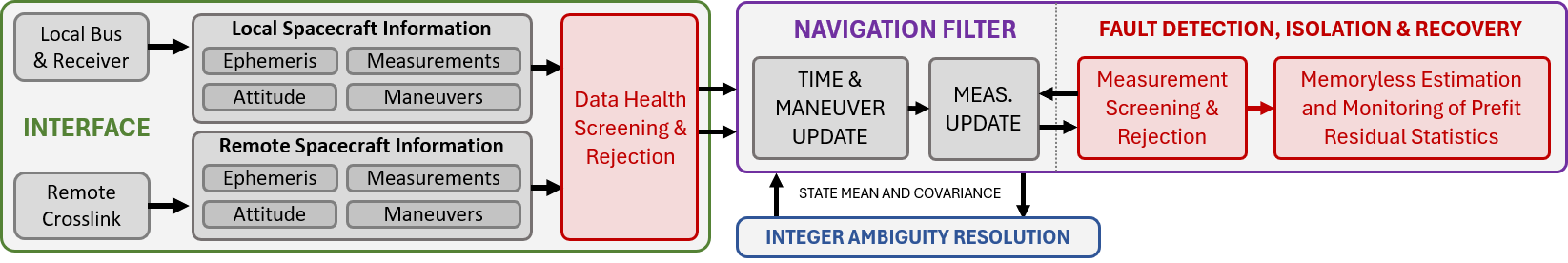}
    \caption{Architecture of the Fault Detection Isolation and Recovery (FDIR) block, expanded from \autoref{fig:architecture}.}
    \label{fig:fdir-block}
\end{figure}

The Fault Detection, Isolation, and Recovery (FDIR) module is designed to ensure continuous and robust navigation against anomalous measurements. Three key sub-functions of FDIR are highlighted in red in the architecture of \autoref{fig:fdir-block},

\vspace{-6mm}

\subsection{Data Health Screening and Rejection}

The health of incoming measurements, maneuvers, attitudes, ephemerides, and telecommands are screened for validity. Packet headers and checksums are inspected. Sanity checks are conducted \textit{e.g.} unexpectedly low or high pseudorange, carrier phase, $C/N_0$ values etc; bad GNSS ephemeris values; time-relevance of maneuvers or measurements in-the-past etc. This includes local (bus) and remote (through crosslink) data. These first-layer checks ensure data integrity down the signal chain in subsequent modules, thereby satisfying Requirement \hyperref[req:2.3]{R2.3}. Potential numerical instabilities are checked, and exceptions are raised if any, as per Requirement \hyperref[req:2.8]{R2.8}. Cycle slip flags from the receiver, if any, are checked too, as per Requirement \hyperref[req:2.9]{R2.9}.

\subsection{Measurement Screening and Rejection}

Measurement outliers are identified by comparing prefit residuals with $\kappa$ multiples of the trace of the innovation variance \autoref{eq:cov-innovation}, where $\kappa$ is a configurable parameter \textit{e.g.} $\kappa = 5$, and $\mathbb{I}$ is simply an indicator function,

\begin{equation}
    \mathbb{I}\left( \ || \Vec{\upsilon} ||_2^2
    > \kappa \cdot \text{tr}(\mathbf{V}) \ \right) = 
    \begin{cases} 
        \text{Accept if } True, \\
        \text{Discard if } False,
    \end{cases}
\label{eq:innovation-threshold}
\end{equation}

The innovation covariance from \ref{eq:cov-innovation} provides an adaptive threshold that accurately reflects the immediate prefit uncertainty, as it sums contributions from both modeled measurement noise plus current state uncertainties projected onto the measurement subspace. Adaptive thresholds more accurately reflects the underlying uncertainty of prefit residuals, and contrasts with fixed gated thresholds that must be manually tuned based on domain knowledge \cite{giralo2019digital, giralo2021digital}. However, careful initialization of the \textit{a priori} state covariance and accurate measurement uncertainty modeling are needed. Overconfident initialization may cause rejection of valid measurements as $\kappa \cdot \text{tr} \left( \mathbf{V} \right)$ provides overly tight bounds, while underconfident initialization may admit outliers.

\subsection{Memoryless Estimation and Monitoring of Prefit Residual Statistics}

While the innovation covariance provides a snapshot of prefit residual uncertainty, a moving-window estimator could capture time-varying statistics.
\hyperref[alg:prefit-stats]{Algorithm 1} provides a memoryless and statistical approach to fault detection,

\begin{algorithm}[H]
\renewcommand{\baselinestretch}{1.0}\selectfont
\caption{Memoryless Estimation and Monitoring of Prefit Residual Statistics}
\label{alg:prefit-stats}
\begin{algorithmic}[1]
    \For{each computed prefit residual $r$} 
        \State $\mu_{k+1} \gets (1-\rho) \mu_k + \rho \cdot r$
        \Comment{Fading memory update of estimated mean of prefit residuals}
        \State $\sigma_{k+1}^2 \gets (1-\rho) \sigma_k^2 + \rho \cdot (r-\mu_k)$
        \Comment{Fading memory update of estimated sigma of prefit residuals}
        \State $S_{k+1} \gets (1-\rho) S_k$
        \Comment{Decay the score of the number of outliers}
        \If{ $||r - \mu_{k+1}||^2 > \eta \cdot \sigma_k^2$ }
            \State $S_{k+1} \gets S_{k+1} + 1$
            \Comment{If prefit residual $r$ lies outside $\eta \cdot \sigma_k^2$ bounds, increase outlier count by $+1$}
        \EndIf
    \EndFor
    \If{ $\left( 1 - \frac{S_{k+1}}{M} \right) > g(\mu_{k+1} , \sigma_{k+1}^2)$ }
        \vspace{2mm}
        \State \textbf{do} emit navigation incapability flag
        \Comment{Flag incapable if number of outliers violate concentration inequalities}
    \EndIf
\end{algorithmic}
\end{algorithm}

where $M$ is the `moving window' size; $\rho \equiv 1/M$ is the forget-factor; $\mu_k$ and $\sigma_k^2$ are the empirically estimated sample mean and covariance of the prefit residuals at time $k$; $S_k$ is a fading count of outliers residing beyond the tail bounds of $\eta \sigma_k^2$ at time $k$; $\eta$ is a hyper-parameter; and $g(\mu_i, \sigma_i^2)$ returns a concentration inequality upper bound for the outlier fraction outside the $\eta \sigma_k^2$ tails.

\begin{figure}[H]
	\centering
    \includegraphics[trim={0cm 0cm 0cm 1cm},clip,width=1.0\textwidth]{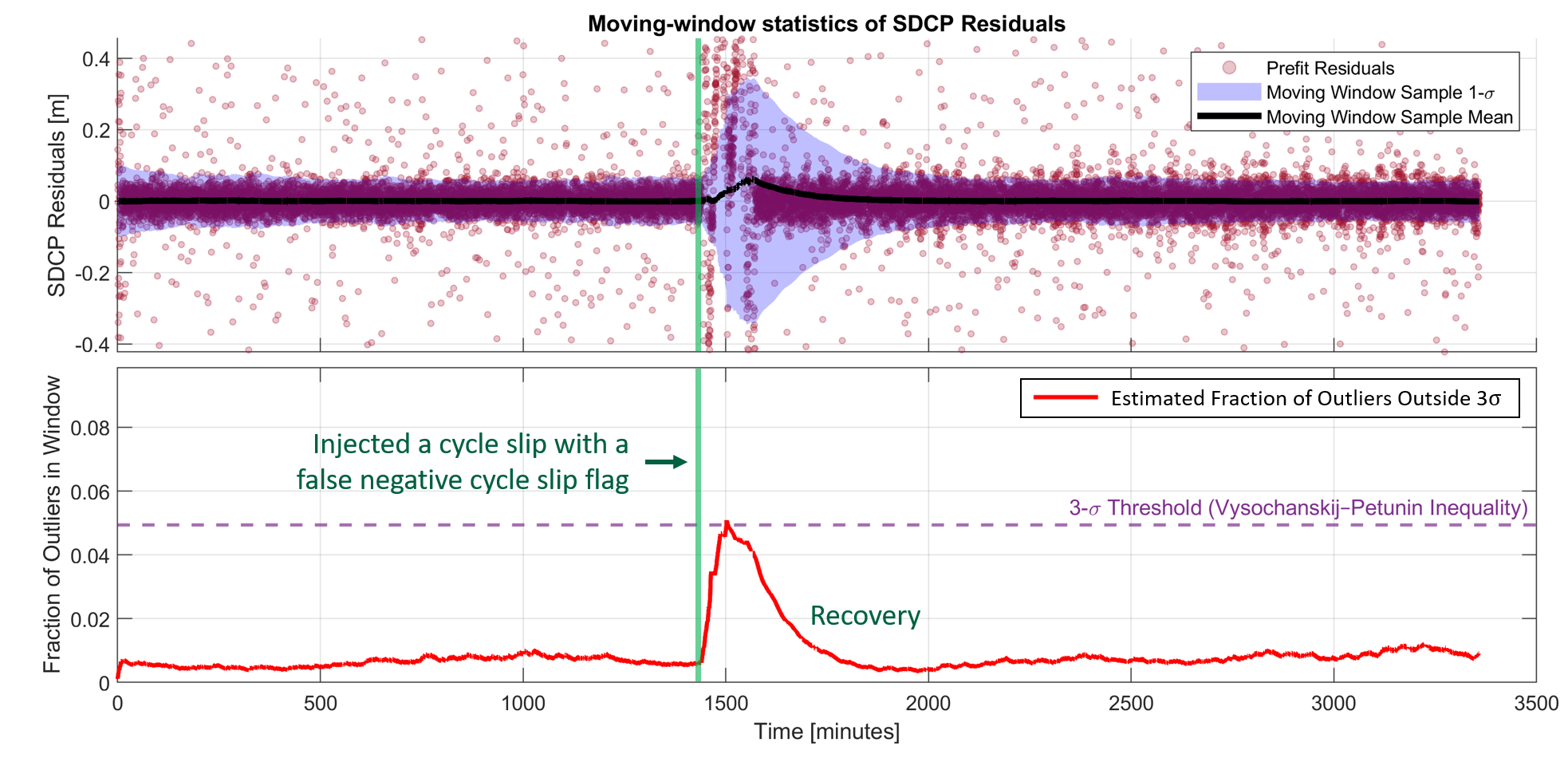}
    \caption{Plot of moving-window sample mean and variance of SDCP measurement prefit residuals (top) and moving-window estimate of number of outliers outside $\eta \sigma^2$ (bottom) for $\eta = 3$ and $M = 6000$; cycle slips were injected at $t = 24$ hours for GPS vehicles currently tracked in 6/12 active channels of one spacecraft. The software detected a fault once outlier counts violated the Vysochanskij-Petunin inequality. The filter auto-recovered after 1 orbit. Default recovery action re-initializes the mean and covariances of float ambiguities.}
    \label{fig:fdir-score}
\end{figure}

The core idea of \hyperref[alg:prefit-stats]{Algorithm~1} is to model prefit residuals as a stationary stochastic process with non-zero decay, implying a time-invariant mean and variance \cite{gallager2013stochastic}. These are estimated empirically using a memoryless moving window. If enough anomalous measurements occur such that the estimated outlier fraction exceeds $g(\mu_i, \sigma_i^2)$, the algorithm deems the current distribution invalid and raises an incapability flag. For general distributions, $g$ may use Chebyshev's Inequality \cite{gallager2013stochastic}; for unimodal ones, the Vysochanskij-Petunin inequality applies \cite{vysochanskij1980inequality, pukelsheim1994threesigma}. Importantly, the moving window does not retain time-series history of $M$ samples, making its memoryless design well suited for flight computers. \autoref{fig:fdir-score} gives the reader a preview ahead of the results, to the FDIR flight software response to and recovery from a deliberately injected cycle slip.

\autoref{fig:fdir-score} shows a time-series of SDCP measurement prefit residuals, with cycle slips injected at $t = 24$ hours, from a two-spacecraft campaign initialized with a 200m along-track separation over $\approx 2.5$ days. Measurements are received every 10s. Each slip was modeled as a Gaussian with bias and noise $\approx 1500 \pm 150$ cycles. Two observations followed: (1) most measurements were rejected by \autoref{eq:innovation-threshold}, and (2) accepted measurements had large enough residuals to bump up the outlier count until it exceeded $g$, triggering an anomaly flag. The system responded by resetting the mean and covariance of $\Tilde{N}_{ZD}$ and $\Tilde{N}_{SD}$ to their \textit{a priori} distributions. Post-reset, the prefit residuals returned to normal values after 1 orbit, with gradual recovery in the outlier score. This exhibits robust navigation and successful FDIR by leveraging statistical guarantees in a memoryless algorithm suited for flight use.

% ========================================
% 5. NAVIGATION SOFTWARE ARCHITECTURE
% ========================================

\section{Navigation Software Design}
\label{section5}

% ========================================
% 5.1 NAVIGATION SOFTWARE: OVERVIEW
% ========================================

\subsection{Overview of Software Design and Interface}

\begin{figure}[H]
	\centering
    \includegraphics[width=1.0\textwidth]{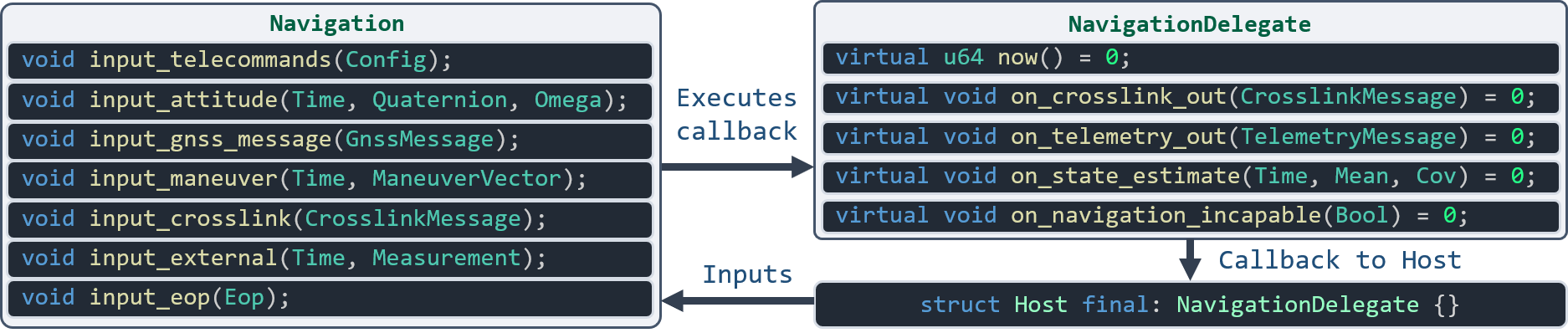}
    \caption{Software interface reflecting the implementation of \autoref{fig:interfaces}. Function signatures use C++ style syntax for illustration.}
    \label{fig:software}
\end{figure}

The \code{Navigation} class executes timely, accurate, and robust state estimation by fusing local and remote (crosslink) measurements with auxiliary data (e.g., maneuvers, attitude, ephemerides). A decision logic determines whether each incoming measurement triggers a filter update. On events \textit{e.g.} state estimates, crosslink, telemetry etc, \code{Navigation} calls back to a \code{NavigationDelegate} inherited by the host class. The software emphasizes encapsulation, modularity, robustness, and efficiency. It is designed for turn-key integration into any host software (referred to here as the \code{Host}), without additional dependencies. The \code{Host} communicates via a single header interface, shown in \autoref{fig:software}. While written in C++17, the architecture generalizes to any object-oriented language. Best practices are followed to ensure development efficiency, ease of testing, and fast runtimes.

\textbf{Encapsulation:} Implementation details of \code{Navigation} are hidden behind a Pointer-to-Implementation design (PImpl) to achieve encapsulation with separation of concerns. This reduces compile-time dependencies, as changes to \code{Navigation} will not require a re-compilation of the \code{Host} files, since the interface remains unchanged. This facilitates faster development overall.

\textbf{Modularity and Flexibility of Response Logic:} The callback-based interface in \autoref{fig:software} allows the \code{Host} to interface callbacks from \code{Navigation} through class inheritance, restricting method overrides solely to the host. Method overriding allows the \code{Host} to customize its behavior in response to callbacks, allowing for flexible response logic and ease of integration.

\textbf{Robustness to Race Conditions:} Ensuring data integrity of \code{Navigation} in real-time is critical to its successful operation. Only a single thread is ever employed in \code{Navigation}. The use of lock guards or mutexes is implemented to ensure that only one thread at a time may interact with \code{Navigation}, which protects states and data from being executed on by several concurrent operations simultaneously. This preserves state consistency, protects data from corruption, and prevents race conditions or deadlock.

\textbf{Maximizing Software Efficiency with Sparsity:} Sparse matrix operations are used wherever possible to reduce runtime, especially given the quadratic growth of covariance size with state dimension. A useful metric is the sparsity ratio $Sp(\mathbf{A}) = 1 - M_0 / mn$ for any matrix $\mathbf{A} \in \mathbb{R}^{n \times m}$, where $M_0$ denotes the number of zero elements. In the time update, the process noise matrix $\mathbf{Q}_t$ is highly sparse, as empirical accelerations do not affect clock offsets or ambiguity states. During the measurement update, the Jacobians $\mathbf{H}_t^{(x)} = \partial h / \partial \Vec{x} \in \mathbb{R}^{m \times n}$ and $\mathbf{H}_t^{(w)} = \partial h / \partial \Vec{w} \in \mathbb{R}^{m \times \ell}$ are also sparse, as measurements depend only on a subset of state variables. Dynamic programming techniques are adopted also for reuse of intermediate results in repeated matrix operations, reducing redundant computations and improving memory efficiency.

% ========================================
% 5.2 NAVIGATION SOFTWARE: DATA STREAMING
% ========================================

\subsection{Data Streaming and Sequencing}

\begin{figure}[H]
  \centering
  \includegraphics[width=0.8\textwidth]{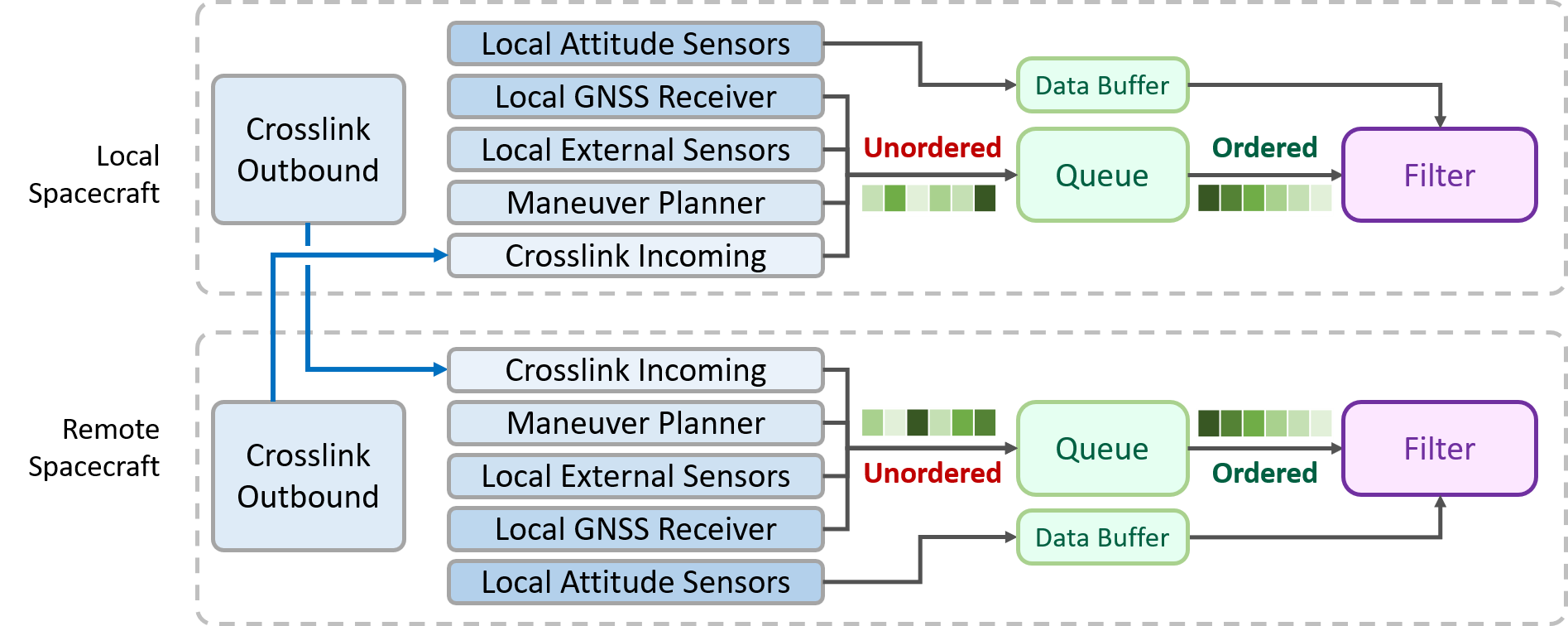}
  \caption{\code{Queue} implemented as a streaming-sorting intermediary between the Interface and the Navigation Filter.}
  \label{fig:software-queue}
\end{figure}

A challenge observed during software integration is that data updates from sensor and actuator streams into \code{Navigation} are not guaranteed to arrive with a time-of-arrival that is in the same sequence as their time-stamps. This is because data are generated by several independent hardware sources per spacecraft (see \autoref{fig:software-queue}), each operating with distinct clocks and communicating over separate signal pathways. Since the recursive navigation filter progresses only forward in time, the \code{Navigation} interface employs a specialized \code{Queue} object to manage these incoming disordered data streams. The \code{Queue} buffers, reorders, and emits the updates in order of time stamps for processing by the filter.

The \code{Queue} is essentially an object executing a streaming-sorting algorithm subjected to temporal constraints. Unlike batch sorting algorithms, it cannot reorder updates that have already been emitted to the filter, nor can it preemptively sort updates not yet received. This limitation means that if no assumptions are made about the maximum delay of an update entering the queue, it is impossible to perform a perfect streaming sort. For example, if the filter's current time is $t=0$ and a GNSS measurement update with a time stamp of $t=2$ arrives, a perfect queue would face a dilemma: it cannot emit the $t=2$ time-stamped update immediately, as there remains the possibility of receiving an earlier $t=1$ time-stamped update later due to latency, which would need to be emitted first. This uncertainty underscores the need for a pragmatic balance in queuing strategy. A set of critical parameters in the design of the \code{Queue} are its buffer periods, which are duration for which updates are held before being forwarded to the filter. Short buffer periods enable faster integration of new information, ensuring the filter's estimate remains timely. Conversely, long buffer periods allow the system to accommodate out-of-order messages streaming in later, reducing the likelihood of discarding latent updates. In the extreme case, a zero-second buffer period would immediately emit any updates as they arrive, and any incoming out-of-order updates will be dropped. The \code{Queue} makes the following assumptions:

\begin{enumerate}
    \item Packets from the same hardware source will arrive in order with respect to each other.
    \item Packets from any source on the local spacecraft arrive at most 2 seconds late.
    \item Packets from any source on the remote spacecraft arrive at most 4 seconds late.
\end{enumerate}

With these assumptions, an `optimal' \code{Queue} can be designed, with the understanding that any received updates that violates these assumptions will be dropped. The caveat is that \code{Navigation} then requires the ability to schedule (via a callback) any state updates in the future, past the expiration of the buffer period, if data is presently unavailable due to latency.
The \code{Queue} also intentionally facilitates seamless transitions between absolute and relative state estimation, ensuring robust operation during crosslink outages. If remote updates are unavailable beyond the 4-second buffer, the Queue continues filtering with local data alone and automatically resumes relative navigation once crosslink is restored. This transition mechanism is verified in \autoref{section6} to meet Requirement \hyperref[req:2.6]{R2.6}.

% ========================================
% 5.3 NAVIGATION SOFTWARE: DYNAMIC RESIZING
% ========================================

\subsection{Dynamic State Resizing}

\begin{figure}[H]
	\centering
    \includegraphics[width=1.0\textwidth]{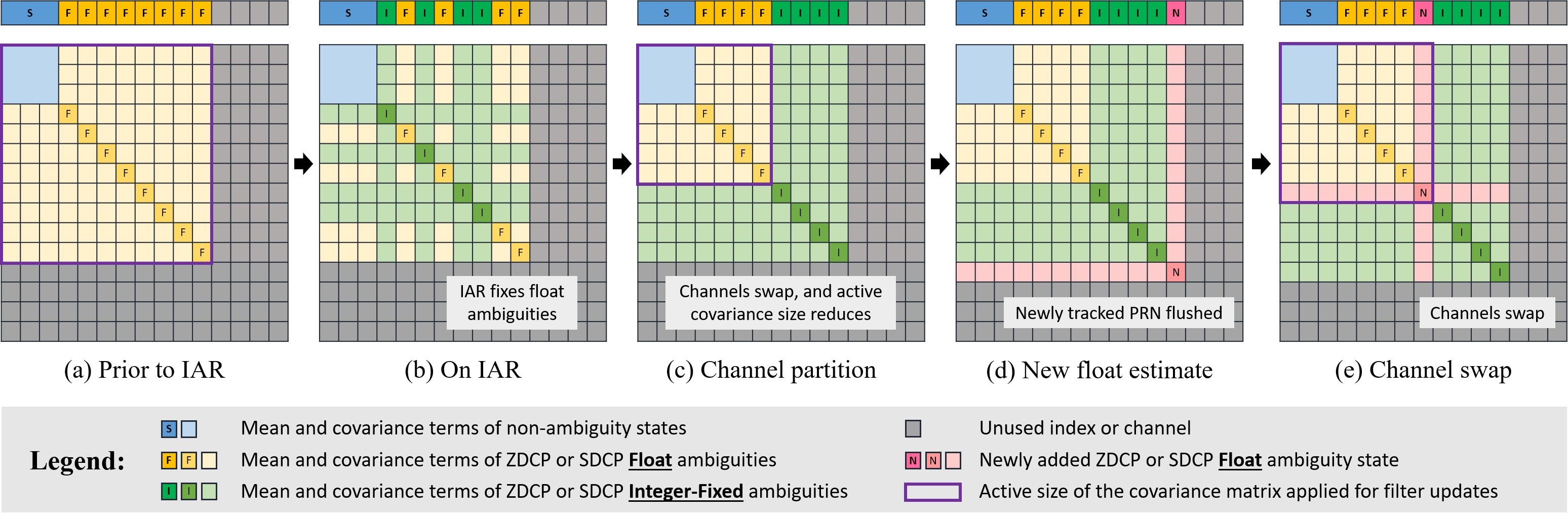}
    \caption{Dynamic resizing and channel flushing of the active covariance matrix.}
    \label{fig:covariance-dynamic}
\end{figure}

Onboard compute efficiency is essential for timely, accurate state estimates that support guidance, control, and decision-making downstream. Beyond exploiting sparsity, this work introduces dynamic state resizing to reduce computational load, especially since matrix multiplications scale as $\mathcal{O}(n^3)$. Resizing is achievable because not all channels are active. After IAR, fixed integers can be treated as known constants with zero uncertainty. Dynamic resizing removes such inactive states, reducing the state mean and covariance to only active states. Floats are flushed first, integers last, as shown in \autoref{fig:covariance-dynamic},

In \autoref{fig:covariance-dynamic}, (a) exemplifies the state mean and covariance prior to IAR in which channels may be ZDCP or SDCP float ambiguities; (b) shows the states after successful IAR, which could be performed on any arbitrary channel index; (c) applies channel flushing so that integer-fixed channels are contiguous with unused channels; (d) illustrates the next time instance where a previously untracked float ambiguity from a carrier phase measurement enters the state and occupies the first unused channel in the state vector; (e) performs a channel swap to flush the new float ambiguities channels with existing ones, in order to ensure that the active states remain contiguous. The final covariance size is hence bounded only by the last float ambiguity channel. Channel flushes can be executed by a series of channel swaps. Channel swaps involve swapping the $i$-th and $j$-th indices of the state mean and channel meta-data (PRN ID, GNSS system and signal, lock times etc), followed by swapping the $i$-th and $j$-th rows, and then the $i$-th and $j$-th columns of the state covariance $\mathbf{\Sigma}$. This can be represented mathematically using the permutation matrix $\mathbf{\Pi}$ as in $\Vec{x} \longleftarrow \mathbf{\Pi} \Vec{x}$ and $\mathbf{\Sigma} \longleftarrow \mathbf{\Pi} \mathbf{\Sigma} \mathbf{\Pi}^\top$.

% ========================================
% 6. TEST RESULTS
% ========================================

\section{Test Campaigns and Results}
\label{section6}

% ========================================
% 6.1 TEST RESULTS: SIM MODELING
% ========================================

\subsection{Testbed Architecture and Modeling Assumptions}

\begin{figure}[H]
    \centering
    \includegraphics[width=0.85\linewidth]{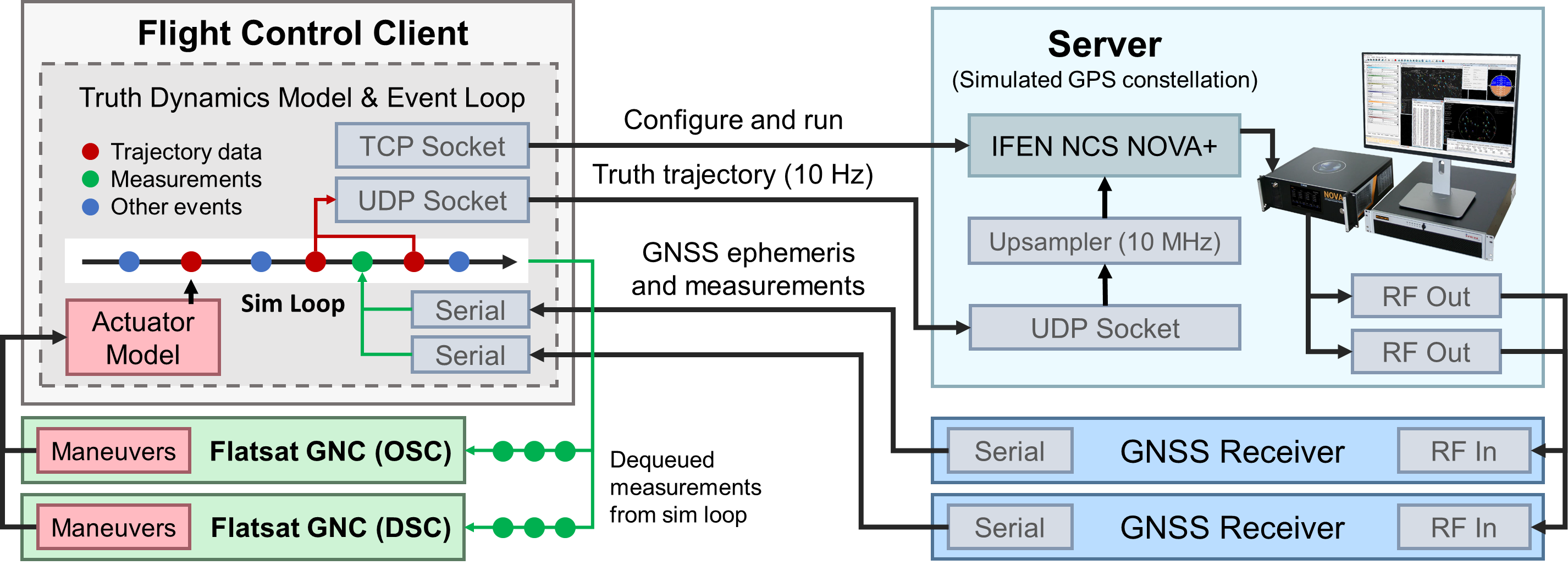}
    \caption{Hardware-in-the-loop (HIL) testbed for closed-loop evaluation of the DiGiTaL v2 software.}
    \label{fig:gnss-ifen-client}
\end{figure}

This section describes the end-to-end testbed, referenced to \autoref{fig:gnss-ifen-client}, for evaluating the navigation flight software, as well as modeling assumptions in the dynamics, measurements, and simulation environment. The flow of events is scheduled in real-time via a hybrid discrete-time continuous-time event loop \cite{toby2025sim}.
Uncertainty modeling with regards to ground truth dynamics, measurement generation, and spacecraft-specific contributions are summarized in \autoref{tab:test-dynamics}, \autoref{tab:test-meas}, and \autoref{tab:test-spacecraft} respectively. The client executes a dynamics-driven event loop in real-time \cite{toby2025sim}, transmitting truth trajectory and attitude at 10-Hz to an IFEN NCS NOVA+ GNSS signal simulator (server). GNSS RF signals then stimulate two Novatel OEM628 GNSS receivers, whose measurements are streamed via serial to two DiGiTaL software instances. Crosslink between each DiGiTaL object is software-emulated with latencies.

\begin{table}[H]
\centering
\caption{Dynamics modeling for both the simulation truth and flight software.}
\label{tab:test-dynamics}
\renewcommand{\arraystretch}{1.05} % adjust row spacing
\small
\begin{tabular}{lll}
    \hline
    & \textbf{Ground truth dynamics} & \textbf{Filter dynamics} \\
    \hline
    \textbf{State representation} & Quasi-nonsingular orbital elements & Cartesian \\
    \grayhline
    \textbf{State equations} & Gauss’ variational equations & Fundamental orbital differential equations \\
    \grayhline
    \textbf{Integrator} & RK45 (Dormand-Prince) & RK4 (non-adaptive step size) \\
    \grayhline
    \textbf{Geopotential} & GGM05S (60x60) & GGM01S (20x20) \\
    \grayhline
    \textbf{Drag model} & \begin{tabular}[c]{@{}l@{}}NRLMSISE 00, F10.7 = 184.6, Ap = 18.4\end{tabular} & Harris-Priester \\
    \grayhline
    \textbf{Solar radiation pressure} & Analytical sun ephemeris with cylindrical shadow & -- \\
    \grayhline
    \textbf{Third body gravity} & Analytical Sun \& Moon ephemeris & -- \\
    \hline
\end{tabular}
\end{table}

\vspace{-10mm}

\begin{minipage}[t]{0.46\textwidth}
\begin{table}[H]
\footnotesize
\centering
\renewcommand{\arraystretch}{1.25} % adjust row spacing
\caption{GNSS signal-in-space conditions modeling}
\label{tab:test-meas}
\begin{tabular}{p{3.5cm} p{3.2cm}}
    \hline
    \textbf{GNSS signals} & GPS L1 only, Block IIR-M \\ 
    \grayhline
    \textbf{GNSS aperture} & 170 degrees \\
    \grayhline
    \textbf{GNSS gain/phase pattern} & Spherical only \\
    \grayhline
    \textbf{GNSS ephemeris source} & Broadcast (no corrections) \\
    \grayhline
    \textbf{Ionospheric model} & Klobuchar (delay only) \\
    \grayhline
    \textbf{Receiver Earth horizon mask} & -20 degrees from horizon \\
    \grayhline
    \textbf{Receiver aperture mask} & 190 degrees \\
    \grayhline
    \textbf{Receiver simulated LNA} & 33 dBW \\
    \grayhline
    \textbf{Receiver gain/phase pattern} & ANTCOM 1.9G1215A \\
    \hline
\end{tabular}
\end{table}
\end{minipage}
\hfill
\begin{minipage}[t]{0.5\textwidth}
\begin{table}[H]
\footnotesize
\centering
\renewcommand{\arraystretch}{1.25} % adjust row spacing
\caption{Spacecraft uncertainty parameters.}
\label{tab:test-spacecraft}
\begin{tabular}{p{3.2cm} p{0.8cm} p{1.2cm} p{0.75cm}}
    \hline
    \textbf{Parameter} & \textbf{Unit} & \textbf{Dist.} & \textbf{Value} \\
    \hline
    \textbf{Spacecraft mass} & kg & Gaussian & 0.04 \\
    \grayhline
    \textbf{Spacecraft COM} & mm & Uniform & 3 \\
    \grayhline
    \textbf{GNSS antenna PCO bias} & mm & - & 50 \\
    \grayhline
    \textbf{GNSS antenna PCO noise} & mm & Gaussian & 1 \\ 
    \grayhline
    \textbf{GNSS antenna direction} & arcsec & Gaussian & 30 \\ 
    \grayhline
    \textbf{Attitude knowledge} & arcsec & Gaussian & 20 \\ 
    \grayhline
    \textbf{Maneuver magnitude} & $\%$ & Gaussian & 5 \\ 
    \grayhline
    \textbf{Maneuver direction} & arcsec & Gaussian & 60 \\ 
    \hline
\end{tabular}
\end{table}
\end{minipage}

\begin{table}[ht]
\centering
\renewcommand{\arraystretch}{1.1}
\caption{Klobuchar ionospheric model parameters used in ground truth measurement emulation.}
\label{tab:klobuchar}
\begin{tabular}{llllllll}
    \hline
    $\alpha_1$ & $\alpha_2$ & $\alpha_3$ & $\alpha_4$ & $\beta_1$ & $\beta_2$ & $\beta_3$ & $\beta_4$ \\
    \grayhline
    2.794E-8 & -7.451E-9 & -1.192E-7 & 1.788E-7 & 137200 & -98300 & 65540 & -393200 \\
    \hline
\end{tabular}
\end{table}

% ========================================
% 6.2 TEST RESULTS: COMPLIANCE TESTING
% ========================================

\subsection{Requirements Verification and Compliance Tests}

This section outlines compliance test scenarios in \autoref{tab:test-campaigns}, with graduating fidelity. It provides traceability by mapping surveyed requirements of \autoref{tab:reqs} to compliance tests, via the RVM in \autoref{tab:test-rvm}. Wide test coverage stresses weak points in the software, revealing where software or navigation logic may fail. This uncovers hidden vulnerabilities, across modes of \autoref{fig:modes}, and ensures the system can handle realistic off-nominal conditions. Test campaigns adopt the VISORS mission concept \cite{visors2023aas} for the full-campaign. For brevity, only scenarios \textbf{T.4.1}, \textbf{T.4.7}, and \textbf{T.7.2} (highlighted in green) are discussed with detailed time-series results in this paper. 

\begin{table}[H]
\centering
\small
\caption{Full set of compliance test campaigns.}
\label{tab:test-campaigns}
\renewcommand{\arraystretch}{1.2} % adjust row spacing
\begin{tabular}{@{}p{0.32\textwidth} p{0.7\textwidth}@{}}
\toprule
\textbf{Test Campaign} & \textbf{Test Scenarios} \\
\midrule

\multirow{1}{=}{\textbf{T.1} - Unit Tests} 
    & \textbf{T.1.1} - Software built-in-tests to verify numerical correctness and function behaviour \\
\grayhline

\multirow{3}{=}{\textbf{T.2} - Initialization and Interfacing} 
    & \textbf{T.2.1} - Receiver cold start test (no prior almanac) \\
    & \textbf{T.2.2} - Receiver warm start test (with prior almanac) \\
    & \textbf{T.2.3} - Sample packet test for crosslink, attitude, telecommands and propulsion \\
\grayhline

\multirow{3}{=}{\textbf{T.3} - Software Emulated GNSS in the Loop Nominal Scenarios} 
    & \textbf{T.3.1} - Standby, uncontrolled, 12 hours \\
    & \textbf{T.3.2} - Full campaign (standby-transfer-operations), 36 orbits \\
    & \textbf{T.3.3} - Full mission, 10 campaigns with 1 week of Standby between with role switches \\
\grayhline

\multirow{5}{=}{\textbf{T.4} - Software Emulated GNSS in the Loop for Degraded Scenarios} 
    & \cgl \textbf{T.4.1} - Standby, uncontrolled, 3 orbits. Total crosslink outage for 1 orbit \\
    & \textbf{T.4.2} - Intermittent crosslink drop-rate of 1\%, with 99\% persistence \\
    & \textbf{T.4.3} - GNSS drop-rate of 1\%, with 75\% persistence \\
    & \textbf{T.4.4} - Maneuvers successfully executed but with 100\% packet drop-rate to navigation \\
    & \textbf{T.4.5} - Elevated noise in measurements, dynamics, and maneuvers to 300\% nominal noise \\
    & \textbf{T.4.6} - Float ambiguity resolution only, no IAR \\
    & \cgl \textbf{T.4.7} - Software-emulated cycle slip injection \\
\grayhline

\multirow{2}{=}{\textbf{T.5} - Software Emulated GNSS in the Loop for Contingency Scenarios} 
    & \textbf{T.5.1} - Science/operations mode, but force-interrupted with triggered escape \\
    & \textbf{T.5.2} - Science/operations mode, but force-interrupted with bus safe mode \\
\grayhline

\multirow{1}{=}{\textbf{T.6} - Software Emulated Sweep Tests} 
    & \textbf{T.6.1} - Full campaign, 36 orbits, with a variable sweep over uncertainty parameters \\
\grayhline

\multirow{2}{=}{\textbf{T.7} - Hardware Emulated Real-Time GNSS in the Loop Nominal Scenarios} 
    & \textbf{T.7.1} - Standby, no station-keeping, sun-pointing attitude, 1 day \\
    & \cgl \textbf{T.7.2} - Full campaign, 36 orbits, in real-time (standby-transfer-operations), with role-switch \\
\grayhline

\multirow{2}{=}{\textbf{T.8} - Hardware Earth-Fixed Live-Sky Antenna Test} 
    & \textbf{T.8.1} - Live-sky daytime navigation test \\
    & \textbf{T.8.2} - Live-sky night time navigation test \\

\bottomrule
\end{tabular}
\end{table}

All test scenarios in \autoref{tab:test-campaigns} are periodically run throughout development, and were presented in the wider VISORS Flight Software review held in Stanford University, in November 2024.
This was paneled by notable reviewers in the space flight community from across multiple universities, aerospace companies, and public science agencies.

\vspace{1mm}

\textbf{T.1}: Unit tests form the foundation of software validation, with test coverage on navigation functions, state transformations, dynamics and measurement modeling, time scale conversions, coordinate rotations etc. A Continuous-Integration and Continuous-Delivery (CI/CD) pipeline integrates these tests after every software commit to ensure functional correctness and catch edge cases, bugs, or unreachable states, prior to any integrated tests.

\vspace{1mm}

\textbf{T.2}: Interface testing validates the accuracy and resilience of the flight software's interaction with data sources. These tests begin with software-emulated GPS receiver packets, crosslink packets, and telecommands. Fidelity progresses to hardware-emulated packets from a NovAtel OEM628 driven by the IFEN NOVA+ GNSS signal simulator as in \autoref{fig:gnss-ifen-client}, and then actual live-sky GNSS packets via rooftop antennas for a zero-baseline test. Live-sky tests validate interfacing with both cold (no prior almanac) and warm starts (with almanac) on the receiver. All interfacing algorithms are tested to ensure correct packet parsing, checksum validation, and reliable operation across all conditions.

\vspace{1mm}

\textbf{T.3 - T.6}: Software-in-the-loop GNSS emulation is a critical enabler for rapid development and deployment, due to faster-than-real-time testing. Modeling realism remains consistent with Tables \ref{tab:test-dynamics}, \ref{tab:test-meas}, \ref{tab:test-spacecraft} and \ref{tab:klobuchar}. Furthermore, it permits testing \code{Navigation} with other emulated subsystems (e.g., Guidance, Control, Safety) for full mission simulation of VISORS \cite{visors2023aas}. This framework allows thorough performance evaluation across performance degraded scenarios (\textbf{T.4}), contingency scenarios such as safe modes or during an escape trajectory (\textbf{T.5}), and robustness to variations in uncertainty parameters via sweep tests (\textbf{T.6}). These campaigns graduates the software readiness ahead of an integrated hardware-in-the-loop testing.

\vspace{1mm}

\textbf{T.7 - T.8}: Hardware-in-the-loop (HIL) GNSS testing represents the culmination of the test campaign, validating the full analog-digital signal chain with the DiGiTaL v2 navigation flight software at its core. It exercises all components of the software: interfacing, functionality, and performance, in real-time with latencies, and with other software subsystems \textit{e.g.} Guidance, Controller, and Safety, in closed-loop. GNSS measurements generated by the hardware in \autoref{fig:gnss-ifen-client} rely on ground truth states, which are driven by the event loop \cite{toby2025sim}. The event loop coordinates the dynamics and the timing asynchronously into a single thread across all threads on each hardware unit (the IFEN NOVA+, the local host running the event loop, and each instance of the \code{Navigation} software. This achieves mission-representative validation of DiGiTaL v2's navigation performance under realistic and real-time operational conditions.

\begin{table}[H]
\centering
\setlength{\tabcolsep}{3pt} % adjust column spacing
\caption{Requirements Verification Matrix (RVM) mapping compliance test in \autoref{tab:test-campaigns} items to navigation requirements in \autoref{tab:reqs}}
\label{tab:test-rvm}
\renewcommand{\arraystretch}{1} % adjust row spacing
\scriptsize
\begin{tabular}{|c|c|c|c|c|c|c|c|c|c|c|c|c|c|c|c|c|c|c|c|c|c|c|c|}
\hline

\cgr & \hyperref[req:1.1]{R1.1} & \hyperref[req:1.2]{R1.2} & \hyperref[req:1.3]{R1.3} & \hyperref[req:1.4]{R1.4} & \hyperref[req:1.5]{R1.5} & \hyperref[req:1.6]{R1.6} & \hyperref[req:2.1]{R2.1} & \hyperref[req:2.2]{R2.2} & \hyperref[req:2.3]{R2.3} & \hyperref[req:2.4]{R2.4} & \hyperref[req:2.5]{R2.5} & \hyperref[req:2.6]{R2.6} & \hyperref[req:2.7]{R2.7} & \hyperref[req:2.8]{R2.8} & \hyperref[req:2.9]{R2.9} & \hyperref[req:3.1]{R3.1} & \hyperref[req:3.2]{R3.2} & \hyperref[req:3.3]{R3.3} & \hyperref[req:3.4]{R3.4} & \hyperref[req:3.5]{R3.5} & \hyperref[req:3.6]{R3.6} & \hyperref[req:3.7]{R3.7} & \hyperref[req:3.8]{R3.8} \\ \hline

T.1.1 & & & & & & & & & & & & & & \cg & & & & & & & & & \\ \hline

T.2.1 & \cg & & & & & & & & \cg & & & & & & & & & & & & & & \\ \hline

T.2.2 & \cg & & & & & & & & \cg & & & & & & & & & & & & & & \\ \hline

T.2.3 & & \cg & \cg & \cg & \cg & & & & & & & & & & & & & & & & & & \\ \hline

T.3.1 & \cg & \cg & & \cg & & \cg & \cg & \cg & \cg & & & & \cg & & & \cg & \cg & \cg & & & \cg & & \\ \hline

T.3.2 & \cg & \cg & \cg & \cg & \cg & \cg & \cg & \cg & \cg & \cg & & & \cg & & & \cg & \cg & \cg & \cg & & \cg & & \\ \hline

T.3.3 & \cg & \cg & \cg & \cg & \cg & \cg & \cg & \cg & \cg & \cg & \cg & & \cg & & & \cg & \cg & \cg & \cg & & \cg & & \\ \hline

T.4.1 & \cg & \cg & & \cg & & \cg & \cg & \cg & \cg & & & \cg & \cg & & & \cg & \cg & \cg & & \cg & \cg & & \\ \hline

T.4.2 & \cg & \cg & & \cg & & \cg & \cg & \cg & \cg & & & \cg & \cg & & & \cg & \cg & \cg & & \cg & \cg & & \\ \hline

T.4.3 & \cg & \cg & & \cg & & \cg & \cg & \cg & \cg & & & \cg & \cg & & & \cg & \cg & \cg & & \cg & \cg & & \\ \hline

T.4.4 & \cg & \cg & & \cg & & \cg & \cg & \cg & \cg & & & \cg & \cg & & & \cg & \cg & \cg & & & \cg & \cg & \\ \hline

T.4.5 & \cg & \cg & & \cg & & \cg & \cg & \cg & \cg & & & & \cg & & & \cg & \cg & \cg & & & \cg & \cg & \cg \\ \hline

T.4.6 & \cg & \cg & & \cg & & \cg & \cg & \cg & \cg & & & & \cg & & & \cg & \cg & & & & \cg & & \\ \hline

T.4.7 & \cg & \cg & & \cg & & \cg & \cg & \cg & \cg & & & & \cg & & \cg & \cg & \cg & \cg & & & \cg & & \\ \hline

T.5.1 & \cg & \cg & \cg & \cg & \cg & \cg & \cg & \cg & \cg & \cg & \cg & & \cg & & & \cg & \cg & \cg & \cg & & \cg & & \\ \hline

T.5.2 & \cg & \cg & \cg & \cg & \cg & \cg & \cg & \cg & \cg & \cg & \cg & & \cg & & & \cg & \cg & \cg & \cg & & \cg & & \\ \hline

T.6.1 & \cg & \cg & \cg & \cg & \cg & \cg & \cg & \cg & \cg & \cg & & & \cg & & & \cg & \cg & \cg & \cg & & \cg & & \\ \hline

T.7.1 & \cg & \cg & & \cg & & \cg & \cg & \cg & \cg & \cg & & \cg & \cg & & \cg & \cg & \cg & \cg & \cg & & \cg & & \cg \\ \hline

T.7.2 & \cg & \cg & \cg & \cg & \cg & \cg & \cg & \cg & \cg & \cg & \cg & \cg & \cg & & \cg & \cg & \cg & \cg & \cg & & \cg & & \cg \\ \hline

T.8.1 & \cg & & & \cg & & \cg & \cg & & \cg & & & \cg & & & \cg & \cg & \cg & \cg & & & \cg & & \cg \\ \hline

T.8.2 & \cg & & & \cg & & \cg & \cg & & \cg & & & \cg & & & \cg & \cg & \cg & \cg & & & \cg & & \cg \\ \hline

\end{tabular}
\end{table}

% ========================================
% 6.2.A TEST RESULTS: VISORS FULL CAMPAIGN
% ========================================

\paragraph{Hardware-in-the-Loop Real-Time Test:}
Here, results from the full VISORS campaign scenario \textbf{T.7.2} with GNSS hardware-in-the-loop as per \autoref{fig:gnss-ifen-client} in real-time are presented. VISORS is a distributed telescope mission for high-resolution imaging of the Sun in the extreme ultraviolet spectrum, using two 6U CubeSats flying in formation in a Sun-synchronous low-Earth orbit. An Optics Spacecraft (OSC) carries a photon sieve lens, while signals passing through the lens is focused on a Detector Spacecraft (DSC). Observations happen at a 40m baseline with sub-cm alignment requirements. Elaborated details of the concept of operations, as well as specifics on the other subsystems can be found in references \cite{visors2021koenig, visors2023aas}. Relative orbit configurations are in \autoref{tab:orbit-config}, with epoch set to 12:00AM, 1 Oct 2024 GPST.

\begin{table}[H]
\centering
\renewcommand{\arraystretch}{0.9} % adjust row spacing
\small
\caption{Reference orbit and quasi-nonsingular relative orbital elements per mode for test conditions.}
\begin{tabular}{@{}lllrr@{}}
\toprule
\multicolumn{2}{c}{\textbf{Reference orbit}} & \multicolumn{3}{c}{\textbf{Relative orbit modes}} \\
\cmidrule(r){1-2} \cmidrule(l){3-5}
\textbf{Element} & \textbf{Sun Sync Parameters} & \textbf{Element} & \textbf{Standby (m)} & \textbf{Science (m)} \\
\midrule
$a$ (km)     & $R_E + 500$ & $a\delta a$      & 0     & $-2.62$ \\
$e$ (-)      & 0.004       & $a\delta\lambda$ & 0     & 45.21 \\
$i$ (deg)    & 97.8        & $a\delta e_x$    & 0     & $-34.51$ \\
$\Omega$ (deg) & 157.5     & $a\delta e_y$    & 200   & 4.78 \\
$u(t_0)$ (deg) & 0         & $a\delta i_x$    & 0     & $-18.72$ \\
LTAN         & 10:00AM     & $a\delta i_y$    & 200   & 2.72 \\
\bottomrule
\end{tabular}
\label{tab:orbit-config}
\end{table}

\textbf{T.7.2} conducts a full test campaign that initializes the DSC and OSC in Standby, transitions to Transfer, and then to Science mode, performing one observation per Science orbit. A Laser Rangefinder (LRF) on the OSC and a Reflector on the DSC enable alignment verification via range measurements. Up to 10 Science orbits, and thus 10 alignment attempts, are executed. Midway, a role-switch is executed to transfer relative orbit control from the one spacecraft to the other, to validate that the \code{Navigation} software instances on both spacecraft remain uninterrupted and performant.

\begin{figure}[H]
    \centering
    \begin{subfigure}[htp]{0.49\textwidth}
        \centering
        \includegraphics[width=\textwidth]{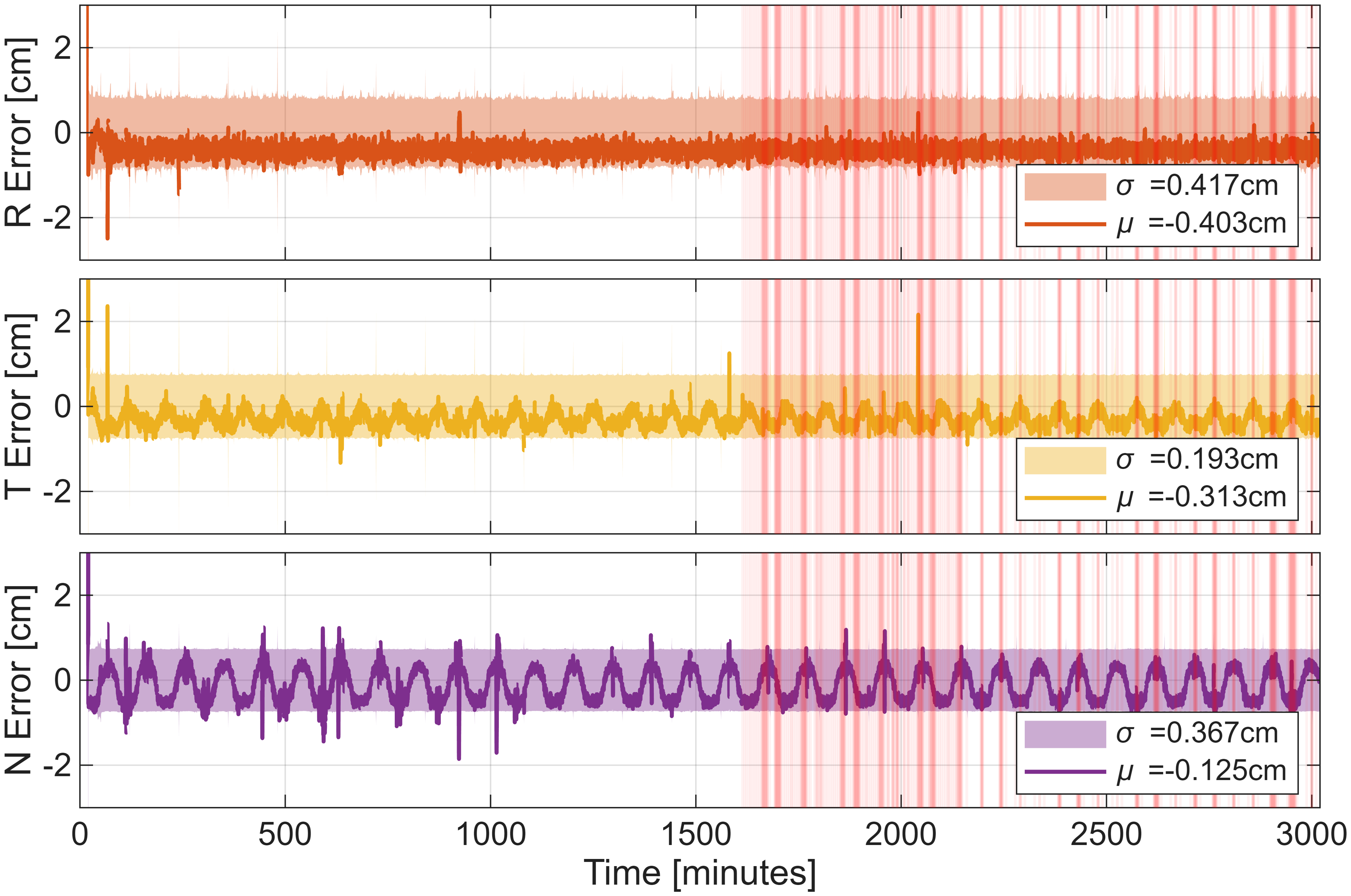}
        \caption{Relative position error and covariance estimated by DSC}
        \label{fig:results-sc0-pos}
    \end{subfigure}
    \hfill
    \begin{subfigure}[htp]{0.49\textwidth}
        \centering
        \includegraphics[width=\textwidth]{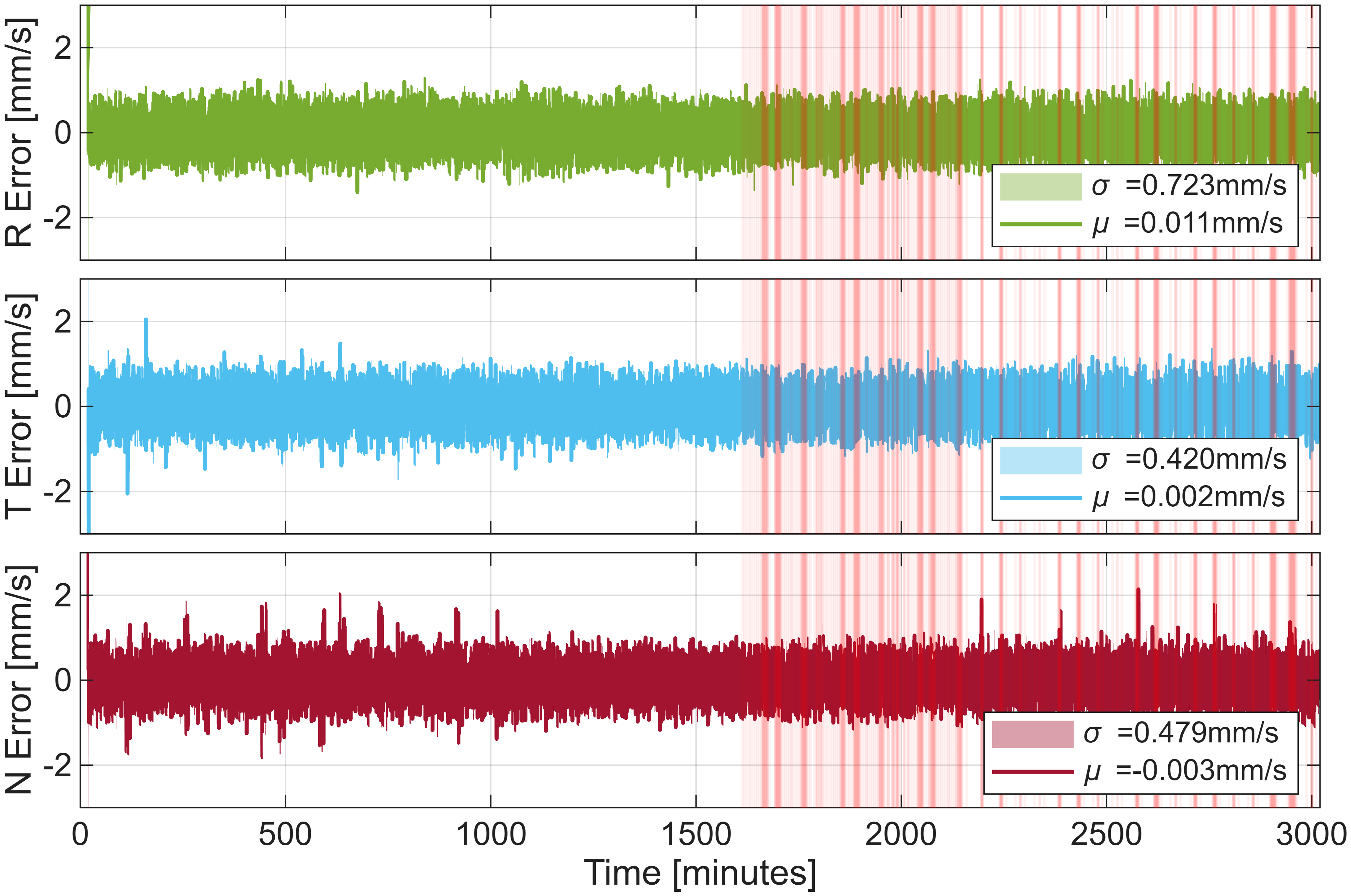}
        \caption{Relative velocity error and covariance estimated by DSC}
        \label{fig:results-sc0-vel}
    \end{subfigure}
    \hfill
    \begin{subfigure}[htp]{0.49\textwidth}
        \centering
        \includegraphics[width=\textwidth]{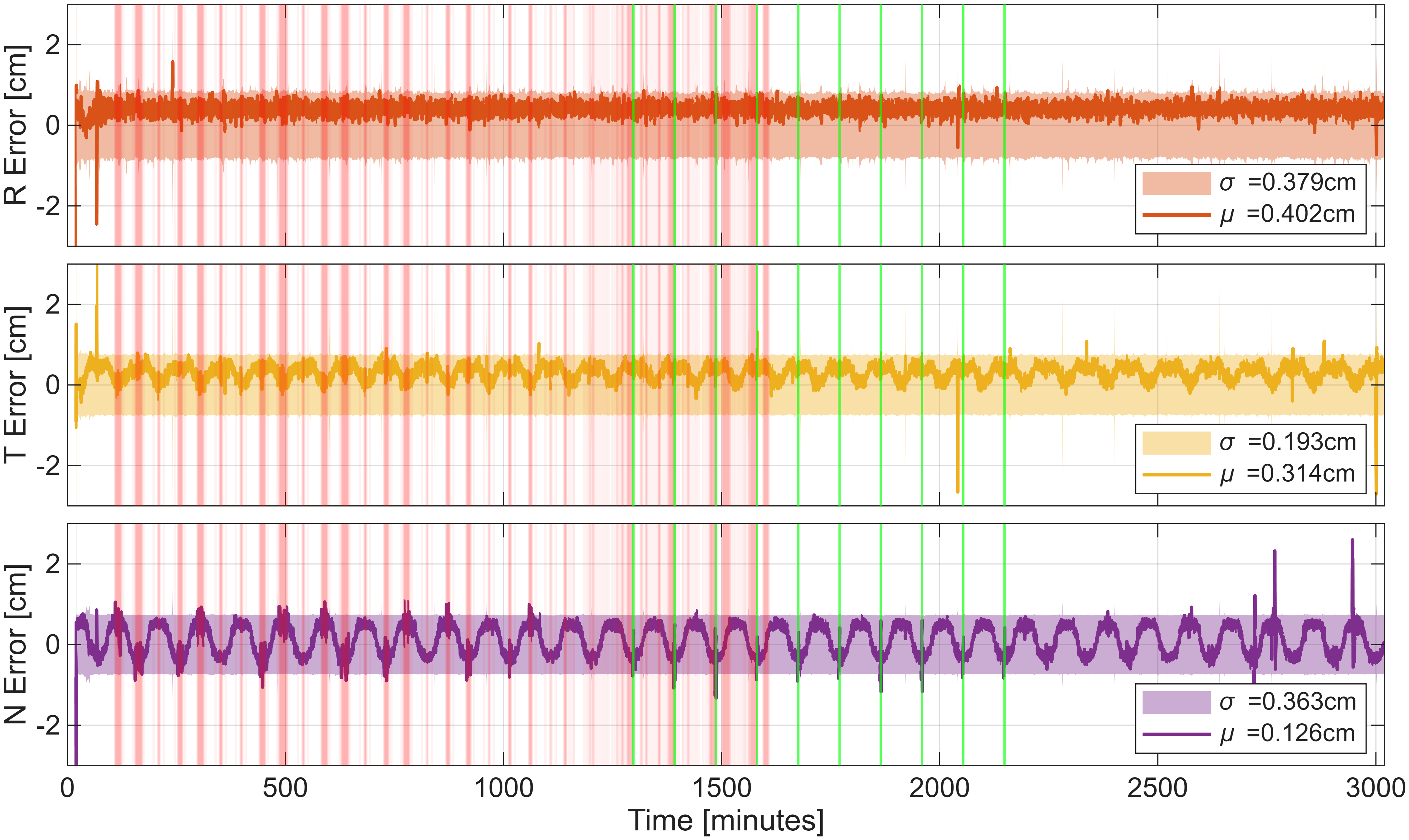}
        \caption{Relative position error and covariance estimated by OSC}
        \label{fig:results-sc1-pos}
    \end{subfigure}
    \hfill
    \begin{subfigure}[htp]{0.49\textwidth}
        \centering
        \includegraphics[width=\textwidth]{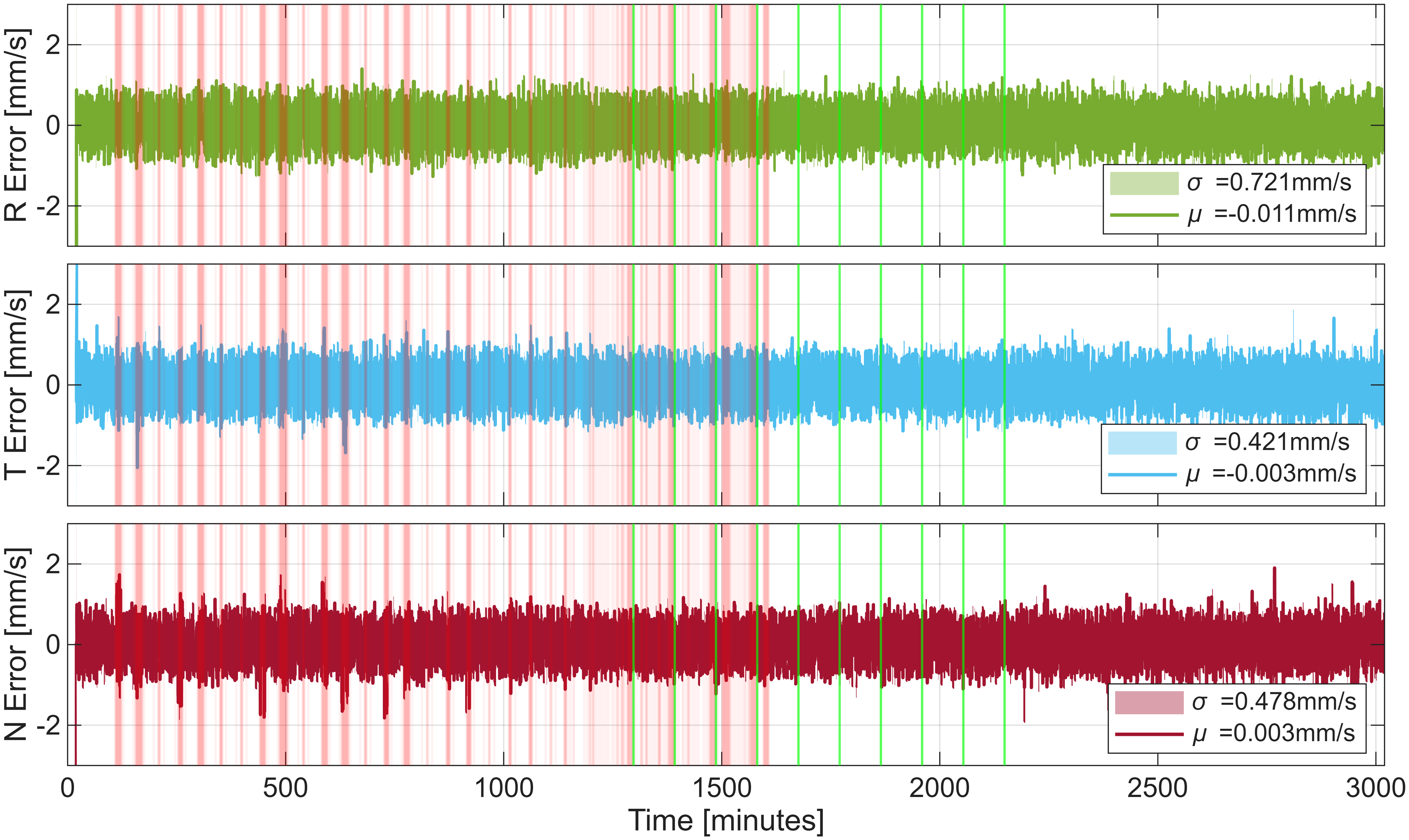}
        \caption{Relative velocity error and covariance estimated by OSC}
        \label{fig:results-sc1-vel}
    \end{subfigure}
    \caption{Relative trajectory errors and covariance of a full VISORS campaign (36 orbits). Translucent \textcolor{red}{red lines} indicate maneuvers, each with impulse bit of $<2$mm/s; while \textcolor{ForestGreen}{green lines} indicate successful LRF alignment.}
    \label{fig:results-sc01-posvel}
\end{figure}

\begin{figure}[H]
    \centering
    \begin{subfigure}[htp]{0.49\textwidth}
        \centering
        \includegraphics[width=\textwidth]{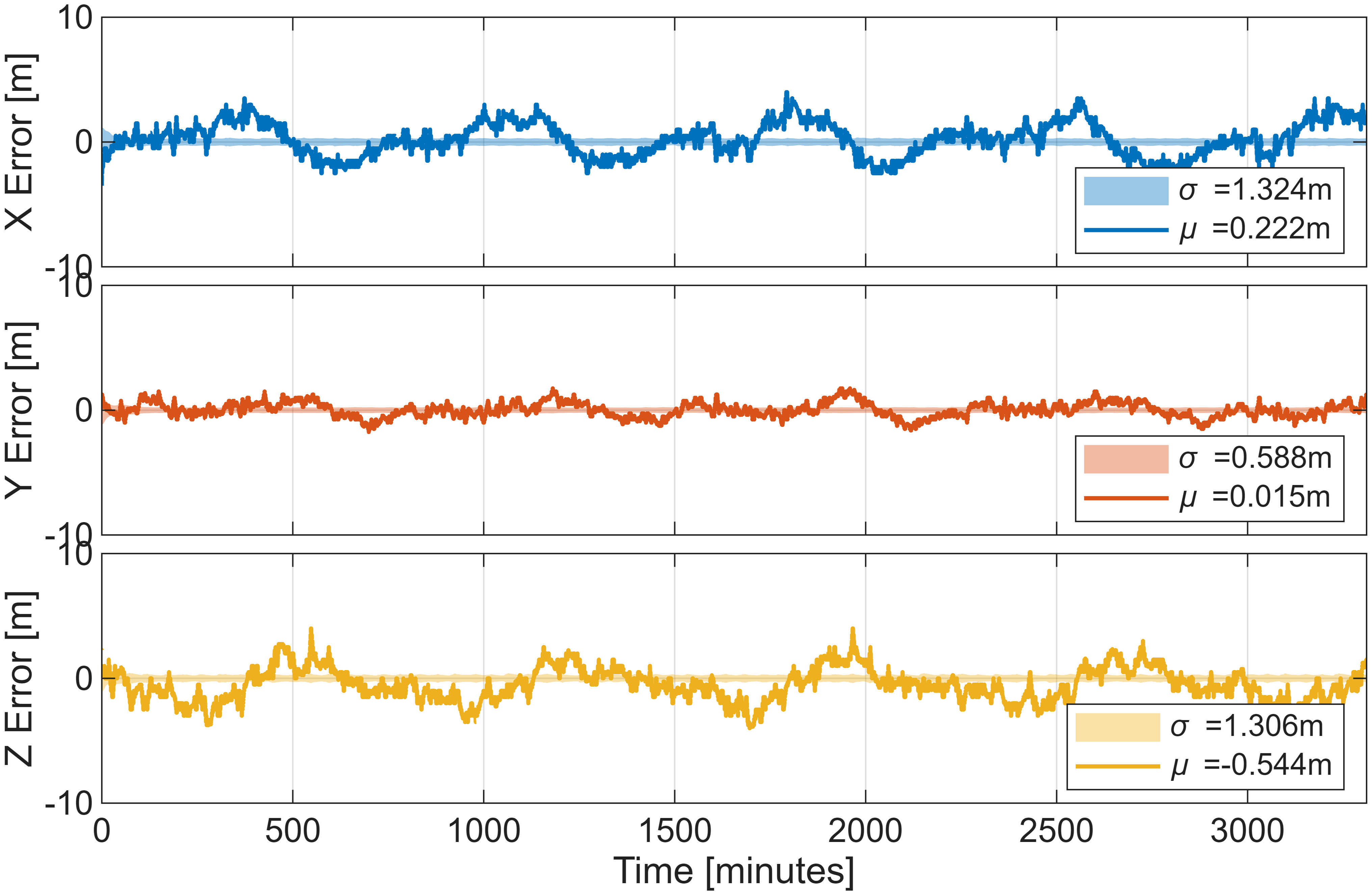}
        \caption{Absolute ECI position errors of the DSC}
        \label{fig:results-abs-pos}
    \end{subfigure}
    \hfill
    \begin{subfigure}[htp]{0.49\textwidth}
        \centering
        \includegraphics[width=\textwidth]{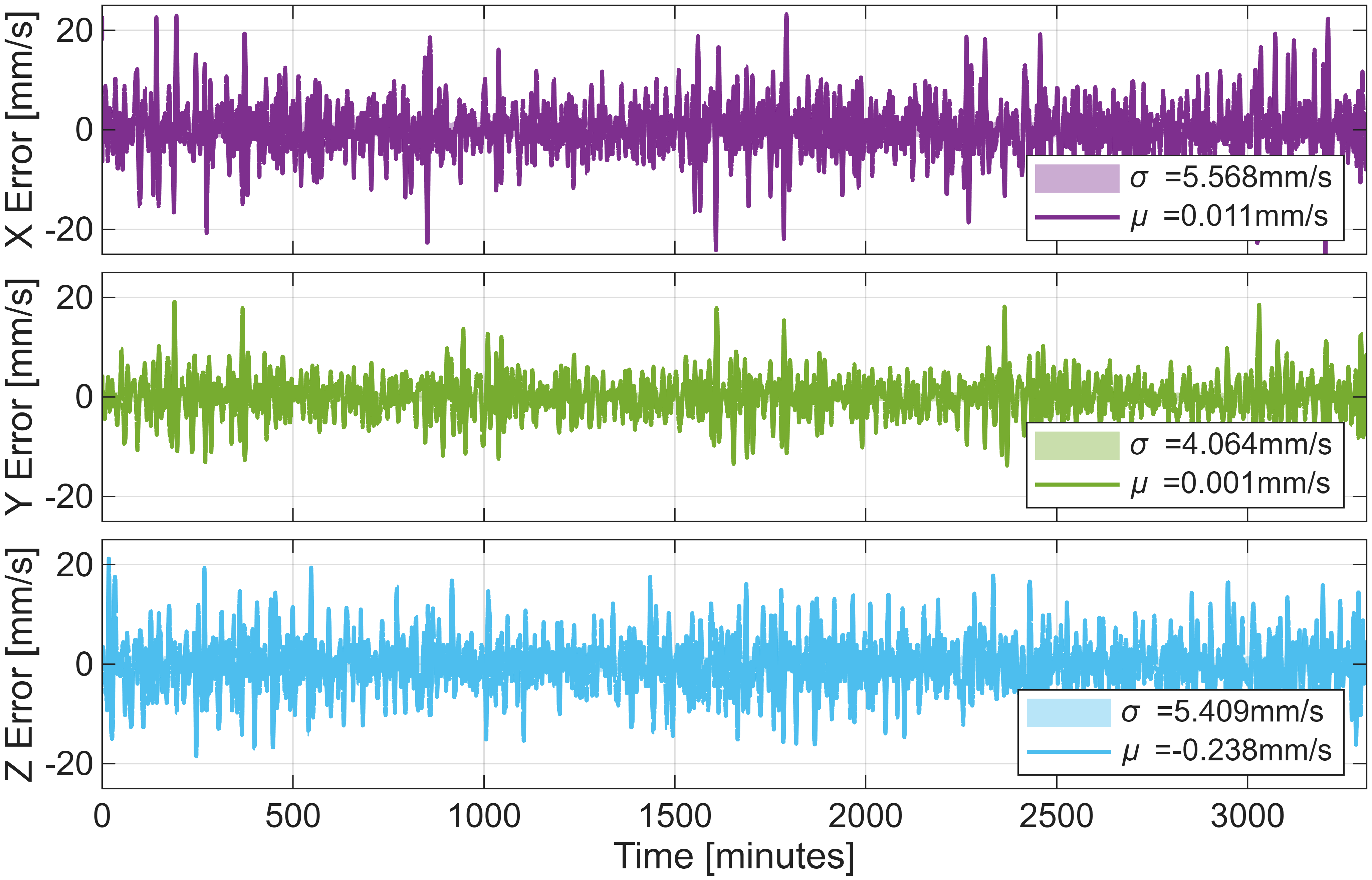}
        \caption{Absolute ECI velocity errors of the DSC}
        \label{fig:results-abs-vel}
    \end{subfigure}
    \vspace{-3mm}
    \caption{Absolute ECI state errors of the DSC. OSC results are similar and omitted for brevity.}
    \label{fig:results-abs-posvel}
\end{figure}

Out of 10 alignment attempts, 4 met observation requirements.
The time of first IAR fix is at $t = 8$ minutes.
Oscillations in relative position errors arise due to the presence of unobservable COM-to-PCO variations simulated, according to \autoref{tab:test-spacecraft}.
This is despite stress-testing the filter by adopting only a 5mm per-axis magnitude in consider state uncertainty, but applying a 5cm actual bias, as per \autoref{tab:test-spacecraft} (one order of magnitude larger).
\autoref{fig:results-sc0-pos} and \autoref{fig:results-sc1-pos} show the filter’s covariance envelope converging conservatively around the true error. The relative positive error mean exhibits only a bias of 4mm, 3mm, and 1.25mm in the $\hat{R}$, $\hat{T}$, and $\hat{N}$ axes.
This desirable result arises, not only due to the presence of consider states, but also because the state vector provides degrees of freedom to absorb static measurement biases (\textit{e.g.} clock offsets and float ambiguity states).
This validates robust state estimation under poorly observed biases.
Next, the relative trajectory, LRF scatter grouping, and measurement residuals are presented.

\vspace{-3mm}
\begin{figure}[H]
    \centering
    \begin{subfigure}[htp]{0.6\textwidth}
        \centering
        \includegraphics[width=\textwidth]{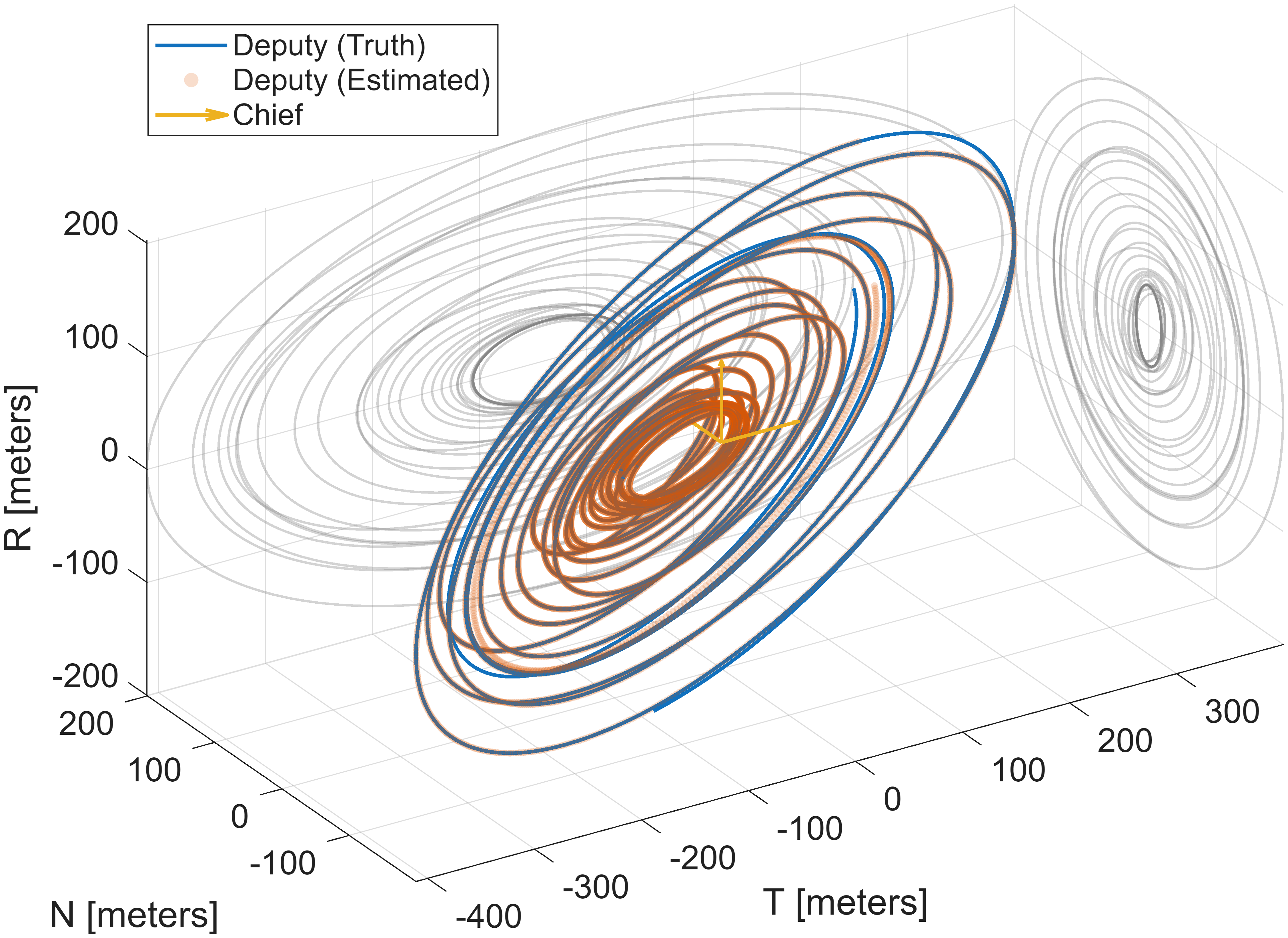}
        \caption{Plot of true and estimated relative orbit, with DSC at the origin.}
        \label{fig:results-orbit}
    \end{subfigure}
    \begin{subfigure}[htp]{0.24 \textwidth}
        \centering
        \includegraphics[width=\textwidth]{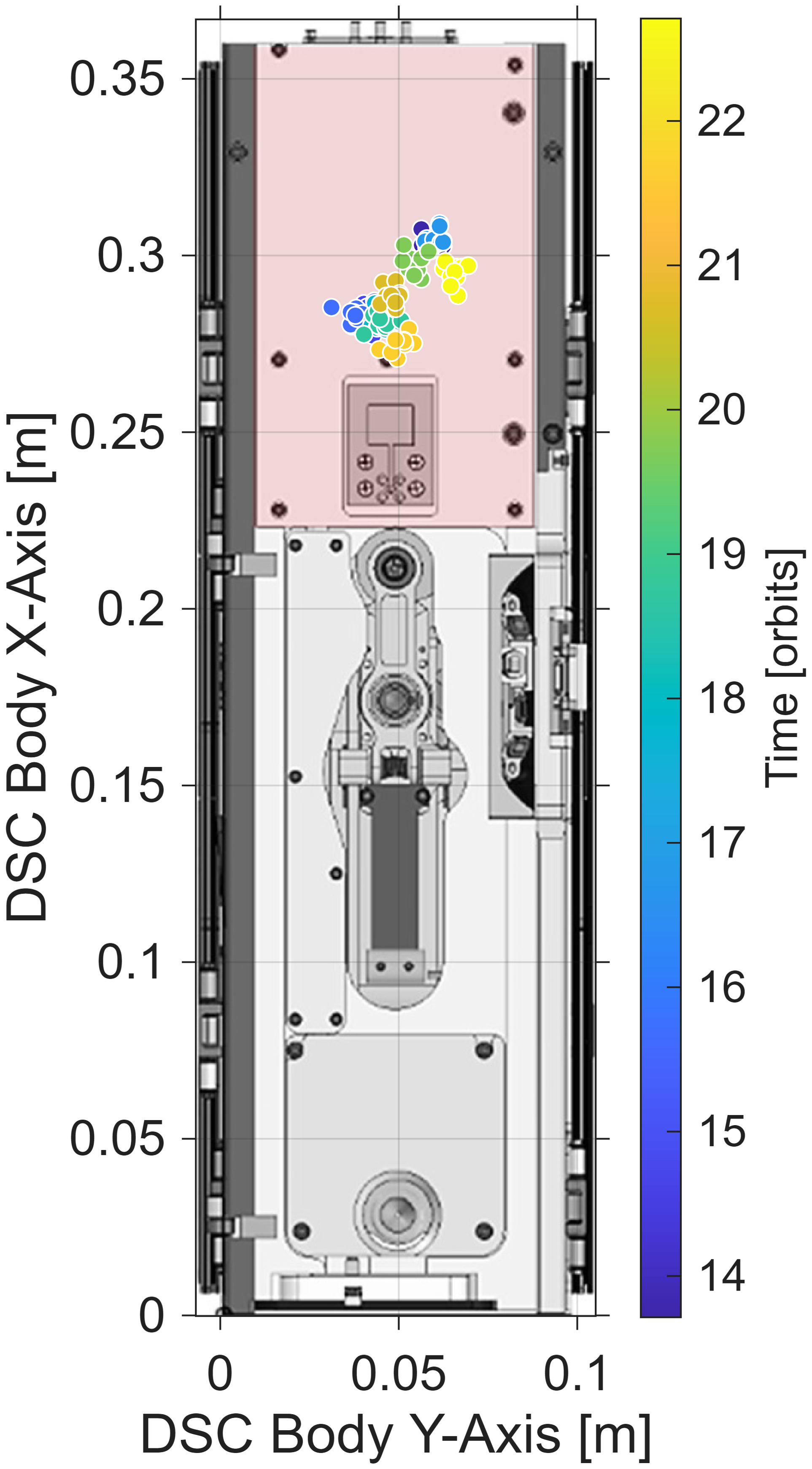}
        \caption{LRF scatter grouping}
        \label{fig:results-lrf}
    \end{subfigure}
    \vspace{-3mm}
    \caption{True vs estimated relative trajectory, with LRF scatter on the DSC reflector panel in the body frame.}
    \label{fig:results-orbit-lrf}
\end{figure}

\vspace{-3mm}
\begin{figure}[H]
    \centering
    \begin{subfigure}[htp]{0.49\textwidth}
        \centering
        \includegraphics[width=\textwidth]{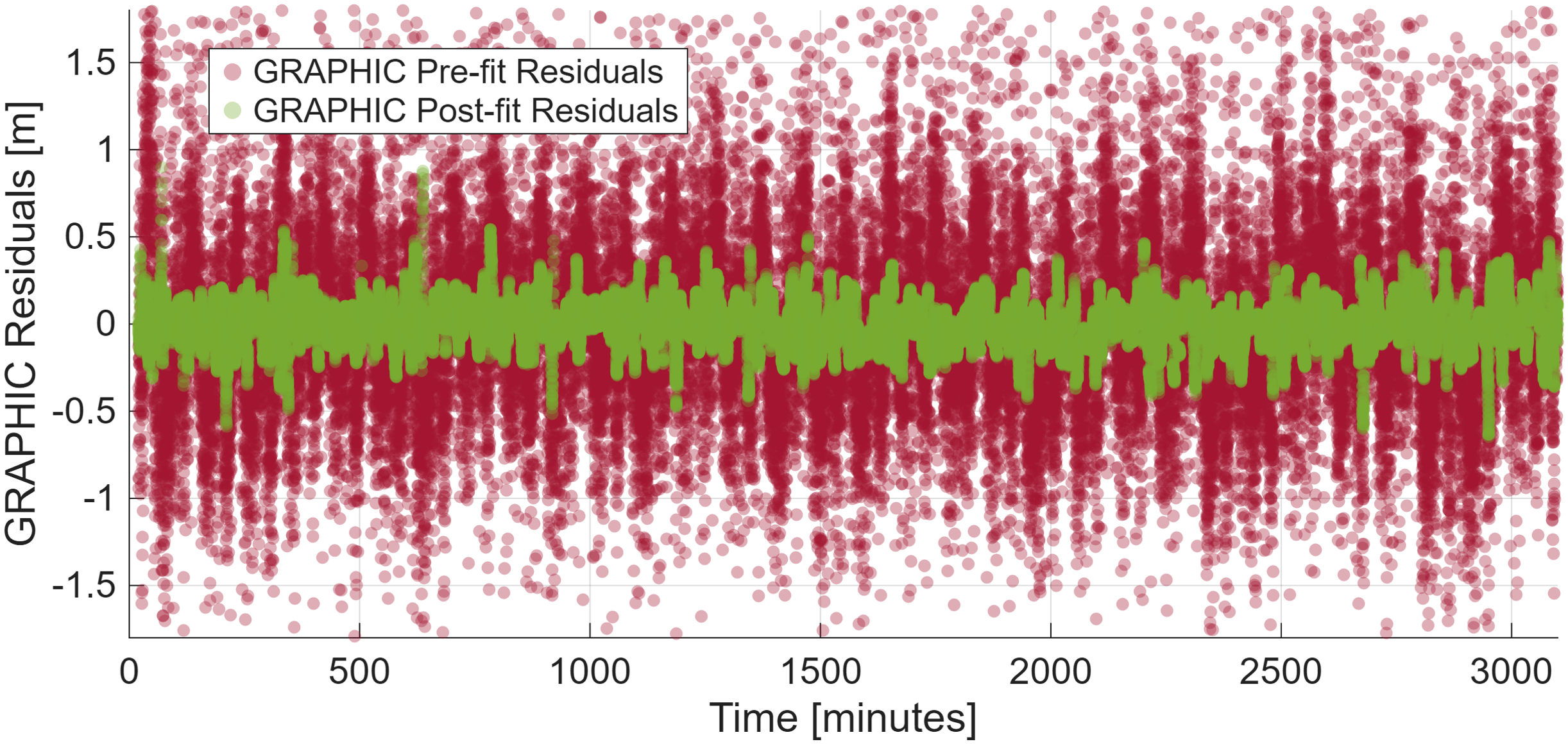}
        \label{fig:results-res-grph}
    \end{subfigure}
    \hfill
    \begin{subfigure}[htp]{0.49\textwidth}
        \centering
        \includegraphics[width=\textwidth]{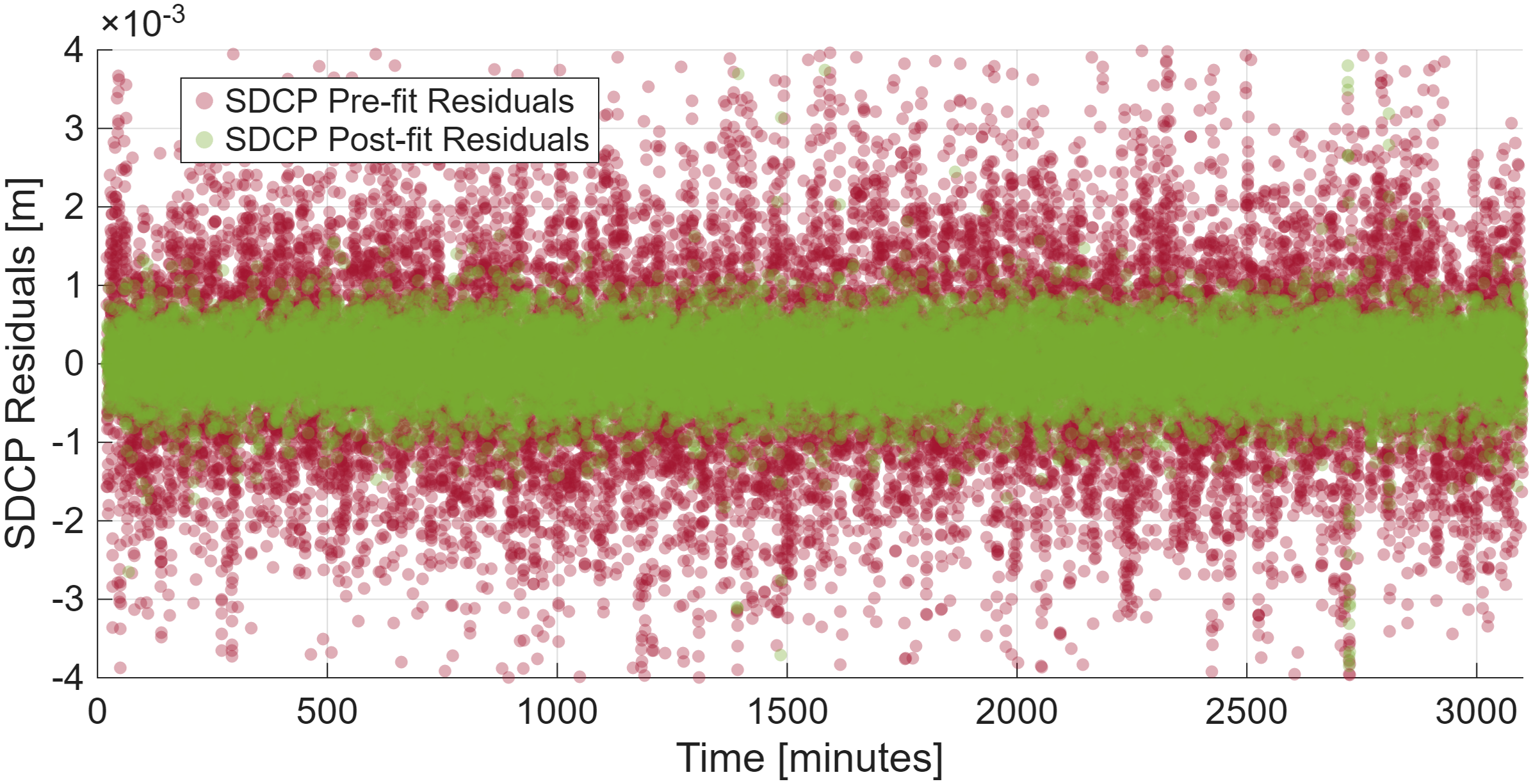}
        \label{fig:results-res-sdcp}
    \end{subfigure}
    \vspace{-3mm}
    \caption{Pre- and post-fit residuals for GRAPHIC (left) and SDCP (right) for the DSC.}
    \label{fig:results-residuals}
\end{figure}

% ========================================
% 6.2.B TEST RESULTS: VISORS CROSSLINK OUTAGE
% ========================================

\begin{figure}[h]
    \centering
    \begin{subfigure}[htp]{0.51\textwidth}
        \centering
        \includegraphics[width=\textwidth]{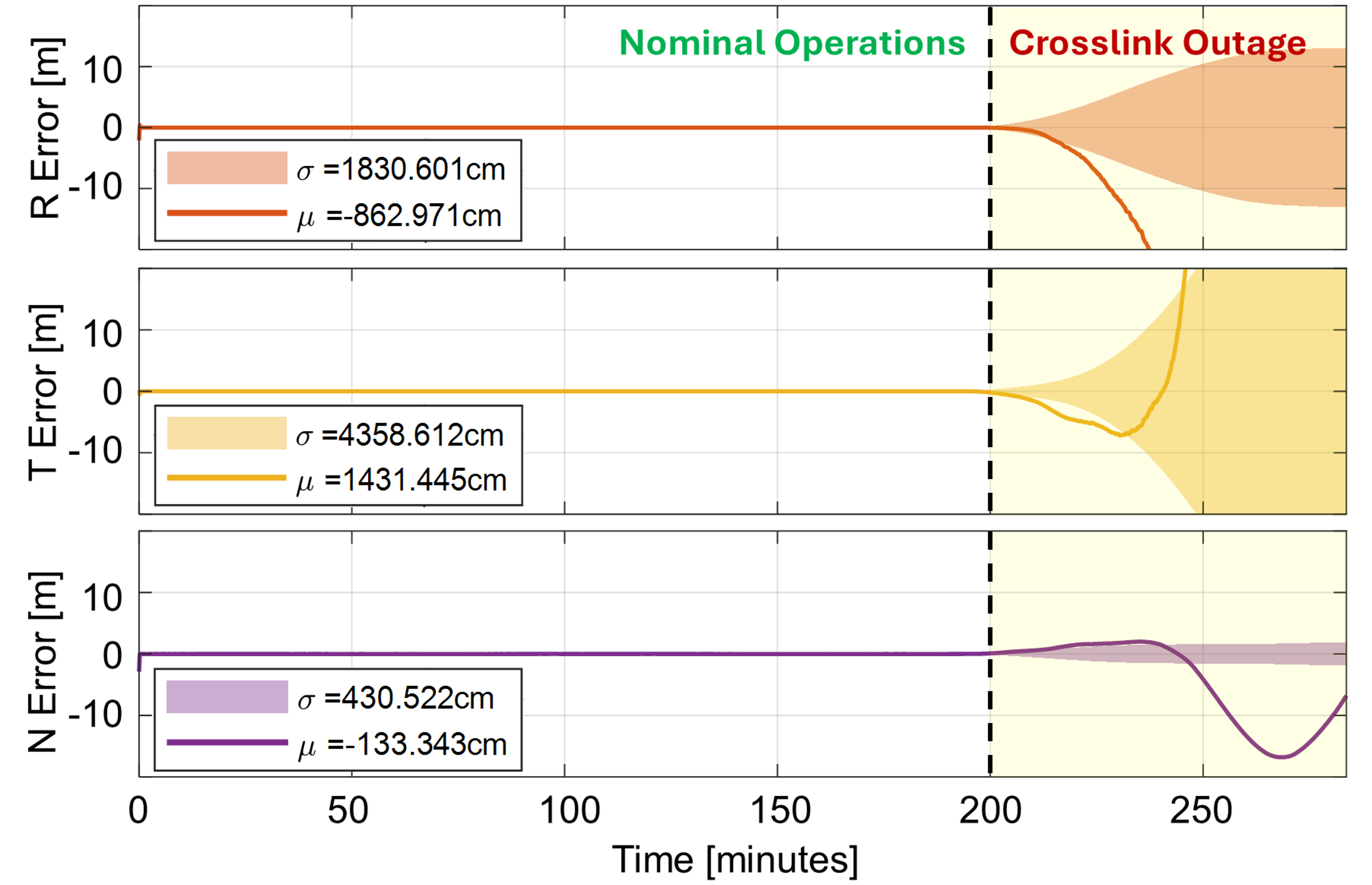}
        \caption{Relative position error in outage, diagonal-only $\mathbf{Q}_t$}
        \label{fig:results-outage-pos-old}
    \end{subfigure}
    \hfill
    \begin{subfigure}[htp]{0.47\textwidth}
        \centering
        \includegraphics[width=\textwidth]{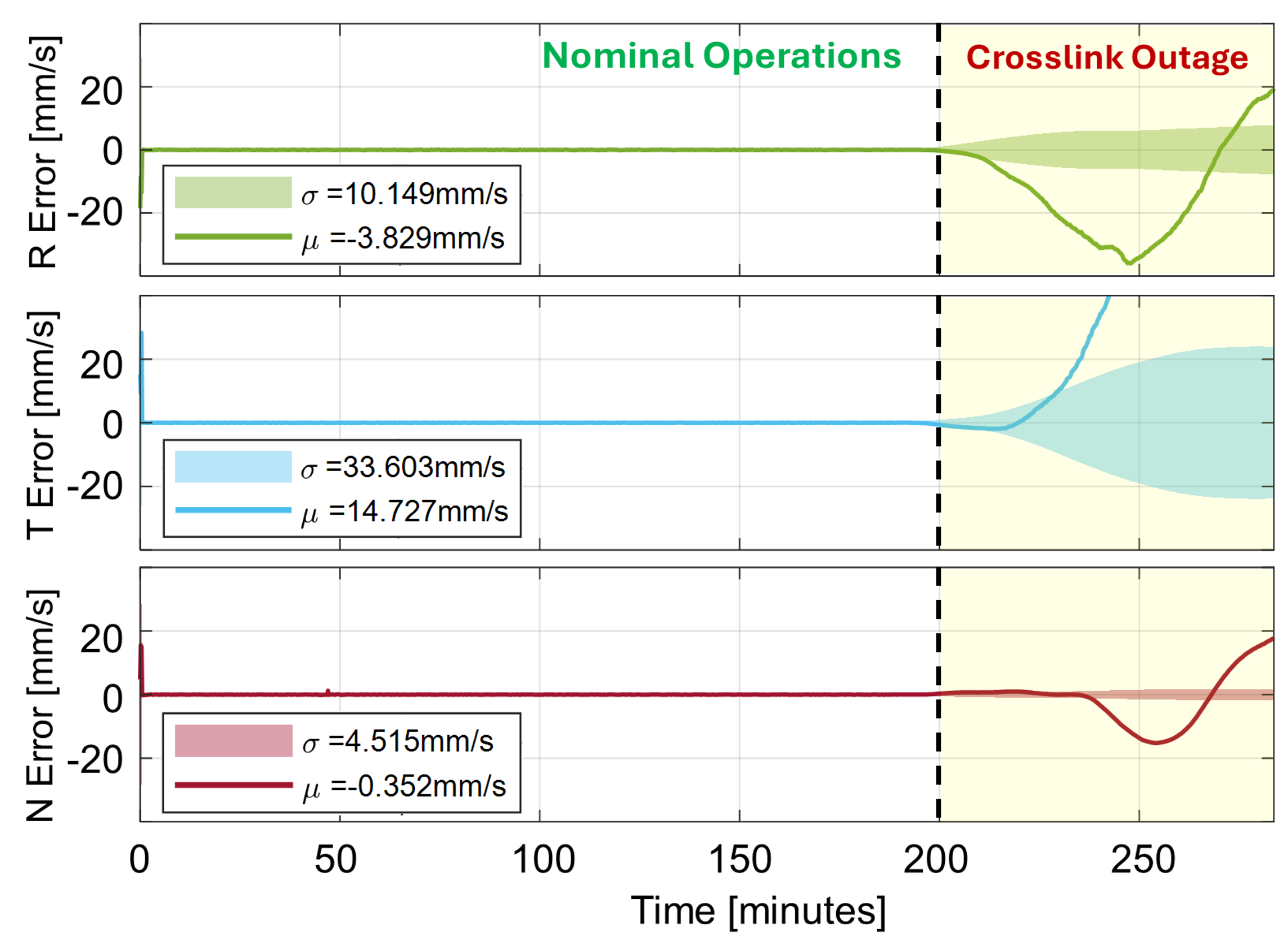}
        \caption{Relative velocity error in outage, diagonal-only $\mathbf{Q}_t$}
        \label{fig:results-outage-vel-old}
    \end{subfigure}
    \hfill
    \begin{subfigure}[htp]{0.51\textwidth}
        \centering
        \includegraphics[width=\textwidth]{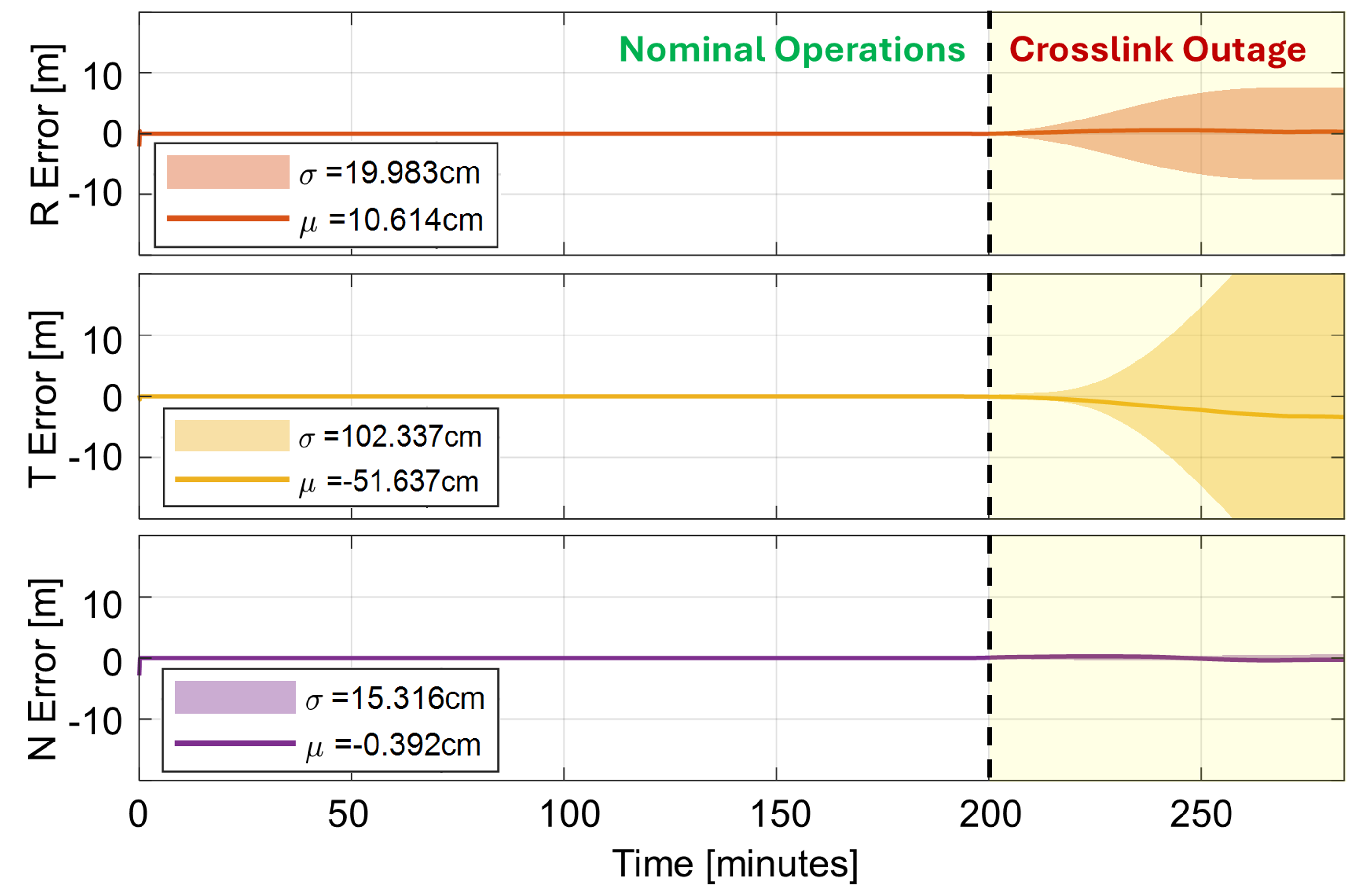}
        \caption{Relative position error in outage, correlated $\mathbf{Q}_t$ [\ref{eq:proc-noise-full}]}
        \label{fig:results-outage-pos-new}
    \end{subfigure}
    \hfill
    \begin{subfigure}[htp]{0.47\textwidth}
        \centering
        \includegraphics[width=\textwidth]{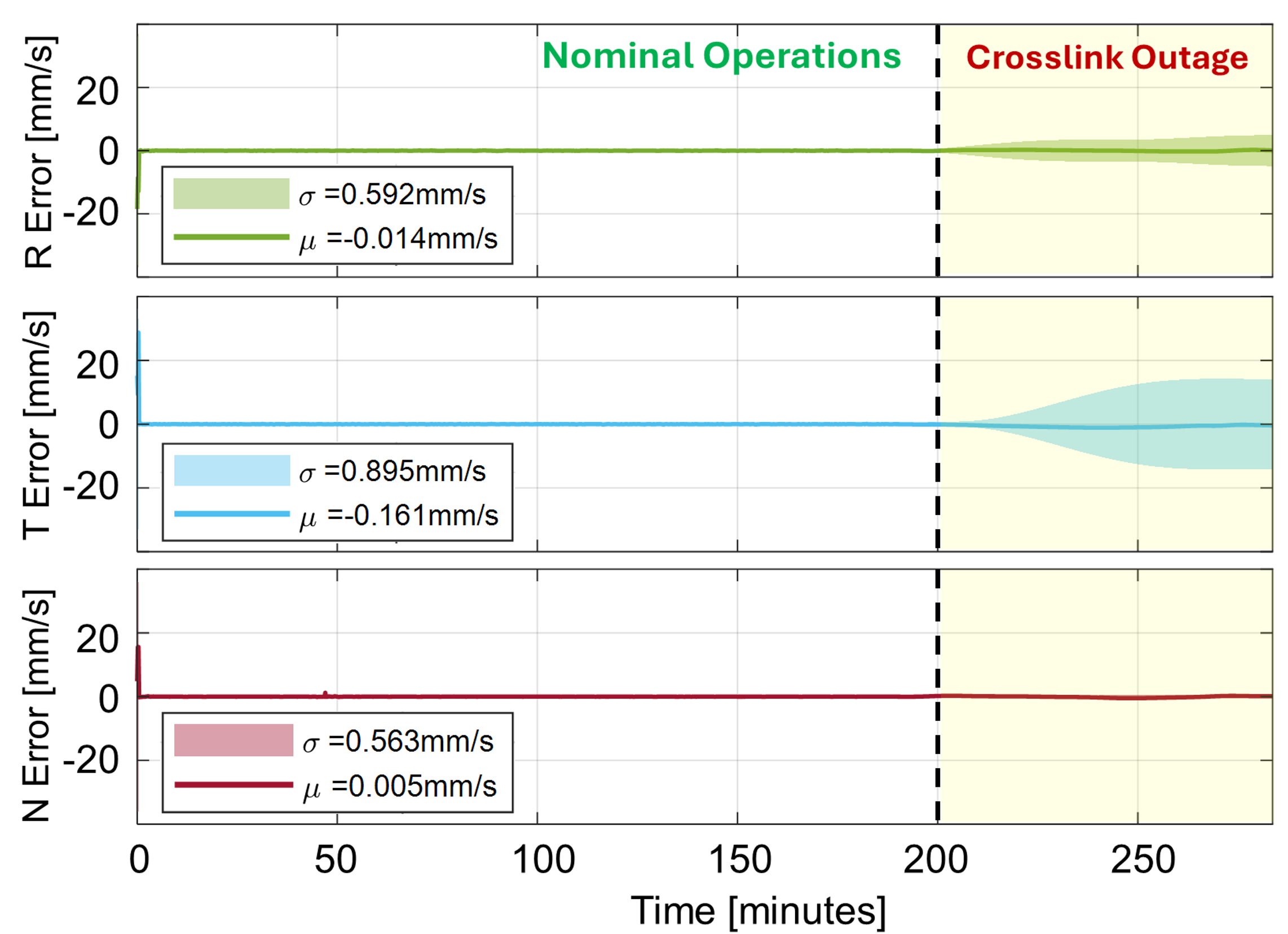}
        \caption{Relative velocity error in outage, correlated $\mathbf{Q}_t$ [\ref{eq:proc-noise-full}]}
        \label{fig:results-outage-vel-new}
    \end{subfigure}
    \caption{Plots of relative navigation performance using the legacy manually tuned diagonal-only process noise in DiGiTaL v1 \cite{giralo2019digital} \textit{(top)}, versus the proposed correlated analytical process noise model of DiGiTaL v2 \textit{(bottom)}.}
    \label{fig:results-outage}
\end{figure}

\paragraph{Software-in-the-Loop Simulated Crosslink Outage:}
Next, scenario \textbf{T.4.1} is presented. This evaluates the impact of the proposed correlated process noise model on orbit prediction accuracy during crosslink outages. Results from the improved DiGiTaL v2 (\autoref{fig:results-outage-pos-new} and \ref{fig:results-outage-vel-new}) are compared with the legacy diagonal-only process noise model from DiGiTaL v1 (\autoref{fig:results-outage-pos-old} and \ref{fig:results-outage-vel-old}) \cite{giralo2019digital}.
The formation is initialized as per \autoref{tab:orbit-config}, under healthy simulated crosslink for 2 orbits. On the 3rd orbit, a crosslink outage is triggered. As shown in \autoref{fig:results-outage}, the new correlated process noise model significantly improves relative position and velocity errors during the outage. By accurately preserving the state covariance correlation structure, the filter time update mitigates state error growth from uncompensated dynamics using the correlation between remote and local spacecraft motion -- even when remote measurements are lost. This demonstrates that the physically consistent process noise model developed in this work significantly enhances prediction fidelity under degraded navigation. This test scenario also demonstrates seamless transition from relative to absolute-only navigation modes when the crosslink becomes unavailable.

% ========================================
% 6.2.B TEST RESULTS: VISORS CROSSLINK OUTAGE
% ========================================

\begin{figure}[H]
    \centering
    \begin{subfigure}[htp]{0.49\textwidth}
        \centering
        \includegraphics[width=\textwidth]{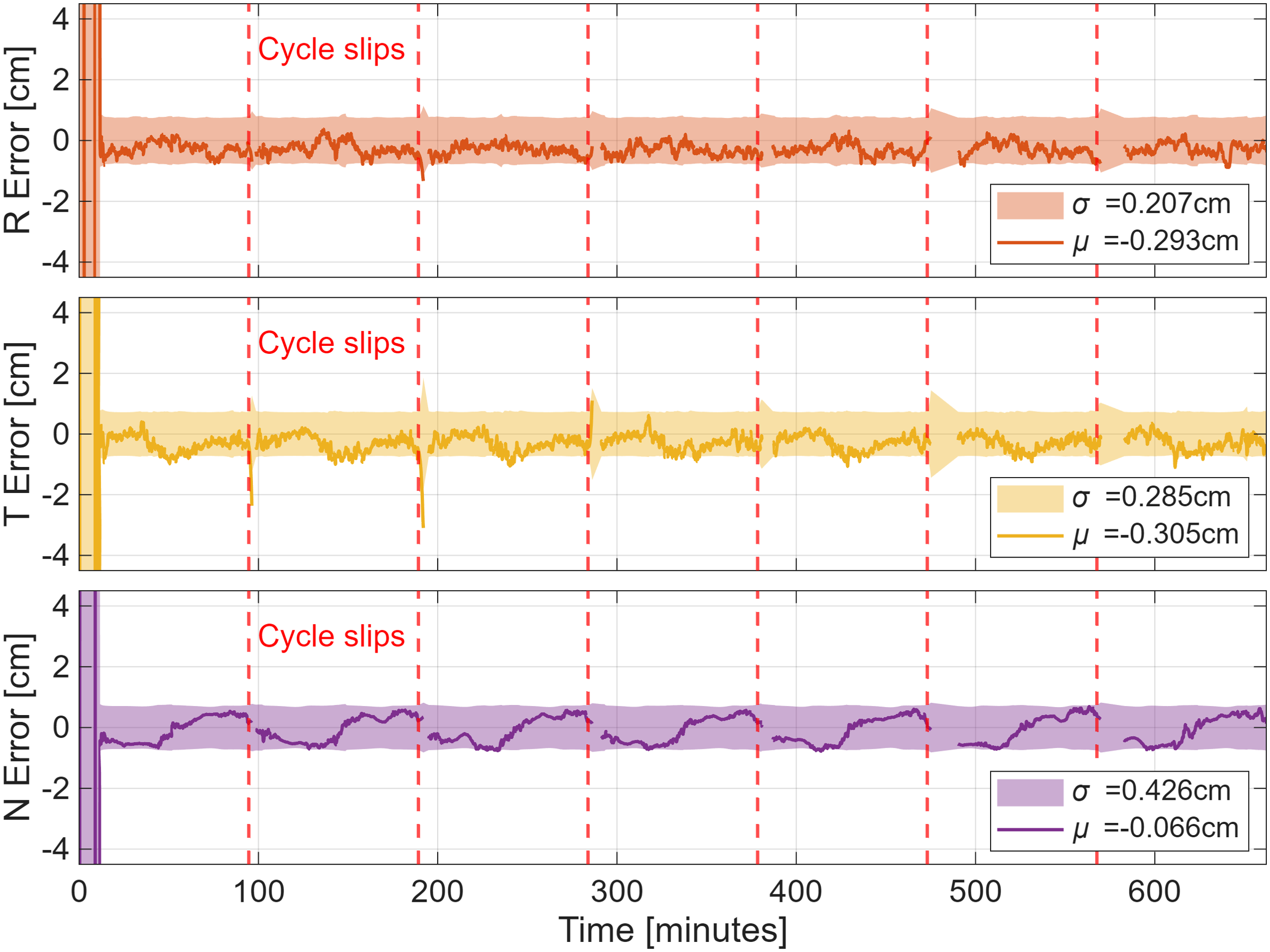}
        \caption{Relative position error, with cycle slips in dashed red}
        \label{fig:results-cycleslips-rel-pos}
    \end{subfigure}
    \hfill
    \begin{subfigure}[htp]{0.49\textwidth}
        \centering
        \includegraphics[width=\textwidth]{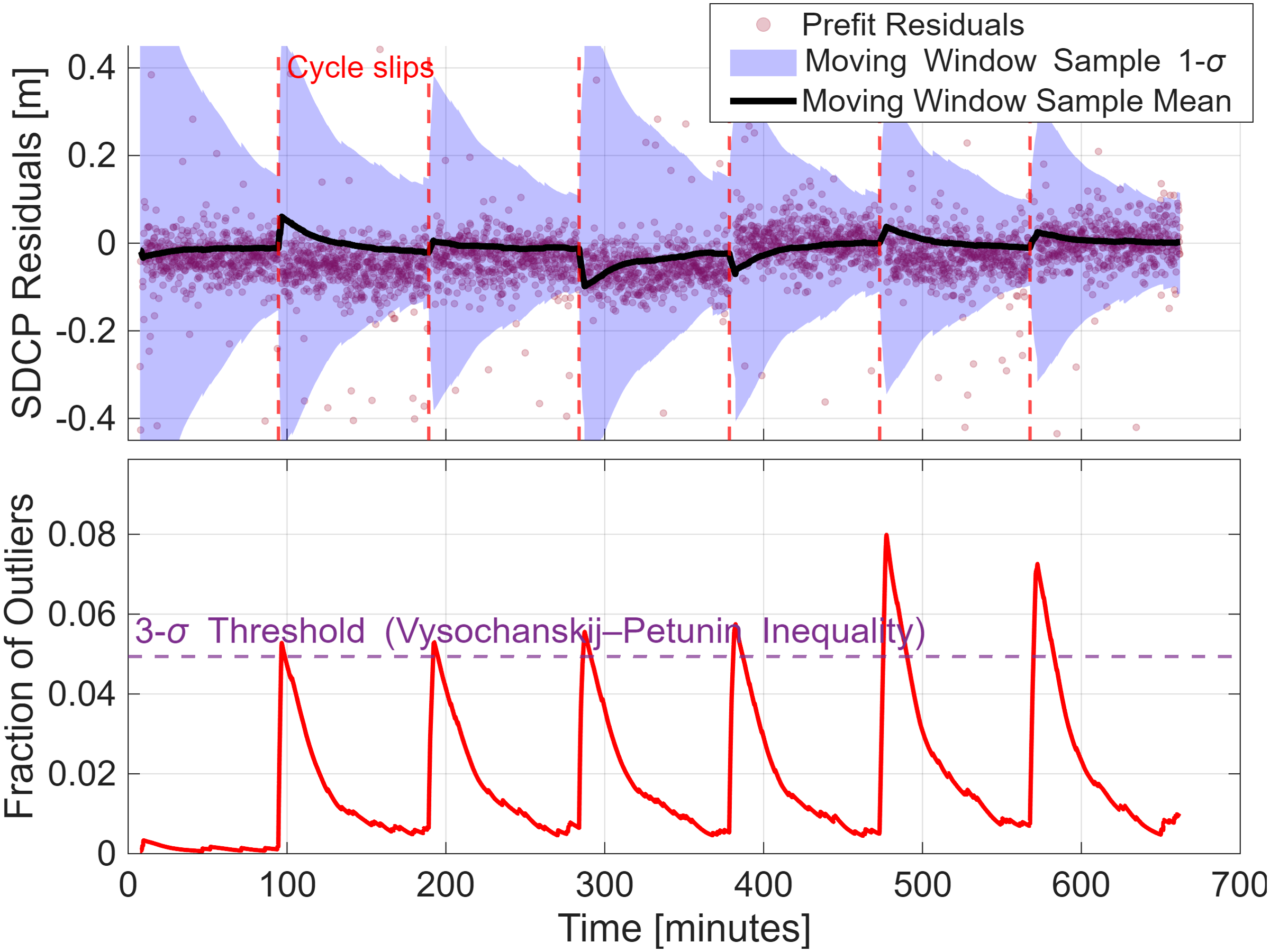}
        \caption{SDCP prefit residuals with variances (top), and moving-window outlier scores (bottom), logged by FDIR telemetry}
        \label{fig:results-cycleslips-fdir}
    \end{subfigure}
    \caption{Plots of relative navigation and FDIR performance, under full-cycle slip events once every orbit.}
    \label{fig:results-cycleslips}
\end{figure}

\paragraph{Software-in-the-Loop Simulated Cycle Slip Injection:}
Next, scenario \textbf{T.4.7} is presented.
Here, since carrier phase ambiguities are software-emulated with control over the true ambiguity values, repeated cycle slips are emulated by corrupting the ground truth ambiguity values once per orbit.
This causes sudden prefit residual outliers and large sample variances observed in \autoref{fig:results-cycleslips-fdir} by the moving-window.
The FDIR recovery logic executes successfully as soon as the Vysochanskij-Petunin concentration inequality bounds are violated.
This exercises \hyperref[alg:prefit-stats]{Algorithm 1} in simulated flight, as an extension of the FDIR experiment done in \autoref{fig:fdir-score}.
Evidently, the relative position estimates continue meeting the sub-cm accuracy requirements despite a full cycle-slip event on all channels every orbit, at the expense of several minutes of invalid state estimation, due to the filter engaging in recovery.

% ========================================
% 6.4 TEST RESULTS: SUMMARY OF RESULTS
% ========================================

\subsection{Comparative Performance and Runtime Analyses with Predecessor Software}

A comparative analyses of performance (\autoref{tab:summary-perf}) and runtime (\autoref{tab:summary-runtime}) was conducted between DiGiTaL v2 and its predecessor \cite{giralo2021digital}, using Test Campaign \textbf{T.3.2} and \textbf{T.4.1}. To ensure a fair ceteris paribus comparison between two versions of the software, \textit{all} modeling parameters (dynamics, measurements, simulation environment) applied are identical to both. The same Mersenne Twister seeds are also applied in the simulation environment to ensure identical pseudorandom inputs that stimulate the flight software. The computer-under-test, which runs the flight software, is a \textit{Xilinx Zynq-7000} System-on-Chip, an exact representation of the VISORS flight computer. It is a two-core embedded processor, ARM Cortex-A9, 650 MHz, where the operating system is PetaLinux 2016.2, running on Linux Kernel v4.4.

\begin{table}[H]
\centering
\renewcommand{\arraystretch}{1}
\setlength{\tabcolsep}{4pt}
\footnotesize
\caption{Root-mean-square errors (RMSE) computed, with VISORS observation success rates, using legacy (v1) and improved (v2) filter designs. Outage scenario follows \textbf{T.4.1}, with RMSE computed during the outage.}
\label{tab:summary-perf}
\begin{tabularx}{\textwidth}{M{1.25cm} M{3.0cm} M{1.1cm}M{1.1cm} M{1.1cm} M{1.1cm} M{1.8cm} M{1.8cm} M{1.7cm}}
\toprule
    & \textbf{Filter Type}
    & \textbf{Abs. Pos. RMSE [meters]}
    & \textbf{Abs. Vel. RMSE [mm/s]}
    & \textbf{Rel. Pos. RMSE [mm]}
    & \textbf{Rel. Vel. RMSE [mm/s]}
    & \textbf{Rel. Pos. (outage) RMSE [m]}
    & \textbf{Rel. Vel. (outage) RMSE [mm/s]}
    & \textbf{VISORS Observation Success} \\
\midrule
    \multirow{3}{*}{DiGiTaL v1} 
    & EKF
    & 12.409
    & 15.561
    & 1.828
    & 0.041
    & 55.838
    & 45.802
    & 1 / 10 \\
    \arrayrulecolor{gray}
    \cmidrule(l){2-9}
    \arrayrulecolor{black}
    & EKF-UKF Hybrid
    & 10.388
    & 13.551
    & 1.820
    & 0.041
    & 53.436
    & 44.745
    & 1 / 10 \\
\midrule
    \multirow{2}{*}{DiGiTaL v2} 
    & Regularized CEKF \linebreak (Joseph Form)
    & 1.950
    & 8.764
    & 1.790
    & 0.040
    & 1.054
    & 1.204
    & 2 / 10 \\
    \arrayrulecolor{gray}
    \cmidrule(l){2-9}
    \arrayrulecolor{black}
    & Sparse Regularized CEKF \linebreak (Non-Joseph-Form)
    & 1.934
    & 8.762
    & 1.813
    & 0.041
    & 1.196
    & 1.289
    & 2 / 10 \\
\bottomrule
\end{tabularx}
\end{table}

\vspace{-3mm}

\begin{table}[H]
\centering
\renewcommand{\arraystretch}{1}
\setlength{\tabcolsep}{4pt}
\footnotesize
\caption{Analyses of averaged runtimes, units in [ms] for core navigation function calls on the VISORS computers (Xilinx Zynq-7000 SoC). IAR runtime is recorded as the time of largest batch size to fix.}
\label{tab:summary-runtime}
\begin{tabularx}{\textwidth}{M{1.25cm} M{3.0cm} M{2.2cm} M{2cm} M{2cm} M{2cm} M{2cm}}
\toprule
    & \textbf{Filter Type}
    & \textbf{Time Update [ms]}
    & \textbf{GRAPHIC [ms]}
    & \textbf{SDCP [ms]}
    & \textbf{IAR (max) [ms]}
    & \textbf{Total [ms]} \\
\midrule
    \multirow{3}{*}{DiGiTaL v1} 
    & EKF
    & 1.384
    & 14.284
    & 14.697
    & 0.861
    & 31.226 \\
    \arrayrulecolor{gray}
    \cmidrule(l){2-7}
    \arrayrulecolor{black}
    & EKF-UKF Hybrid
    & 1.386
    & 20.806
    & 34.226
    & 0.863
    & 57.281 \\
\midrule
    \multirow{2}{*}{DiGiTaL v2} 
    & Regularized CEKF \linebreak (Joseph Form)
    & 1.414
    & 19.208
    & 19.726
    & 0.908
    & 41.256 \\
    \arrayrulecolor{gray}
    \cmidrule(l){2-7}
    \arrayrulecolor{black}
    & Sparse Regularized CEKF \linebreak (Non-Joseph-Form)
    & 1.405
    & 1.043
    & 1.515
    & 0.884
    & 4.847 \\
\bottomrule
\end{tabularx}
\end{table}

In terms of accuracy, the filter designs of DiGiTaL v2 exceed its predecessor in all areas in both nominal and off-nominal (outage) conditions. Significant improvements were observed in open-loop orbit prediction accuracy and absolute position estimation, by virtue of the improved process noise modeling. Marginal improvements were observed in relative state accuracy.
In terms of runtime, a trade-off exists where the Sparse-Regularized CEKF in non-Joseph form is significantly faster and more memory-efficient but does not provide a mathematical guarantee on a positive definite state covariance. In contrast, the Joseph form does do so, but at the cost of significantly slower execution speed, comparable to its predecessor, because matrix terms in Joseph form are dense. This choice depends on the balance between efficiency and robustness. In light of this, the software has enabled a feature to toggle on or off the Joseph-form of the CEKF, via telecommands. Either way, runtimes still meet state update interval requirements for VISORS.

% ========================================
% 6. CONCLUSION
% ========================================

\vspace{-2mm}
\section{Conclusion}
\label{section7}

This paper presents a complete CDGNSS-based navigation flight software package tailored for high-precision DSS missions. A rigorous, requirements-driven design process led to DiGiTaL v2, an end-to-end architecture emphasizing numerical stability, computational efficiency, and fault detection with isolation and recovery. The modular software framework and optimization strategies facilitate integration into emerging DSS host systems. Performance is validated through a multi-stage test campaign culminating in flight-like results from the VISORS mission. A carefully designed test campaign, with graduated fidelity, enables rapid failure point traceability via bisection. Comprehensive documentation of the design, development, and outcomes of this work aims to position the DiGiTaL v2 as a reference for emerging missions with stringent relative navigation needs. Viable research directions to improve the software in terms of scalability, robustness, and fault tolerance have been identified.

\vspace{2mm}

\textbf{Acknowledgements:} The authors gratefully acknowledge the support of the the NSF (Award No. 1936663), as well as DSO National Laboratories for fellowship support. The authors also extend their gratitude towards members of the VISORS team at large, as well as our reviewers: Adam Koenig (Relativity Space), Bo Naasz (NASA) and Christopher Roscoe (Ten One Aerospace).

% % ========================================
% % 7. APPENDIX
% % ========================================

% \pagebreak
% \section{Appendix}
% \label{appendix}

% % ========================================
% % 7.1 NAVIGATION SOFTWARE: EXECUTION LOGIC
% % ========================================

% \subsection{Navigation Software Execution Logic}

% \begin{figure}[H]
% 	\centering
%     \includegraphics[width=0.8\textwidth]{software-logic-gnss.png}
%     \caption{On-Receive of \textbf{local} GNSS packets, with time-matched \textbf{remote} GNSS packets present in queue.}
%     \label{fig:software-logic-gnss}
% \end{figure}

% \vspace{-5mm}

% \begin{figure}[H]
% 	\centering
%     \includegraphics[width=0.8\textwidth]{software-logic-xlink.png}
%     \caption{On-Receive of \textbf{remote} GNSS packets, with time-matched \textbf{local} GNSS packets present in queue.}
%     \label{fig:software-logic-xlink}
% \end{figure}

% \vspace{-5mm}

% \begin{figure}[H]
% 	\centering
%     \includegraphics[width=0.8\textwidth]{software-logic-gnss-no-xlink.png}
%     \caption{On-Receive of local GNSS packets under a crosslink outage, where remote packets have expired past their buffer period.}
%     \label{fig:software-logic-gnss-no-xlink}
% \end{figure}

% \vspace{-5mm}
% \pagebreak

% ========================================
% BIBLIOGRAPHY
% ========================================

\newgeometry{letterpaper, left=1.5cm, right=1.5cm, top=1.5cm, bottom=2.5cm}
\begin{multicols}{2}
\footnotesize
\bibliography{references}

@inproceedings{corazzini1997dgps,
  title={GPS sensing for spacecraft formation flying},
  author={Corazzini, Tobe’ and Robertson, Andrew and Adams, John Carl and Hassibi, Arash and How, Jonathan P},
  booktitle={Proceedings of the 10th International Technical Meeting of the Satellite Division of The Institute of Navigation (ION GPS 1997)},
  pages={735--744},
  year={1997}
}

@inproceedings{inalhan2000dgps,
  title={Precise formation flying control of multiple spacecraft using carrier-phase differential GPS},
  author={Inalhan, Gokhan and Busse, Franz D and How, Jonathan P},
  booktitle={AAS/AIAA Space Flight Mechanics Meeting, Clearwater, FL},
  pages={23--26},
  year={2000}
}

@article{kroes2005grace,
  title={Precise GRACE baseline determination using GPS},
  author={Kroes, Remco and Montenbruck, Oliver and Bertiger, William and Visser, Pieter},
  journal={GPS Solutions},
  volume={9},
  pages={21--31},
  year={2005},
  publisher={Springer}
}

@phdthesis{gnc2010canx45,
  author={Roth,Niels H.},
  year={2010},
  school={University of Toronto},
  title={Navigation and Control Design for the CanX-4/-5 Satellite Formation Flying Mission},
  journal={ProQuest Dissertations and Theses},
  pages={142},
  isbn={978-0-494-85461-7},
  language={English},
}

@inproceedings{damico2010dgps,
  title={Differential GPS: An enabling technology for formation flying satellites},
  author={D’Amico, Simone and Montenbruck, Oliver},
  booktitle={Small Satellite Missions for Earth Observation: New Developments and Trends},
  pages={457--465},
  year={2010},
  organization={Springer}
}

@phdthesis{damico2010thesis,
  author = {D'Amico, Simone},
  year = {2010},
  school = {TU Delft},
  title = {Autonomous Formation Flying in Low Earth Orbit}
}

@article{montenbruck2011tsx,
  title={Carrier phase differential GPS for LEO formation flying--the PRISMA and TanDEM-X flight experience},
  author={Montenbruck, Oliver and D’Amico, Simone and Ardaens, Jean-Sebastien and Wermuth, Martin},
  journal={Paper AAS},
  pages={11--489},
  year={2011}
}

@article{damico2012safe,
  author = {D'Amico, S. and Ardaens, J.-S. and Larsson, R.},
  title = {Spaceborne Autonomous Formation-Flying Experiment on the PRISMA Mission},
  journal = {Journal of Guidance, Control, and Dynamics},
  volume = {35},
  number = {3},
  pages = {834-850},
  year = {2012},
  doi = {10.2514/1.55638}
}

@article{damico2013prisma,
title = {Autonomous formation flying based on GPS - PRISMA flight results},
  author = {Simone D'Amico and Jean-Sebastien Ardaens and Sergio {De Florio}},
  journal = {Acta Astronautica},
  volume = {82},
  number = {1},
  pages = {69-79},
  year = {2013},
  note = {6th International Workshop on Satellite Constellation and Formation Flying},
  issn = {0094-5765},
  doi = {https://doi.org/10.1016/j.actaastro.2012.04.033},
}

@article{llorente2013proba,
  title={PROBA-3: Precise formation flying demonstration mission},
  author={Llorente, J Salvador and Agenjo, Alfredo and Carrascosa, Carmelo and de Negueruela, Cristina and Mestreau-Garreau, Agnes and Cropp, Alexander and Santovincenzo, Andrea},
  journal={Acta Astronautica},
  volume={82},
  number={1},
  pages={38--46},
  year={2013},
  publisher={Elsevier}
}

@article{ardaens2013proba3gps,
  title={GPS-based relative navigation for the Proba-3 formation flying mission},
  author={Ardaens, Jean-S{\'e}bastien and D'Amico, Simone and Cropp, Alexander},
  journal={Acta Astronautica},
  volume={91},
  pages={341--355},
  year={2013},
  publisher={Elsevier}
}

@inproceedings{koenig2015mdot,
  title={Formation design analysis for a miniaturized distributed occulter/telescope in earth orbit},
  author={Koenig, Adam W and D’Amico, Simone and Macintosh, Bruce and Titus, Charles J},
  booktitle={International Symposium on Space Flight Dynamics (ISSFD)},
  year={2015},
  organization={DLR German Space Operations Center and the European Space Agency},
}

@inproceedings{bowen2015cpod,
  title={Cubesat proximity operations demonstration (cpod) mission update},
  author={Bowen, John and Tsuda, Al and Abel, John and Villa, Marco},
  booktitle={2015 IEEE Aerospace Conference},
  pages={1--8},
  year={2015},
  organization={IEEE}
}

@article{roscoe2018cpod,
  title={Overview and GNC design of the CubeSat Proximity Operations Demonstration (CPOD) mission},
  author={Roscoe, Christopher WT and Westphal, Jason J and Mosleh, Ehson},
  journal={Acta Astronautica},
  volume={153},
  pages={410--421},
  year={2018},
  publisher={Elsevier}
}

@article{kahr2018canx45,
  author = {Kahr, Erin and Roth, Niels and Montenbruck, Oliver and Risi, Ben and Zee, Robert E.},
  title = {GPS Relative Navigation for the CanX-4 and CanX-5 Formation-Flying Nanosatellites},
  journal = {Journal of Spacecraft and Rockets},
  volume = {55},
  number = {6},
  pages = {1545-1558},
  year = {2018},
  doi = {10.2514/1.A34117},
}

@article{koenig2018safety,
  author = {Koenig, Adam W. and D’Amico, Simone},
  title = {Robust and Safe N-Spacecraft Swarming in Perturbed Near-Circular Orbits},
  journal = {Journal of Guidance, Control, and Dynamics},
  volume = {41},
  number = {8},
  pages = {1643-1662},
  year = {2018},
  doi = {10.2514/1.G003249},
}

@inproceedings{enderle2019proba,
  title={Proba-3 precise orbit determination based on GNSS observations},
  author={Enderle, Werner and Gini, Francesco and Sch{\"o}nemann, Erik and Mayer, Volker},
  booktitle={Proceedings of the 32nd International Technical Meeting of the Satellite Division of The Institute of Navigation (ION GNSS+ 2019)},
  pages={1187--1198},
  year={2019}
}

@article{kornfield2019gracefo,
  author = {Kornfeld, Richard P. and Arnold, Bradford W. and Gross, Michael A. and Dahya, Neil T. and Klipstein, William M. and Gath, Peter F. and Bettadpur, Srinivas},
  title = {GRACE-FO: The Gravity Recovery and Climate Experiment Follow-On Mission},
  journal = {Journal of Spacecraft and Rockets},
  volume = {56},
  number = {3},
  pages = {931-951},
  year = {2019},
  doi = {10.2514/1.A34326},
}

@article{giralo2019digital,
  author = {Giralo, Vincent Paul and D’Amico, Simone},
  title = {Distributed multi-GNSS timing and localization for nanosatellites},
  journal = {NAVIGATION},
  volume = {66},
  number = {4},
  pages = {729-746},
  doi = {10.1002/navi.337},
  year = {2019}
}

@article{monnier2019roadmap,
  title={A realistic roadmap to formation flying space interferometry},
  author={Monnier, John D and others},
  journal={arXiv preprint arXiv:1907.09583},
  year={2019}
}

@inproceedings{giralo2020dwarf,
  title={Guidance, navigation, and control for the DWARF formation-flying mission},
  author={Giralo, Vincent Paul and Chernick, Michelle and D’Amico, Simone},
  booktitle={AAS/AIAA Astrodynamics Specialist Conference, South Lake Tahoe, CA},
  year={2020},
  pages={0}
}

@phdthesis{giralo2021digital,
  title={Precision Navigation of Miniaturized Distributed Space Systems using GNSS},
  author={Giralo, Vincent Paul},
  year={2021},
  school={Stanford University},
}

@article{xia2021gracefo,
  title={On GPS data quality of GRACE-FO and GRACE satellites: Effects of phase center variation and satellite attitude on precise orbit determination},
  author={Xia, Yaowei and Liu, Xin and Guo, Jinyun and Yang, Zhouming and Qi, Linhu and Ji, Bing and Chang, Xiaotao},
  journal={Acta geodaetica et geophysica},
  volume={56},
  pages={93--111},
  year={2021},
  publisher={Springer}
}

@inproceedings{giralo2021mdot,
  title={Precise real-time relative orbit determination for large-baseline formations using GNSS},
  author={Giralo, Vincent and D’Amico, Simone},
  booktitle={Proceedings of the 2021 International Technical Meeting of The Institute of Navigation},
  pages={366--384},
  year={2021}
}

@inbook{visors2021koenig,
  author = {Adam Koenig and Simone D'Amico and E Glenn Lightsey},
  title = {Formation Flying Orbit and Control Concept for the VISORS Mission},
  booktitle = {AIAA Scitech 2021 Forum},
  publisher = {AIAA},
  year = {2021},
  chapter = {0},
  pages = {0},
  doi = {10.2514/6.2021-0423},
}

@incollection{scala2021gnss,
  title = {GNSS-based navigation for a remote sensing three-satellite formation flying},
  author = {Scala, F and Colombo, C and Gaias, GVM and Martin-Neira, M and others},
  booktitle = {SpaceOps 2021 Virtual Edition},
  publisher = {16th International Conference on Space Operations},
  pages = {1--18},
  year = {2021}
}

@inproceedings{low2022pointing,
  title={Optimal Pointing Sequences in Spacecraft Formation Flying using Online Planning with Resource Constraints},
  author={Low, Samuel and Kochenderfer, Mykel},
  booktitle={Learning for Dynamics and Control Conference},
  pages={355--365},
  year={2022},
  organization={PMLR}
}

@inproceedings{lowe2022errbudget,
  title={Relative navigation and pointing error budget for an x-ray astronomy formation-flying mission},
  author={Lowe, Shane and Markevitch, Maxim and D’Amico, Simone},
  booktitle={Proceedings of the 44th Annual American Astronautical Society Guidance, Navigation, and Control Conference, 2022},
  pages={1433--1445},
  year={2022},
  organization={Springer}
}

@techreport{damicos2022mdot,
  author = {Macintosh, Bruce and D’Amico, Simone and Koenig, Adam and Bendek, Eduardo and Grogran, Keith and Shaklan, Stuart and Madurowicz, A. and de Rosa, R. and Greene, T. and Debes, J. and Douglas, E. and Jensen-Clem, R. and Duchene, G. and Esposito,  T.},
  title = {Miniature Distributed Occulter Telescope (mDOT) Publicly Released Project Report},
  institution = {Stanford University, Space Rendezvous Laboratory},
  year = {2022},
}

@inproceedings{visors2023aas,
  author = {Guffanti, Tommaso and Bell, Toby and Low, Samuel Y. W. and Murray-Cooper, Mason and D'Amico, Simone},
  title = {Autonomous Guidance, Navigation and Control of the VISORS Formation Flying Mission},
  booktitle = {AAS/AIAA Astrodynamics Specialist Conference},
  year = {2023},
  address = {Big Sky, Montana},
  pages={0},
}

@inproceedings{monnier2024stari,
  title={STARI: starlight acquisition and reflection toward interferometry},
  author={Monnier, John D and Jain, Prachet and Kalluri, Shashank and Cutler, James and D'Amico, Simone and Lightsey, Glenn and Pogorelyuk, Leonid and Vasisht, Gautam and Cahoy, Kerri and Meyer, Michael},
  booktitle={Space Telescopes and Instrumentation 2024: Optical, Infrared, and Millimeter Wave},
  volume={13092},
  pages={1095--1107},
  year={2024},
  organization={SPIE}
}

@inproceedings{kruger2024rpokit,
  title={Adaptive End-to-End Architecture for Autonomous Spacecraft Navigation and Control During Rendezvous and Proximity Operations},
  author={Kruger, Justin J and Guffanti, Tommaso and Park, Tae Ha and Murray-Cooper, Mason and Low, Samuel YW and Bell, Toby and D'Amico, Simone and Roscoe, Christopher W and Westphal, Jason},
  booktitle={AIAA SCITECH 2024 Forum},
  pages={0430},
  year={2024}
}

@article{shim2024precise,
  title={Precise in-orbit relative navigation technique for rendezvous mission of CubeSats using only GPS receivers},
  author={Shim, Hanjoon and Kim, O-Jong and Yu, Sunkyoung and Kee, Changdon and Cho, Dong-Hyun and Kim, Hae-Dong},
  journal={CEAS Space Journal},
  volume={16},
  number={1},
  pages={117--137},
  year={2024},
  publisher={Springer}
}

@article{shim2024rtk,
  title={Highly Efficient Real-Time Kinematic-Based Precise Relative Navigation for Autonomous Rendezvous CubeSat},
  author={Shim, Hanjoon and Kee, Changdon},
  journal={NAVIGATION: Journal of the Institute of Navigation},
  volume={71},
  number={3},
  year={2024},
  publisher={Institute of Navigation}
}

@inproceedings{hwang2025snuglite3,
  title={Propellant-Free Rendezvous Mission of SNUGLITE-III CubeSat: Orbit Control Using Aerodynamic Forces},
  author={Hwang, Jae Woong and Shim, Hanjoon and Bae, Yonghwan and Kee, Changdon and Kim, Jaegang},
  booktitle={2025 IEEE Aerospace Conference},
  pages={1--12},
  year={2025},
  organization={IEEE}
}

@article{ito2025silvia,
  title={SILVIA: Ultra-precision formation flying demonstration for space-based interferometry},
  author={Ito, Takahiro and Izumi, Kiwamu and Kawano, Isao and Funaki, Ikkoh and Sato, Shuichi and Akutsu, Tomotada and Komori, Kentaro and Musha, Mitsuru and Michimura, Yuta and Satoh, Satoshi and others},
  journal={arXiv preprint arXiv:2504.05001},
  year={2025}
}

@book{kapurch2010nasa,
  title={NASA systems engineering handbook},
  author={Kapurch, Stephen J},
  year={2010},
  publisher={Diane Publishing}
}

@incollection{schmidt1966cekf,
  title={Application of state-space methods to navigation problems},
  author={Schmidt, Stanley F},
  booktitle={Advances in control systems},
  volume={3},
  pages={293--340},
  year={1966},
  publisher={Elsevier}
}

@article{schlee1967divergence,
  author = {Schlee, F. H. and Standih, C. J. and Toda, N. F.},
  title = {Divergence in the Kalman filter.},
  journal = {AIAA Journal},
  volume = {5},
  number = {6},
  pages = {1114-1120},
  year = {1967},
  doi = {10.2514/3.4146},
}

@book{jazwinski1970filter,
  title={Stochastic processes and filtering theory},
  author={Jazwinski, Andrew H},
  year={1970},
  publisher={Academic Press, Inc.,}
}

@article{carpenter2005navigation,
  title={Navigation accuracy guidelines for orbital formation flying},
  author={Carpenter, J Russell and Alfriend, Kyle T},
  journal={The Journal of the Astronautical Sciences},
  volume={53},
  pages={207--219},
  year={2005},
  publisher={Springer}
}

@article{jonhow2007sma,
  title={Differential semimajor axis estimation performance using carrier-phase differential global positioning system measurements},
  author={How, Jonathan P and Breger, Louis S and Mitchell, Megan and Alfriend, Kyle T and Carpenter, Russell},
  journal={Journal of guidance, control, and dynamics},
  volume={30},
  number={2},
  pages={301--313},
  year={2007}
}

@article{zanetti2010underweight,
  author = {Zanetti, Renato and DeMars, Kyle J. and Bishop, Robert H.},
  title = {Underweighting Nonlinear Measurements},
  journal = {Journal of Guidance, Control, and Dynamics},
  volume = {33},
  number = {5},
  pages = {1670-1675},
  year = {2010},
  doi = {10.2514/1.50596},
}

@inbook{woodbury2010cekf,
  author = {Drew Woodbury and John Junkins},
  title = {On the Consider Kalman Filter},
  booktitle = {Guidance, Navigation, and Control Conference},
  publisher = {AIAA},
  year = {2010},
  chapter = {0},
  pages = {0},
  doi = {10.2514/6.2010-7752},
}

@techreport{nasa2018filter,
  title={Navigation filter best practices},
  author={Carpenter, J Russell and D’souza, Christopher N},
  year={2018},
  institution={NASA},
}

@article{stacey2021adaptive,
  title={Adaptive and dynamically constrained process noise estimation for orbit determination},
  author={Stacey, Nathan and D’Amico, Simone},
  journal={IEEE Transactions on Aerospace and Electronic Systems},
  volume={57},
  number={5},
  pages={2920--2937},
  year={2021},
  publisher={IEEE}
}

@article{stacey2022noise,
  title={Analytical process noise covariance modeling for absolute and relative orbits},
  author={Stacey, Nathan and D’Amico, Simone},
  journal={Acta Astronautica},
  volume={194},
  pages={34--47},
  year={2022},
  publisher={Elsevier}
}

@misc{low2025cekf,
  author={Low, Samuel YW and D’Amico, Simone},
  title={The Consider Kalman Filter in Joseph Form},
  howpublished={Stanford Space Rendezvous Laboratory},
  year={2025},
  note={Internal technical note.}
}

@article{yunck1993graphic,
  title={Coping with the atmosphere and ionosphere in precise satellite and ground positioning},
  author={Yunck, Thomas P},
  journal={Washington DC American Geophysical Union Geophysical Monograph Series},
  volume={73},
  pages={1--16},
  year={1993}
}

@article{psiaki2007cdgps,
  author = {Psiaki, Mark L. and Mohiuddin, Shan},
  title = {Modeling, Analysis, and Simulation of GPS Carrier Phase for Spacecraft Relative Navigation},
  journal = {Journal of Guidance, Control, and Dynamics},
  volume = {30},
  number = {6},
  pages = {1628-1639},
  year = {2007},
  doi = {10.2514/1.29534},
}

@article{gutsche2024pco,
  title={Addressing Inaccurate Phase Center Offsets in Precise Orbit Determination for Agile Satellite Missions},
  author={Gutsche, Kevin and Hobiger, Thomas and Winkler, Stefan},
  journal={NAVIGATION: Journal of the Institute of Navigation},
  volume={71},
  number={4},
  year={2024},
  publisher={Institute of Navigation}
}

@book{gpstextbook2006,
  title={The Global Positioning System: Signals, Measurements, and Performance (Second Edition)},
  author={Enge, Per K and Misra, Pratap},
  year={2006},
  publisher={Ganga-Jamuna Press}
}

@book{montenbruck2010satellite,
  title={Satellite Orbits: Models, Methods and Applications},
  author={Montenbruck, Oliver and Gill, Eberhard},
  year={2010},
  publisher={Springer}
}

@book{gallager2013stochastic,
  title={Stochastic processes: theory for applications},
  author={Gallager, Robert G},
  year={2013},
  publisher={Cambridge University Press}
}

@inproceedings{teunissen1994lambda,
  title={A new method for fast carrier phase ambiguity estimation},
  author={Teunissen, PJG},
  booktitle={Proceedings of 1994 IEEE Position, Location and Navigation Symposium-PLANS'94},
  pages={562--573},
  year={1994},
  organization={IEEE}
}

@article{teunissen1995invertible,
  title={The invertible GPS ambiguity transformations},
  author={Teunissen, Peter},
  journal = {Manuscripta Geodetica},
  volume = {20},
  pages = {489--497},
  year={1995}
}

@inproceedings{teunissen1997adop,
  title={Ambiguity dilution of precision: definition, properties and application},
  author={Teunissen, Peter JG and Odijk, Dennis},
  booktitle={Proceedings of the 10th International Technical Meeting of the Satellite Division of The Institute of Navigation (ION GPS 1997)},
  pages={891--899},
  year={1997}
}

@article{teunissen1998success,
  title={Success probability of integer GPS ambiguity rounding and bootstrapping},
  author={Teunissen, PJG},
  journal={Journal of geodesy},
  volume={72},
  pages={606--612},
  year={1998},
  publisher={Springer}
}

@article{chang2005mlambda,
  title={MLAMBDA: A modified LAMBDA method for integer least-squares estimation},
  author={Chang, X -W and Yang, X and Zhou, T},
  journal={Journal of Geodesy},
  volume={79},
  pages={552--565},
  year={2005},
  publisher={Springer}
}

@inproceedings{low2024digital,
  title={Precise Distributed Satellite Navigation: Differential GPS with Sensor-Coupling for Integer Ambiguity Resolution},
  author={Low, Samuel YW and D’Amico, Simone},
  booktitle={2024 IEEE Aerospace Conference},
  pages={1--18},
  year={2024},
  organization={IEEE}
}

@inproceedings{toby2025sim,
  title={Event-Driven Simulation for Rapid Iterative Development of Distributed Space Flight Software},
  author={Bell, Toby and D'Amico, Simone},
  booktitle={2025 IEEE Aerospace Conference},
  pages={},
  year={2025},
  organization={IEEE}
}

@inproceedings{shoemake1985slerp,
  title={Animating rotation with quaternion curves},
  author={Shoemake, Ken},
  booktitle={Proceedings of the 12th annual conference on Computer graphics and interactive techniques},
  pages={245--254},
  year={1985}
}

@article{pukelsheim1994threesigma,
  title={The three sigma rule},
  author={Pukelsheim, Friedrich},
  journal={The American Statistician},
  volume={48},
  number={2},
  pages={88--91},
  year={1994},
  publisher={Taylor \& Francis}
}

@article{vysochanskij1980inequality,
  title={Justification of the 3$\sigma$ rule for unimodal distributions},
  author={Vysochanskij, DF and Petunin, Yu I},
  journal={Theory of Probability and Mathematical Statistics},
  volume={21},
  number={25-36},
  year={1980}
}
\end{multicols}
\restoregeometry

\end{document}